\documentclass[11pt,a4paper]{article}
\usepackage{epsfig,amsmath,amssymb,scalefnt,cite}

\usepackage{color}

\long\def\symbolfootnote[#1]#2{\begingroup%
\def\thefootnote{\fnsymbol{footnote}}\footnote[#1]{#2}\endgroup}

\newcommand{\abbrev}{\rm\scalefont{.9}}
\newcommand{\zwat}{{\abbrev ZWA}}
\newcommand{\sm}{{\abbrev SM}}
\newcommand{\bsm}{{\abbrev BSM}}
\newcommand{\thdm}{{\abbrev 2HDM}}
\newcommand{\mssm}{{\abbrev MSSM}}
\newcommand{\nmssm}{{\abbrev NMSSM}}
\newcommand{\cp}{{\abbrev CP}}

\newcommand{\lhc}{{\abbrev LHC}}
\newcommand{\ilc}{{\abbrev ILC}}
\newcommand{\lc}{{\abbrev LC}}
\newcommand{\vbf}{{\abbrev VBF}}
\newcommand{\met}{{\abbrev MET}}
\newcommand{\cms}{{\abbrev cms}}
\newcommand{\lhchxswg}{{\abbrev LHC-HXSWG}}

\newcommand{\smallz}{{\scriptscriptstyle Z}} 
\newcommand{\smallw}{{\scriptscriptstyle W}} %
\newcommand{\smallv}{{\scriptscriptstyle V}} %
\newcommand{\smallH}{{\scriptscriptstyle H}} %

\newcommand{\smallr}{{\scriptscriptstyle R}} %
\newcommand{\smalla}{{\scriptscriptstyle A}} %
\newcommand{\mz}{m_\smallz}
\newcommand{\mv}{m_\smallv}
\newcommand{\mzz}{m_{\smallz\smallz}}
\newcommand{\mzznr}{m_{\smallz_2\smallz_3}}
\newcommand{\mvv}{m_{\smallv\smallv}}
\newcommand{\mww}{m_{\smallw\smallw}}

\newcommand{\mw}{m_\smallw}
\newcommand{\mh}{m_\smallH}
\newcommand{\mr}{m_\smallr}
\newcommand{\mj}{m_{4j}}
\newcommand{\ptj}{p_{T,4j}}
\newcommand{\zh}{{\smallz\smallH}}
\newcommand{\nunuh}{{\nu\bar\nu\smallH}}
\newcommand{\zzz}{{\smallz\smallz\smallz}}
\newcommand{\zzznr}{{\smallz_1\smallz_2\smallz_3}}
\newcommand{\zvv}{{\smallz\smallv\smallv}}
\newcommand{\zww}{{\smallz\smallw\smallw}}
\newcommand{\hvv}{{\smallH\smallv\smallv}}
\newcommand{\nunuzz}{{\nu\bar\nu\smallz\smallz}}
\newcommand{\nunuww}{{\nu\bar\nu\smallw\smallw}}
\newcommand{\nunuvv}{{\nu\bar\nu\smallv\smallv}}

\newcommand{\GaH}{\Gamma_\smallH}
\newcommand{\ma}{m_\smalla}
\newcommand{\zwa}{{\rm\scriptscriptstyle ZWA}}
\newcommand{\lumi}{{\int\!\!\mathcal{L}dt}}

\newcommand{\eqn}[1]{Eq.\,(\ref{#1})}

\newcommand{\fig}[1]{Fig.\,\ref{#1}}

\newcommand{\tab}[1]{Tab.\,\ref{#1}}
\newcommand{\sct}[1]{Section~\ref{#1}}

\newcommand{\citere}[1]{Ref.~\cite{#1}}
\newcommand{\citeres}[1]{Refs.~\cite{#1}}

\begin{document}

\begin{titlepage}

{\flushright{
        \begin{minipage}{2.5cm}
          DESY 14-133
        \end{minipage}        }

}
\renewcommand{\thefootnote}{\fnsymbol{footnote}}
\vskip 2cm
\begin{center}
{\LARGE\bf Off-shell effects in Higgs processes at a linear\\[.3em]
collider and implications for the LHC}
\vskip 1.0cm
{\Large  Stefan Liebler$^{a}$, Gudrid Moortgat-Pick$^{a,b}$, 
Georg Weiglein$^{a}$}
\vspace*{8mm} \\
{\sl ${}^a$
DESY, Notkestra\ss e 85, \\
22607 Hamburg, Germany}
\vspace*{8mm} \\
{\sl ${}^b$
II. Institut f\"ur Theoretische Physik, Universit\"at Hamburg,\\ Luruper Chaussee 149\\
22761 Hamburg, Germany}
\end{center}
\symbolfootnote[0]{{\tt e-mail addresses:}} 
\symbolfootnote[0]{{\tt stefan.liebler@desy.de, gudrid.moortgat-pick@desy.de, georg.weiglein@desy.de}}

\vskip 0.7cm

\begin{abstract}
The importance of off-shell contributions is discussed for $H\rightarrow
VV^{(*)}$ with $V\in\lbrace Z,W\rbrace$ for large invariant masses
$\mvv$ involving a standard model (\sm{})-like Higgs boson with
$\mh=125$\,GeV at a linear collider (\lc{}). Both dominant
production processes $e^+e^-\rightarrow ZH\rightarrow ZVV^{(*)}$ and
$e^+e^-\rightarrow \nu\bar\nu H\rightarrow \nu\bar\nu VV^{(*)}$ 
are taken into account, and
the signal processes are compared with background yielding the same final
state.  The relative size of the off-shell contributions is strongly
dependent on the centre-of-mass energy. These
contributions can have an important impact on the
determination of cross sections and branching ratios. However,
the combination of on- and off-shell contributions can also be
utilised to lift degeneracies allowing to test higher-dimensional operators,
unitarity and light and heavy Higgs interferences in extended Higgs
sectors. The latter is demonstrated
in the context of a 2-Higgs-Doublet model.
We also discuss the impact of these aspects for 
the Large Hadron Collider (\lhc{}) where they are relevant.
The importance of a precise measurement of the Higgs mass for
on-shell contributions in $H\rightarrow VV^{(*)}$ is emphasized. 
A particular focus is put on
methods for extracting the Higgs width 
at a \lc{}. Off-shell
contributions are shown to have a negligible impact on the width
determination at low $\sqrt{s}$ 
when applying the $Z$ recoil method to extract branching ratios 
in combination with an appropriate determination of a partial width. On the
other hand, off-shell contributions can be exploited to constrain the
Higgs width in a similar fashion as in recent analyses at the \lhc{}. It is
demonstrated that this approach, besides relying heavily on theoretical
assumptions, is affected by the negative interference of
Higgs and background contributions that may limit the 
sensitivity that is achievable with the highest foreseeable
statistics at the \lhc{} and a \lc{}.
\end{abstract}
\vfill
\end{titlepage}    

\setcounter{footnote}{0}


\section{Introduction}
\label{sec:intro}

With the spectacular discovery of a signal in the Higgs searches at the Large Hadron
Collider (\lhc{}) \cite{Aad:2012tfa,Chatrchyan:2012ufa} particle physics
entered a new era. Detailed measurements of the mass, spin and CP
properties of the new particle, its couplings
to the other standard model (\sm{}) particles, its total width 
and its self-coupling will be crucial to reveal the underlying mechanism of
electroweak symmetry breaking. The determination of significant deviations
from the properties predicted in the \sm{} would have
profound implications in this context, indicating for instance the presence
of additional states of an extended Higgs sector or a composite nature of
the observed state. The required high-precision measurements will have to
be complemented with sufficiently accurate theory predictions for the
observables accessible with current and future experimental data.

A special role in this endeavour will be played by a future linear 
$e^+e^-$ collider
like the International Linear Collider (\ilc{}), for which a detailed technical
design report~\cite{Behnke:2013xla,Baer:2013cma,Adolphsen:2013jya,Adolphsen:2013kya,Behnke:2013lya,Moortgat-Picka:2015yla} exists.
In contrast to the \lhc{} a linear $e^+e^-$ collider will not only
improve significantly the precision on the Higgs couplings to
the \sm{} gauge bosons and fermions, but moreover will provide
a model-independent measurement of the width of the Higgs boson. For a \sm{} Higgs
boson of $\mh=125$\,GeV the latter is tiny, namely just
$\GaH^{\sm}=4.07$\,MeV~\cite{Dittmaier:2011ti,Dittmaier:2012vm,Heinemeyer:2013tqa}.
Since the width is orders of magnitude below the detector resolution
of high-energy physics experiments, an access to it by
a measurement of the differential distribution with respect to
the invariant mass of the decay products is impossible.

Recently off-shell contributions in $H\rightarrow VV^{(*)}$ with
$V\in\lbrace Z,W\rbrace$ at the \lhc{} attracted significant attention.  The term
``off-shell'' refers to contributions where the invariant mass $\mvv$ of
the two gauge bosons exceeds the Higgs mass, $\mvv\gg \mh$. A recent discussion
of the off-shell contributions for the dominant \lhc{} production
process can be found in
\citeres{Kauer:2012hd,Kauer:2013cga,Kauer:2013qba,Kauer:2015pma},
earlier works on Higgs boson contributions to $gg\rightarrow\gamma\gamma/VV$ in 
\citeres{Glover:1988rg,Dixon:2003yb,Campbell:2011cu}.
It is remarkable that in combination with on-shell contributions -- but
relying heavily on theoretical assumptions -- the off-shell contributions can be
used to constrain the Higgs width~\cite{Caola:2013yja}.
Meanwhile the theoretical knowledge with respect to higher
orders and jet vetoes was significantly improved for various final states in
\citeres{Campbell:2013una,Campbell:2013wga,Chen:2013waa,Moult:2014pja,
Campbell:2014gha,Campbell:2014gua,Campbell:2015vwa}.
An experimental analysis was recently presented by the CMS collaboration
quoting an upper bound of $4.2\cdot\GaH^{\sm}$ on the Higgs
boson width \cite{CMS:2014ala,Khachatryan:2014iha}.  The ATLAS
collaboration obtained a bound of $(4.8-7.7)\cdot \GaH^{\sm}$
\cite{ATLASwidth}.
Moreover off-shell decays of a Higgs boson can be utilised to put
constraints on higher-dimensional operators, as worked out
for the \lhc{} in \citeres{Gainer:2014hha,Ghezzi:2014qpa}.
\citeres{Azatov:2014jga,Cacciapaglia:2014rla,Buschmann:2014sia}
elaborate on the extraction of the top Yukawa coupling from the resolution
to long- and short-distance contributions to Higgs production via gluon fusion.

In this work we want to investigate
the off-shell effects in $H\rightarrow VV^{(*)}$
for a linear collider in different regimes of the centre-of-mass (\cms{}) energy and
as a function of the polarisation of the initial state.
In contrast to gluon fusion
at the \lhc{} the main production mechanisms at an $e^+e^-$ collider
occur already at tree-level, namely Higgsstrahlung, where the Higgs
is radiated of a $Z$~boson, and
vector-boson-fusion (\vbf{}), where the Higgs stems from the fusion
of two weak gauge bosons.
The discussion of off-shell effects in $H\rightarrow VV^{(*)}$ is closely
related to the inadequacy of the zero-width approximation (\zwat{})
for the two production processes.
We quantify the size of the off-shell effects for both processes
and analyse their significance with respect to the background yielding
the same final state.
Based on those results the impact of off-shell contributions
is discussed in various contexts:
\begin{itemize}
\item[$\triangleright$] Depending on the \cms{} energy
we investigate their influence 
on the measurements of cross sections and branching ratios
and on the determination of Higgs-boson properties 
that can be inferred from those measurements. The impact of the 
off-shell contributions on the extraction of
Higgs couplings at low \cms{} energies~$\sqrt{s}$,
i.e.\ close to the production threshold, is small.
At high $\sqrt{s}$, on the other hand, their influence increases, and
they can be utilised to test higher-dimensional operators and to check the
destructive interference between Higgs and background contributions
at high invariant masses $\mvv$.
In extended Higgs sectors light and heavy Higgs contributions
can interfere over a large range of $\mvv$,
which we demonstrate in the context of a 2-Higgs-Doublet model (\thdm).
\item[$\triangleright$] We show the importance of a precise Higgs mass measurement
for the on-shell Higgs contributions in $H\rightarrow VV^{(*)}$, whereas
off-shell contributions are mostly insensitive to the precise numerical
value of the Higgs mass.
\item[$\triangleright$] We also discuss off-shell contributions in the context
of the Higgs width determination at a linear collider.
The state-of-the-art method is based on the detection of
the $Z$ decay products in
Higgsstrahlung ($Z$ recoil method) yielding an absolute measurement of
Higgs branching ratios in combination with an appropriate determination of a
partial width. With this procedure the Higgs width can be
determined in a model-independent way,
see~\citeres{Li:2009fc,Li:2010wu,Li:2012taa,Han:2013kya,Durig:2014lfa},
and a high precision for the Higgs width is achieved. This method
is affected by
off-shell effects for low \cms{} energies only at the sub-permil level.
Bounding the Higgs width from the combination of on-shell and off-shell 
contributions, on the other hand,
is mainly limited by the destructive
interference of Higgs and background contributions.
\end{itemize}
For quantitative statements we use
{\tt MadGraph5\_aMC@NLO}~\cite{Alwall:2014hca} to simulate the
full processes $e^+e^-\rightarrow 6$\,fermions.
In our discussion we also address the choice of the Higgs propagator,
the inclusion of initial state radiation and beamstrahlung, as well
as higher order contributions.

The outline of the paper is as follows:
In \sct{sec:basics} we start our discussion with a short description
of the Higgs propagator structure.
Then we quantify off-shell contributions in $e^+e^-\rightarrow ZH$
and $e^+e^-\rightarrow \nu\bar\nu H$ followed by $H\rightarrow ZZ^{(*)}$
and $H\rightarrow W^\pm W^{\mp(*)}$ in \sct{sec:offVV}.
The issue of gauge dependence is discussed in this context.
The off-shell contributions are quantified with respect to the on-shell
ones and with respect to the background.
Of particular importance is the mainly destructive interference
between Higgs induced contributions and the background.
We investigate different \cms{} energies and the dependence
on the polarisation of the initial state, as well as the dependence
on the precise numerical value of the Higgs mass. In \sct{sec:practical} we
address the impact on the $Z$ recoil method and the extraction
of Higgs to gauge boson couplings at an $e^+e^-$ collider.
After a brief discussion of the tests of
unitarity and the sensitivity to higher-dimensional operators
we subsequently describe the two processes $e^+e^-\rightarrow
\nu\bar\nu+4$\,jets and $e^+e^-\rightarrow \mu^+\mu^-+4$\,jets
and their dependence on Higgs induced contributions in \sct{sec:example1}.
We discuss the inclusion
of initial state radiation and beamstrahlung as well as higher
order effects in \sct{sec:ISRNLO}. The latter should be taken
into account in future analyses. We then
investigate the sensitivity of the two above example processes
on the Higgs width in \sct{sec:width} and comment on the limitations
of the method in case of the \lhc{}.
Finally we discuss the interference of an on-shell heavy Higgs
with the off-shell light Higgs contributions in the context
of a 2-Higgs-Doublet model in \sct{sec:2hdm}.
We conclude in \sct{sec:conclusions}.

\section{Relation of the Higgs mass and width to the complex pole of the propagator}
\label{sec:basics}

Before we start our discussion of off-shell effects in $H\rightarrow VV^{(*)}$
in the subsequent section, we shortly elaborate on the relation between the 
mass and total width of the Higgs boson and the complex pole of the propagator.
Denoting with $m_0$ the tree-level Higgs mass and with $\hat{\Sigma}$ the renormalized
self-energy of the Higgs propagator, the complex pole is obtained
through the relation $\mathcal{M}^2-m_0^2+\hat{\Sigma}(\mathcal{M}^2)=0$,
where the complex pole can be written
in the form $\mathcal{M}^2=\mh^2-i\mh\GaH$.
Therein $\mh$ is the physical Higgs mass and $\GaH$ the total width of the
Higgs boson. Expanding the inverse propagator around the complex pole yields
\begin{align}
p^2-m_0^2+\hat{\Sigma}(p^2) \simeq (p^2-\mathcal{M}^2)\left\lbrace
1+\hat{\Sigma}'(\mathcal{M}^2)\right\rbrace 
\end{align}
in the vicinity of the complex pole.
Accordingly, the Higgs propagator in the vicinity of the complex pole can be expressed
in the well-known form of a Breit-Wigner propagator with constant width~$\GaH$,
\begin{align}
\Delta_H(p^2)=
\frac{i}{p^2-\mathcal{M}^2} =
\frac{i}{p^2-\mh^2+i\mh\GaH} .
\end{align}
Away from the pole, i.e.\ in the far off-shell region with $p^2\gg \mh^2$, the
Higgs width is not of relevance.
For the specific processes that are considered in this paper
our choice is equivalent to the
complex-mass scheme \cite{Denner:1999gp,Denner:2006ic}, which
is known to provide gauge-independent results. 
Differences with respect to the scheme
defined in \citeres{Actis:2006rc,Passarino:2010qk,Goria:2011wa} are
expected to be small, in particular since the constant width $\GaH$ is close
to the width therein \cite{Goria:2011wa}.
For our subsequent discussion we fix $\mh=125$\,GeV and $\GaH^{\sm}=4.07\cdot 10^{-3}$\,GeV,
the latter in accordance with the prescription of the
\lhc{} Higgs cross section working group
(\lhchxswg{}) \cite{Dittmaier:2011ti,Dittmaier:2012vm,Heinemeyer:2013tqa}.

\section{Off-shell contributions in $H\rightarrow ZZ^{(*)}$ and $H\rightarrow W^\pm W^{\mp(*)}$}
\label{sec:offVV}

\begin{figure}[ht]
\begin{center}
\begin{tabular}{cc}
\includegraphics[width=0.3\textwidth]{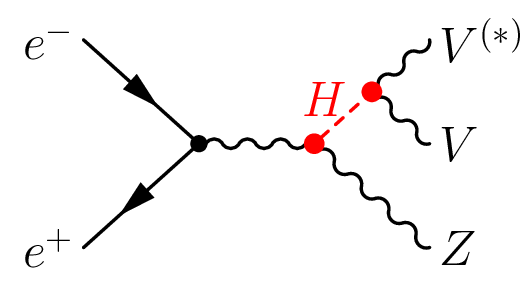} &
\includegraphics[width=0.3\textwidth]{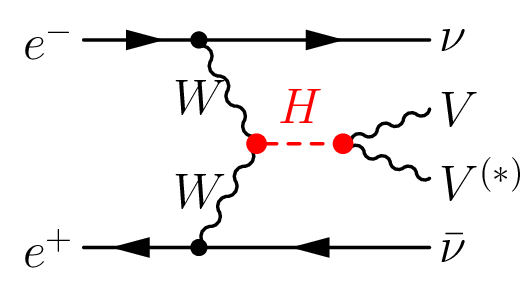} \\[-0.5cm]
 (a) & (b)
\end{tabular}
\end{center}
\vspace{-0.6cm}
\caption{Feynman diagrams for (a) $e^+e^-\rightarrow ZH\rightarrow ZVV^{(*)}$;
(b) $e^+e^-\rightarrow \nu\bar\nu H\rightarrow \nu\bar\nu VV^{(*)}$.}
\label{fig:feynmanZZ} 
\end{figure}

Given the two dominant production processes for a Higgs boson~$H$ at a linear
collider, $e^+e^-\rightarrow ZH$ and $e^+e^-\rightarrow \nu\bar\nu H$,
we discuss the validity of the zero-width approximation (\zwat{})
for the Higgs decays $H\rightarrow WW^{(*)}$ and $H\rightarrow ZZ^{(*)}$ within this section.
The relevant Feynman diagrams are presented in \fig{fig:feynmanZZ}.
Our discussion follows \citeres{Kauer:2012hd,Kauer:2013cga,Kauer:2013qba}, which are
specific to the dominant production process at the \lhc{}, gluon fusion.

Supplementing the \zwat{} for the production and the decay part of the
process with a Breit-Wigner propagator, the differential cross section
$e^+e^-\rightarrow ZH\rightarrow ZVV$ can be written as (see e.g.\
\citere{Kauer:2012hd})
\begin{align}
\label{eq:ZWA}
\left(\frac{d\sigma^{\zvv}_{\zwa}}{d\mvv }\right)
=\sigma^{\zh}(\mh)\frac{2\mvv}{(\mvv^2-\mh^2)^2+(\mh\GaH)^2}\frac{\mh\Gamma_{H\rightarrow
VV}(\mh)}{\pi}\quad,
\end{align}
with $\mvv^2=p_H^2$ being the invariant mass of the two gauge bosons
originating from the intermediate Higgs $H$. The production cross
sections $e^+e^-\rightarrow ZH$ is represented by $\sigma^{\zh}$, the
partial width of an on-shell Higgs boson into two gauge bosons by
$\Gamma_{H\rightarrow VV}(\mh)$, given in the restframe of the Higgs boson~$H$.  Integrating over the invariant mass
$\mvv$ in the limit where $\mh\GaH\rightarrow 0$ results in the well-known formula
\begin{align}
\sigma_{\zwa}^\zvv
=\sigma^{\zh}(\mh)\frac{\Gamma_{H\rightarrow VV}(\mh)}{\GaH}
=\sigma^{\zh}(\mh)\text{BR}_{H\rightarrow VV}(\mh)\quad.
\end{align}
Going beyond the \zwat, one can define (see e.g.\ \citere{Kauer:2012hd})
an off-shell production cross section according to
\begin{align}
\label{eq:sigmaoff}
\left(\frac{d\sigma^\zvv_{\text{off}}}{d\mvv}\right)
=\sigma^{\zh}(\mvv)\frac{2\mvv}{(\mvv^2-\mh^2)^2+(\mh\GaH)^2}\frac{\mvv\Gamma_{H\rightarrow VV}(\mvv)}{\pi}\quad.
\end{align}
The result is identical to the explicit calculation of the Higgs induced
production cross section $e^+e^-\rightarrow ZH\rightarrow ZVV$ with $\mvv>2\mv$
at tree-level.
The formulas can be directly translated to the production
process $e^+e^-\rightarrow \nu\bar\nu H\rightarrow \nu\bar\nu VV$.
Our own code, generated with {\tt FeynArts}~\cite{Kublbeck:1990xc,Hahn:2000kx}
and {\tt FormCalc}~\cite{Hahn:1998yk}, makes use of the 't Hooft-Feynman-gauge
and explicitly takes into account Goldstone boson contributions.
{\tt MadGraph5\_aMC@NLO}~\cite{Alwall:2014hca} by default uses unitary gauge, allowing us to check the
gauge independence. For the considered case of lowest order in perturbation theory
within the \sm{} the choice of all diagrams involving a Higgs is gauge independent, since
the Goldstone bosons do not couple to the initial state (in good 
approximation the electron and positron can be treated as massless).
The dependence on the Higgs mass $\mh$, which is an independent parameter
of the \sm{}, is suppressed for high invariant masses of the two 
gauge bosons~$\mvv$.
The leading (Higgs-mass independent) contributions for high $\mvv$
cancel with the corresponding background contributions involving gauge
bosons, as it is required in order to restore unitarity for
the scattering of longitudinally polarised gauge bosons.

\begin{figure}[ht]
\begin{center}
\includegraphics[width=0.3\textwidth]{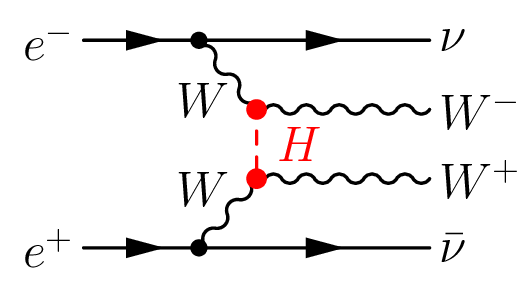}
\end{center}
\vspace{-0.6cm}
\caption{$t$-channel Feynman diagram for 
$e^+e^-\rightarrow \nu\bar\nu W^+W^-$.}
\label{fig:feynmannunuWWtchannel} 
\end{figure}

Two processes need particular attention:
For $e^+e^-\rightarrow \nu\bar\nu W^+W^-$ we present the result just involving
the $s$-channel Higgs boson in the off-shell region corresponding to the
approximation of \eqn{eq:sigmaoff}, 
$e^+e^-\rightarrow \nu\bar\nu H\rightarrow \nu\bar\nu W^+W^-$. In a second
step we add the contribution
induced by $t$-channel Higgs boson exchange between the two gauge bosons,
see \fig{fig:feynmannunuWWtchannel}.

In case of $e^+e^-\rightarrow ZH\rightarrow ZZZ$ it is at first sight unclear which
two out of the three $Z$ bosons originate from a Higgs boson.
For the most conservative
approach we average over the three possible invariant mass combinations $\mzz$.
If only one out of the two gauge bosons to calculate the invariant
mass $\mzz$ stems from the Higgs boson, the invariant mass $\mzz$ cannot be related
to the on/off-shellness of the Higgs boson, and contributions with an on-shell
Higgs boson enter the differential cross section over
a wide range of values of $\mzz$, also in the region $\mzz>2\mz$.
Necessarily the peak in the on-shell region $\mzz\approx \mh$
is mainly induced by contributions with the correct assignment of the two gauge bosons
to the Higgs boson and thus the averaging over the three invariant
mass combinations approximately corresponds to a division of the cross section
obtained by \eqn{eq:sigmaoff} by a factor of~$3$.
For $\mzz>2\mz$ all assignments of $ZZ$ pairs are of relevance and
interferences between the three possible options need
to be taken into account, which we calculate separately by our code.
Taking the average of possible $ZZ$ pairings
effectively increases the relevance of the off-shell
contributions with respect to the on-shell region for this specific process
due to the possibly ``wrong'' assignment of gauge bosons, which as mentioned
already causes on-shell Higgs events to contribute in the region $\mzz>2\mz$.
In a more optimistic approach, using different weights for the $ZZ$ pairings
a discrimination between the different $ZZ$ pairs could be achieved.
In our notation the latter case is indicated by additional indices, $e^+e^-\rightarrow Z_1Z_2Z_3$,
in order to emphasize the distinction of the $Z$ bosons.
To summarise, for $e^+e^-\rightarrow \nu\bar\nu H\rightarrow \nu\bar\nu W^+W^-$
and $e^+e^-\rightarrow Z_1H\rightarrow Z_1Z_2Z_3$ we do not only present the results obtained
by the usage of \eqn{eq:sigmaoff}, but also add results obtained
by taking into account $t$-channel contributions and averaging over the
different $ZZ$ pairs, $e^+e^-\rightarrow ZH\rightarrow ZZZ$, respectively.
Lastly we note that the final states  $e^+e^-\rightarrow \nu\bar\nu VV$ can also
be obtained via $e^+e^-\rightarrow ZH\rightarrow \nu\bar\nu VV$. However, Higgsstrahlung
is sub-dominant in regions where vector boson fusion is of relevance and can
be moreover suppressed by appropriate cuts. It is thus
addressed in the background discussion, see \sct{sec:backgroundVV}.

When using \eqn{eq:sigmaoff} we calculate the production
cross sections $\sigma^\zh$ and $\sigma^\nunuh$ using our own code
and obtain the partial width $\Gamma_{H\rightarrow VV}$
from the values for the branching ratio $\text{BR}_{H\rightarrow VV}$
and the Higgs width $\GaH$ given by
the \lhchxswg{}~\cite{Dittmaier:2011ti,Dittmaier:2012vm,Heinemeyer:2013tqa}.
For $\mvv>2\mv$ the partial width $\Gamma_{H\rightarrow VV}$ is rather close to the
tree-level partial width, which enters our explicit calculation of production cross sections.

\begin{figure}[htp]
\begin{center}
\begin{tabular}{cc}
\includegraphics[width=0.4\textwidth]{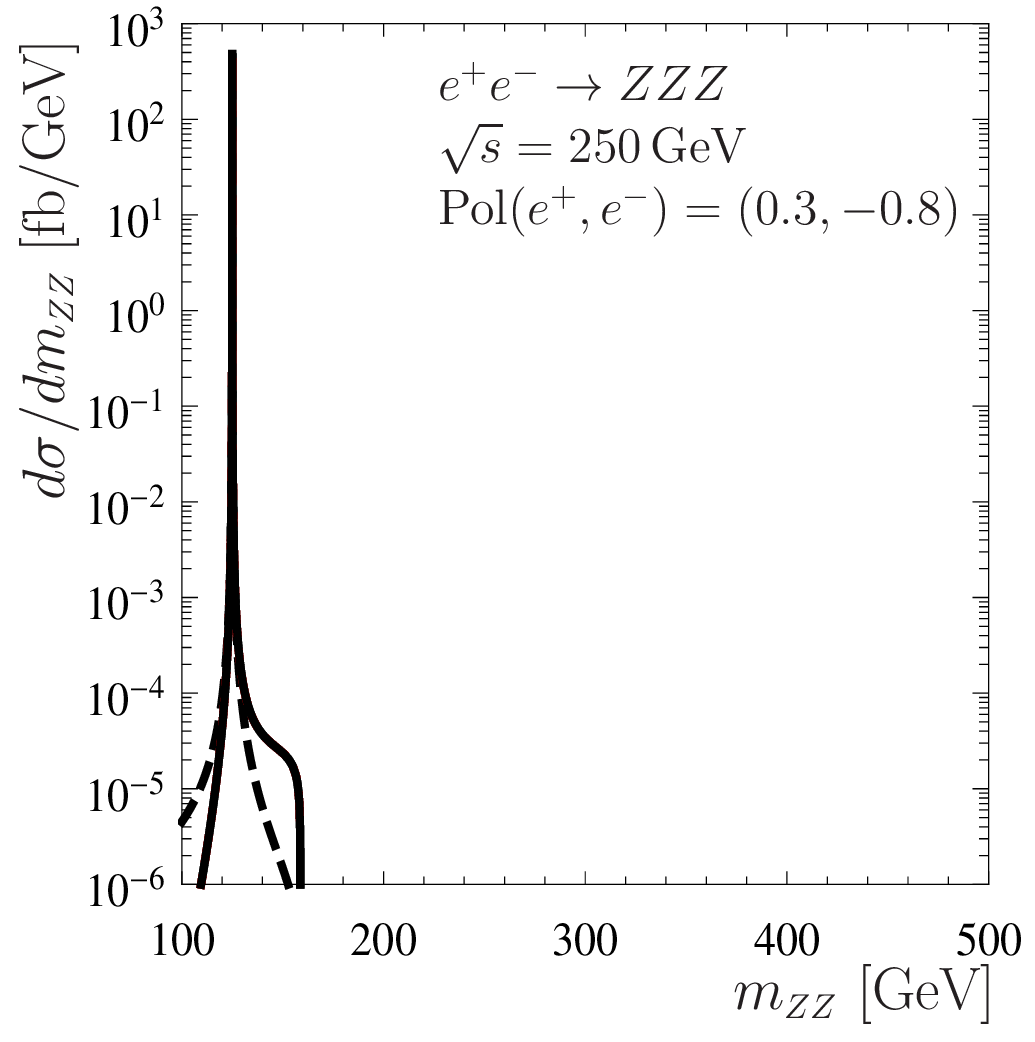} &
\includegraphics[width=0.4\textwidth]{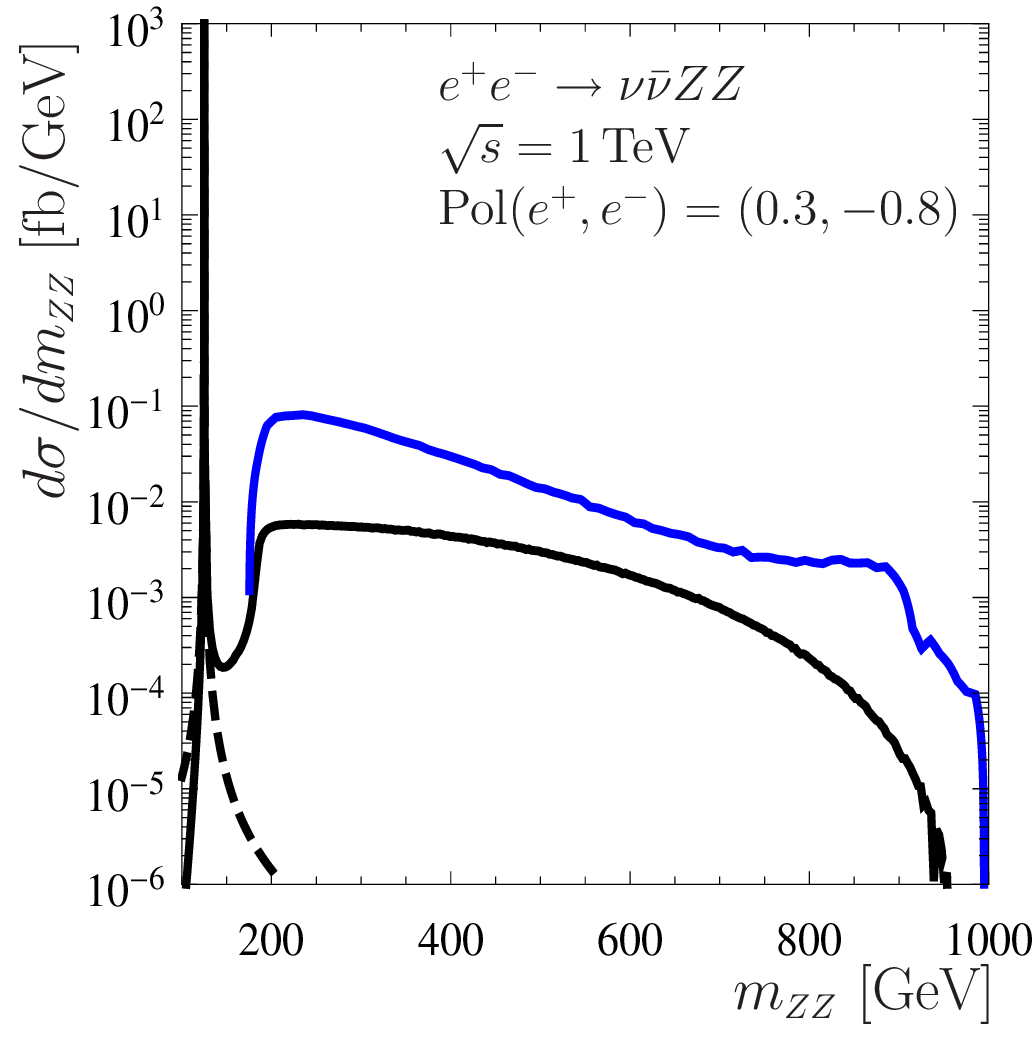} \\[-0.6cm]
 (a) & (b) \\
\includegraphics[width=0.4\textwidth]{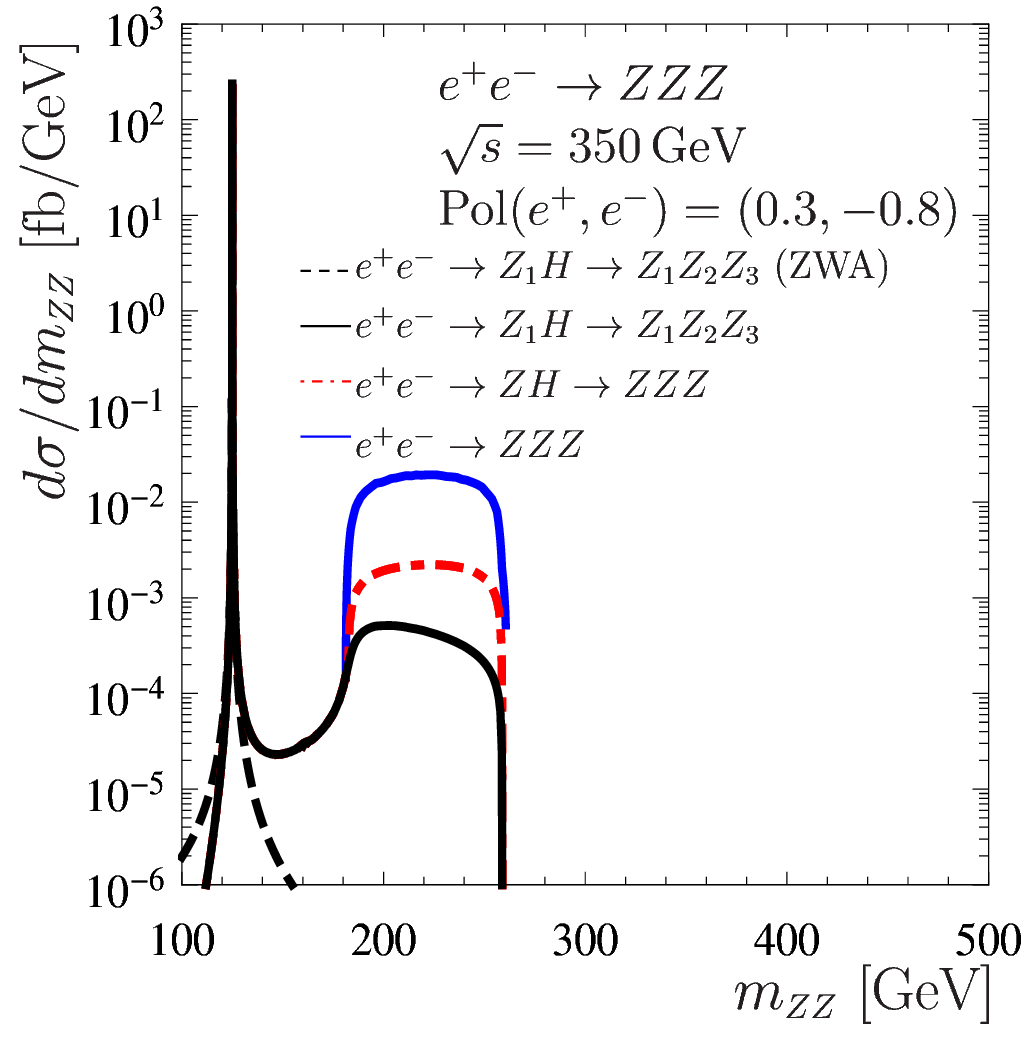} &
\includegraphics[width=0.4\textwidth]{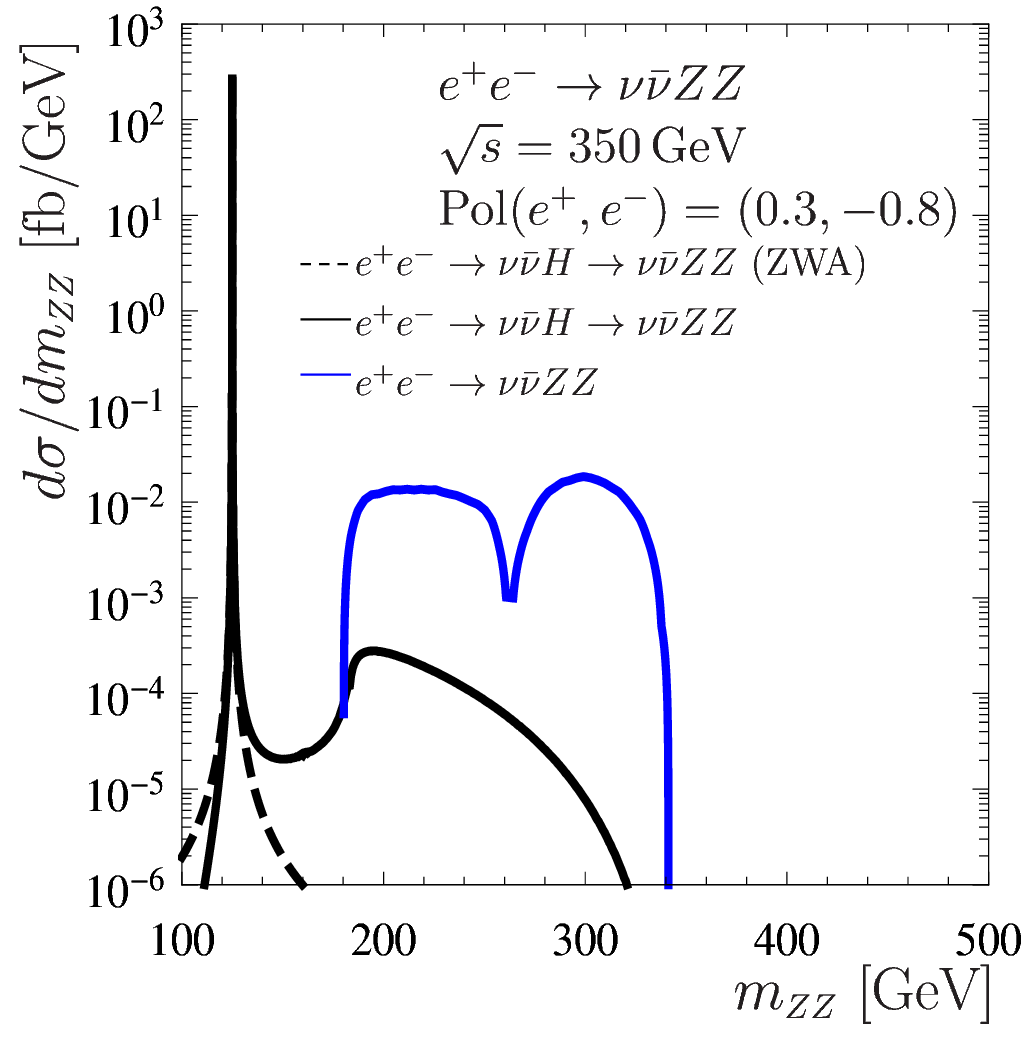} \\[-0.6cm]
 (c) & (d) \\
\includegraphics[width=0.4\textwidth]{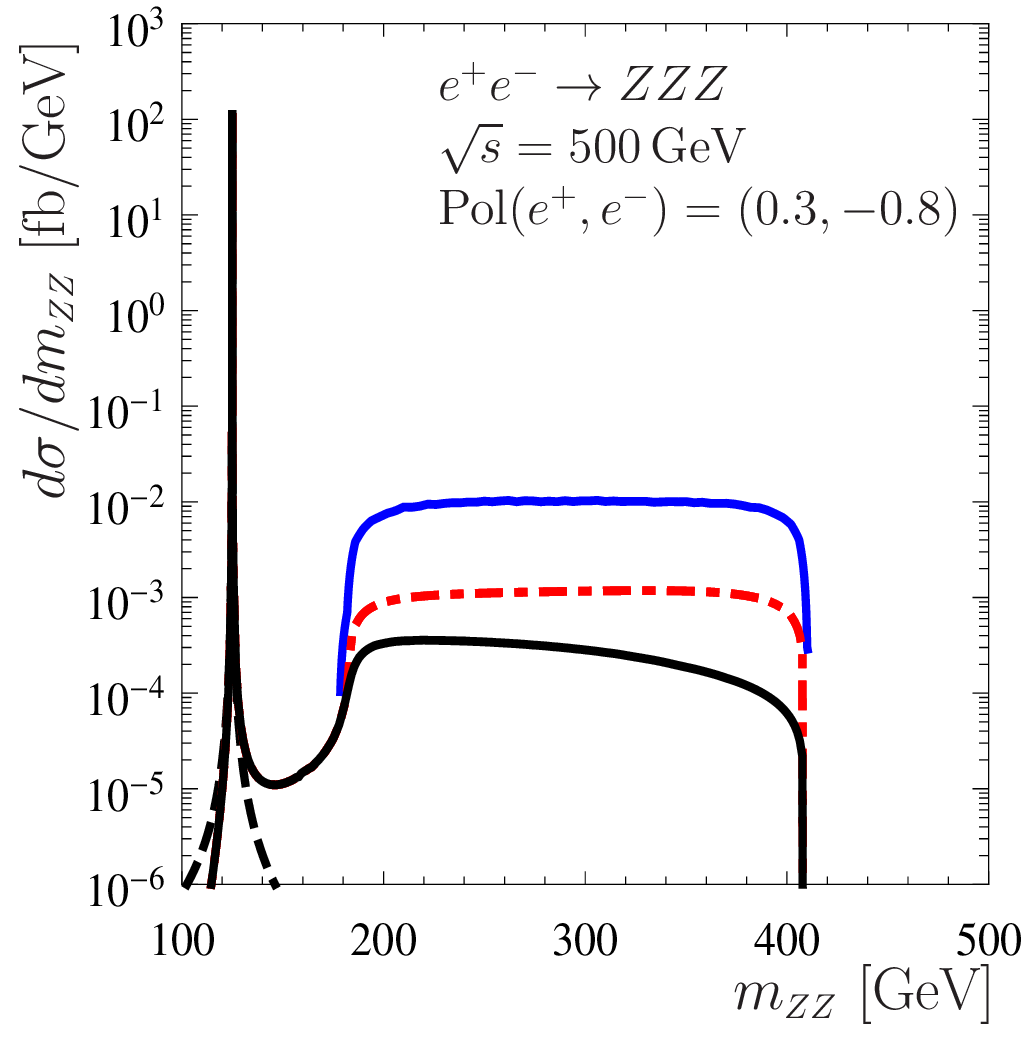} &
\includegraphics[width=0.4\textwidth]{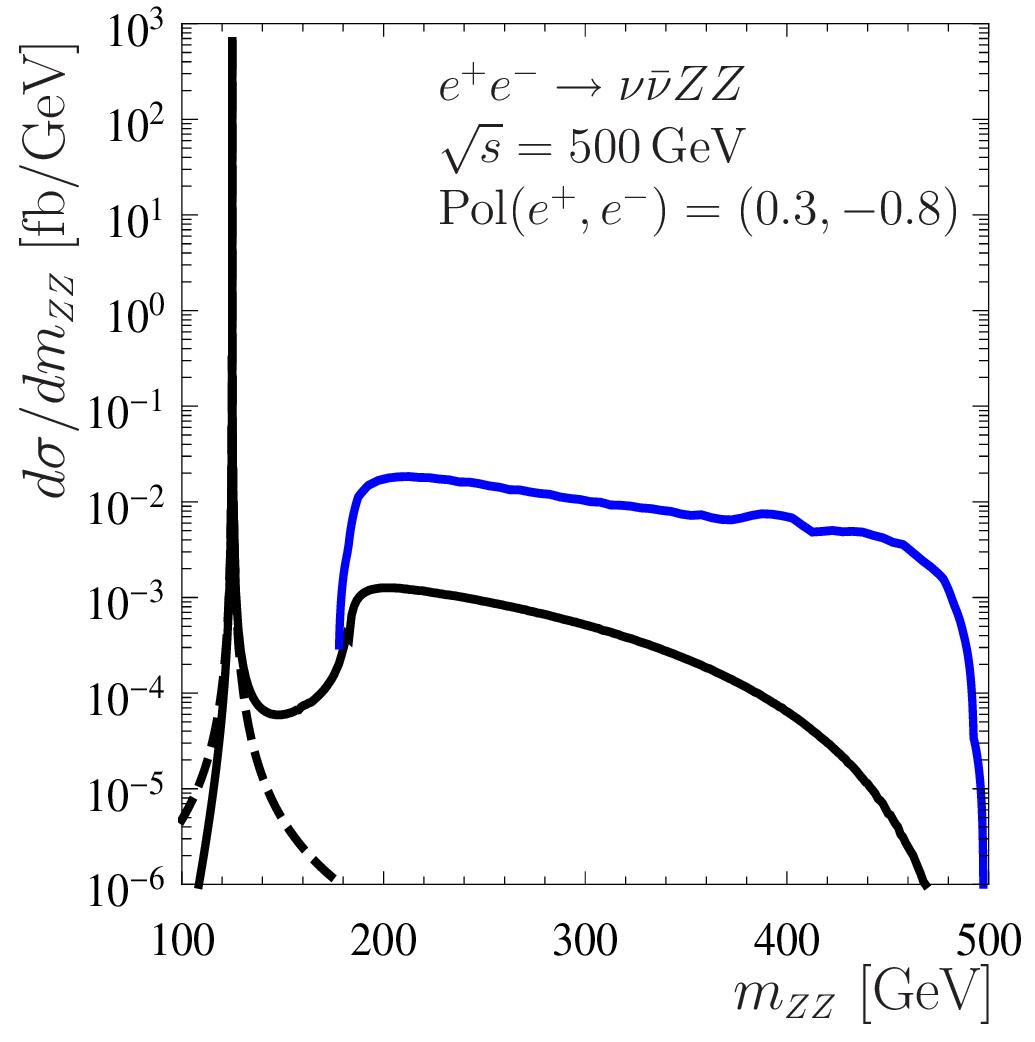} \\[-0.6cm]
 (e) & (f) \\
\end{tabular}
\end{center}
\vspace{-0.7cm}
\caption{$d\sigma/d\mzz$ in fb/GeV as a function of $\mzz$ in GeV defined in \eqn{eq:ZWA} (\zwat{}) (black, dashed)
 and \eqn{eq:sigmaoff} (black, solid) for a fixed polarisation Pol$(e^+,e^-)=(0.3,-0.8)$ for
 (a,c,e) $e^+e^-\rightarrow Z_1H\rightarrow Z_1Z_2Z_3$ for \cms{} energies $\sqrt{s}=250,\, 350,\, 500$\,GeV (top to bottom)
 and (b,d,f) $e^+e^-\rightarrow\nu\bar\nu H\rightarrow \nu\bar\nu ZZ$
 for \cms{} energies $\sqrt{s}=1000,\, 350,\, 500$\,GeV (top to bottom).
 The black curves in (a,c,e) are shown as a function of $\mzznr$.
 The red, dot-dashed curve shows the calculation of $e^+e^-\rightarrow ZH\rightarrow ZZZ$ with averaging over the $ZZ$ pairs.
 As blue curve we add the complete calculation including background
 contributions $e^+e^-\rightarrow ZZZ/\nu\bar\nu ZZ$, see \sct{sec:backgroundVV}.
 The legend of the centered figures is valid for the upper and lower figures as well.
}
\label{fig:ZZinvmass} 
\end{figure}

\begin{figure}[htp]
\begin{center}
\begin{tabular}{cc}
\includegraphics[width=0.4\textwidth]{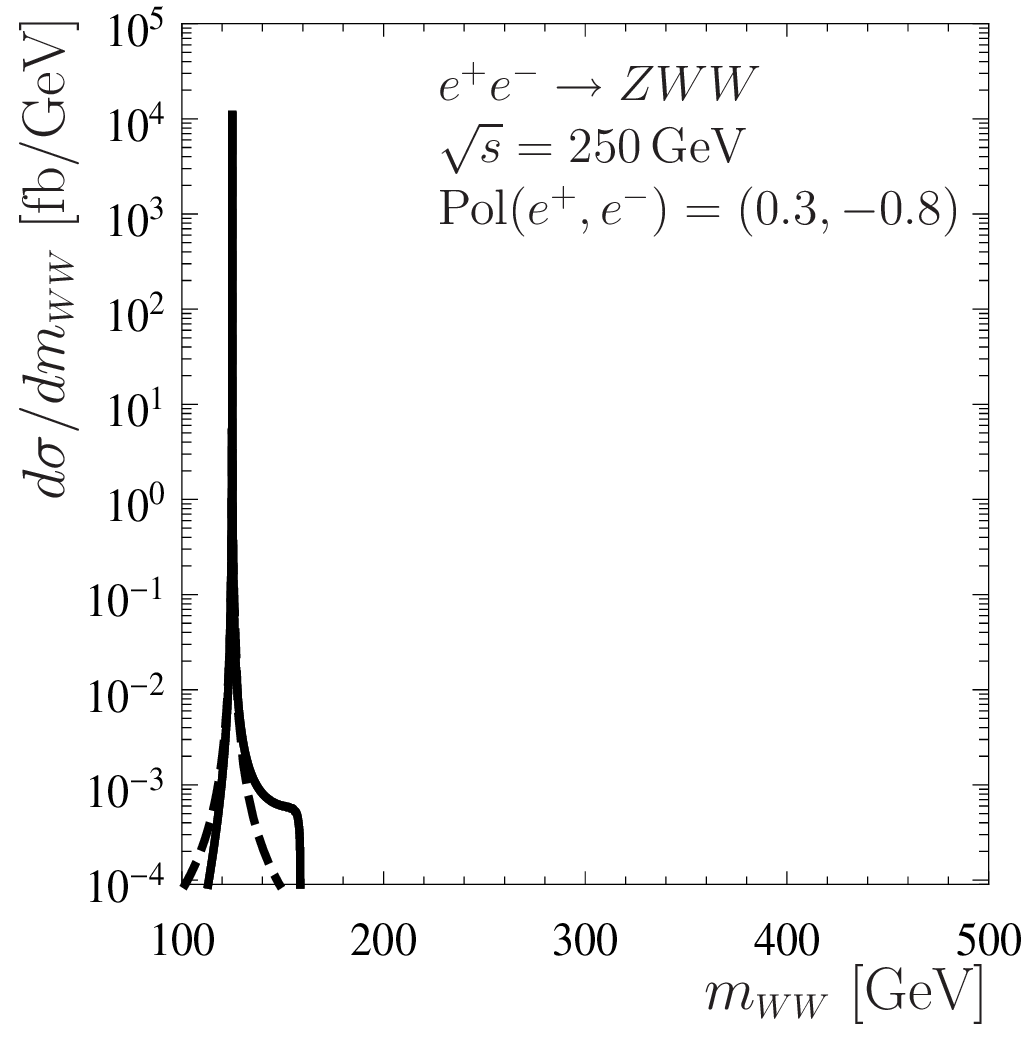} &
\includegraphics[width=0.4\textwidth]{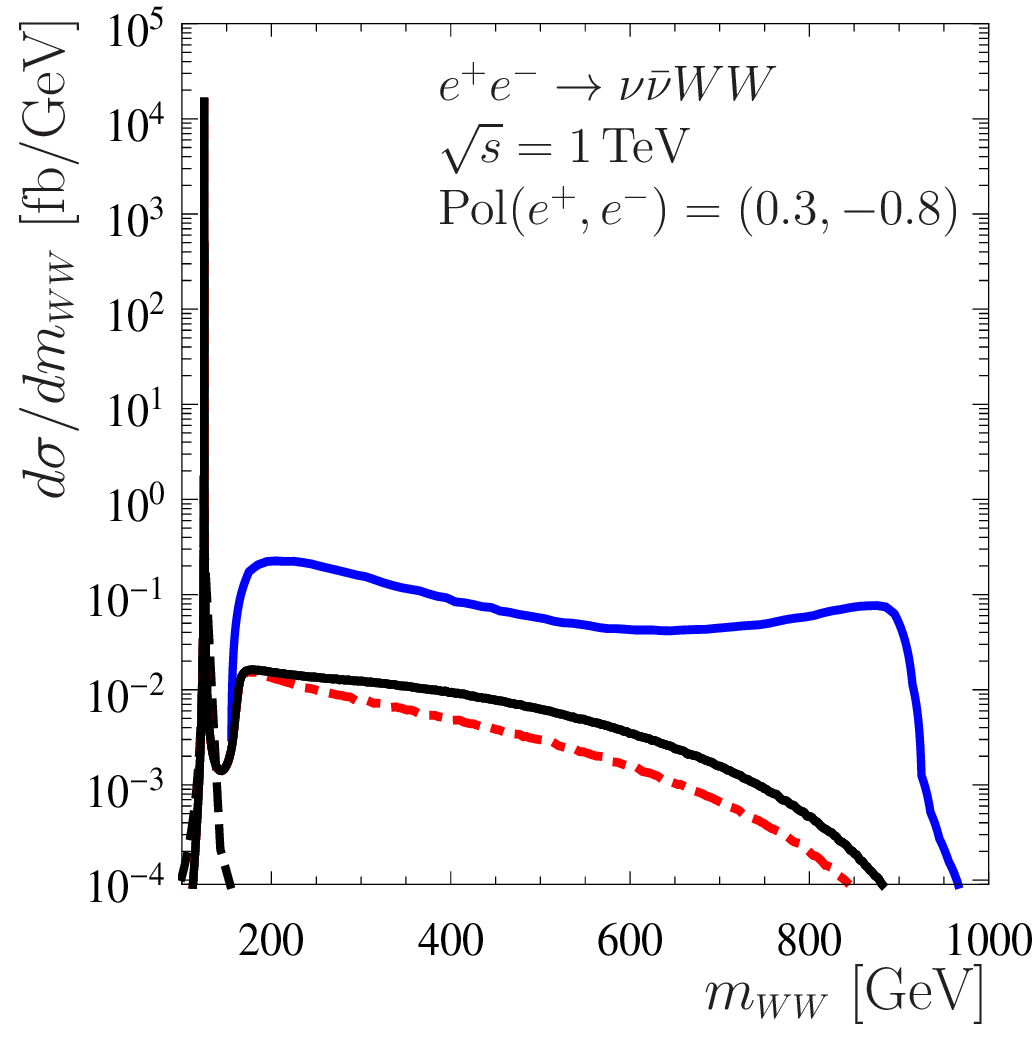} \\[-0.6cm]
 (a) & (b) \\
\includegraphics[width=0.4\textwidth]{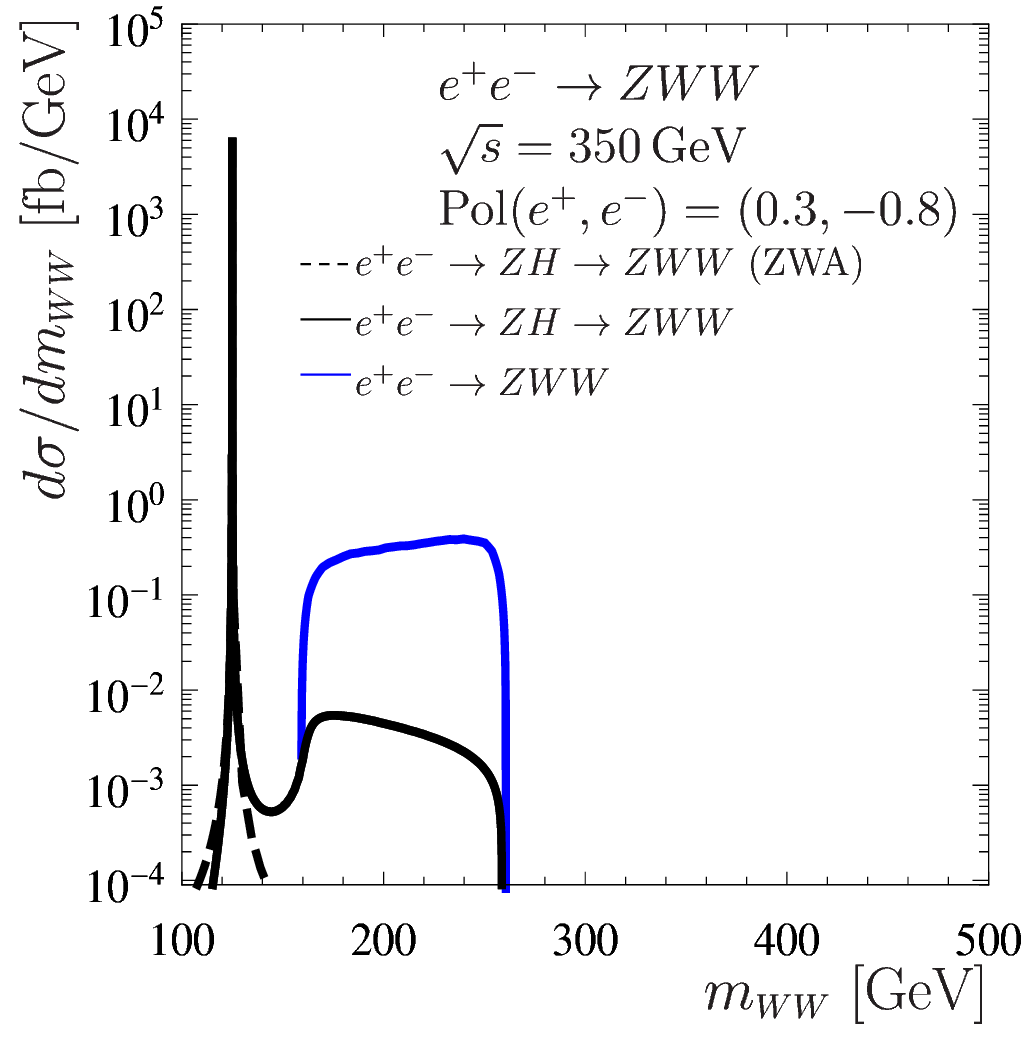} &
\includegraphics[width=0.4\textwidth]{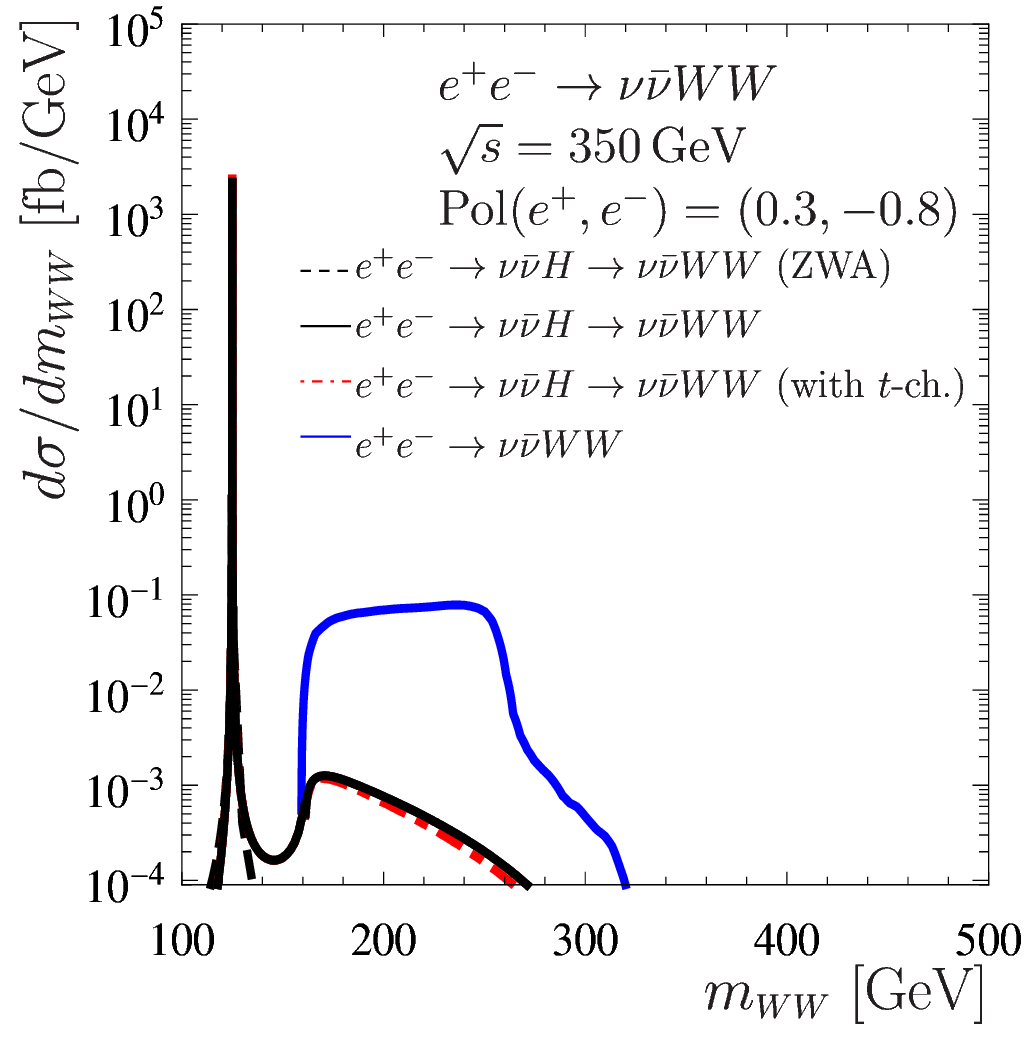} \\[-0.6cm]
 (c) & (d) \\
\includegraphics[width=0.4\textwidth]{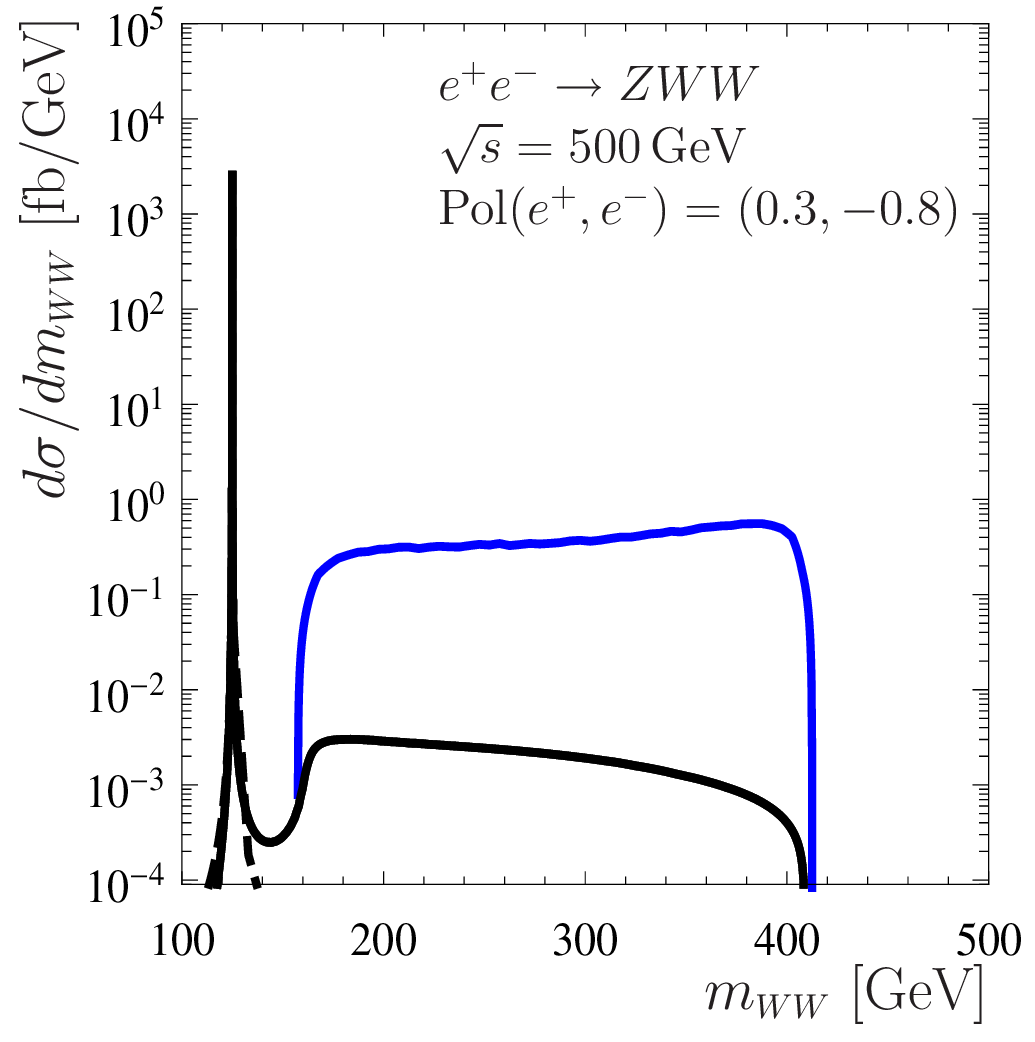} &
\includegraphics[width=0.4\textwidth]{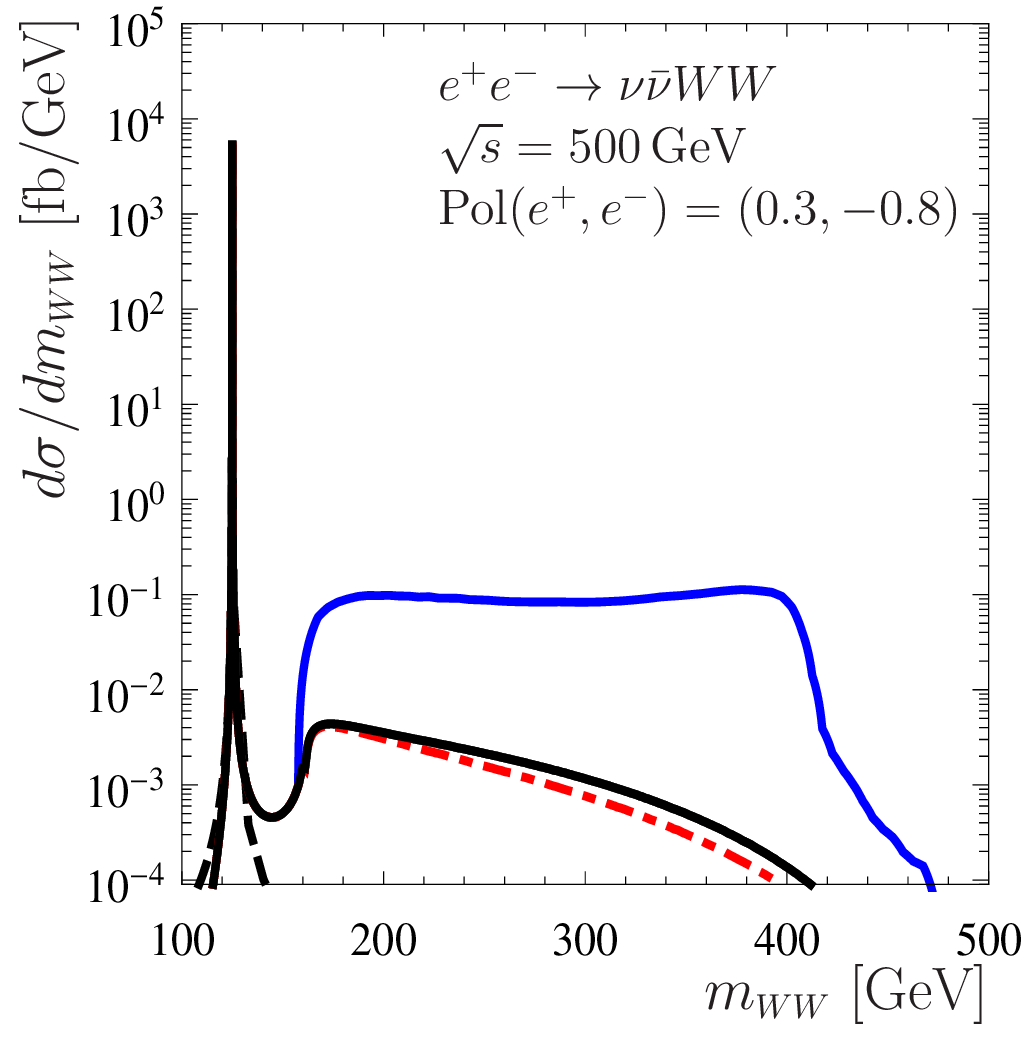} \\[-0.6cm]
 (e) & (f) \\
\end{tabular}
\end{center}
\vspace{-0.7cm}
\caption{$d\sigma/d\mww$ in fb/GeV as a function of $\mww$ in GeV defined in \eqn{eq:ZWA} (\zwat{}) (black, dashed)
 and \eqn{eq:sigmaoff} (black, solid) for a fixed polarisation Pol$(e^+,e^-)=(0.3,-0.8)$ for
 (a,c,e) $e^+e^-\rightarrow ZH\rightarrow ZWW$ for \cms{} energies $\sqrt{s}=250,\, 350,\, 500$\,GeV (top to bottom)
 and (b,d,f) $e^+e^-\rightarrow \nu\bar\nu H\rightarrow \nu\bar\nu WW$
 for \cms{} energies $\sqrt{s}=1000,\, 350,\, 500$\,GeV (top to bottom).
 The red, dot-dashed curve shows the calculation of $e^+e^-\rightarrow \nu\bar\nu WW$ including s- and $t$-channel
 Higgs induced contributions, see text.
 As blue curve we add the complete calculation including background contributions
 $e^+e^-\rightarrow ZWW/\nu\bar\nu WW$, see \sct{sec:backgroundVV}. The legend of
 the centered figures is valid for the upper and lower figures as well.
}
\label{fig:WWinvmass} 
\end{figure}

The resulting differential cross sections $d\sigma/d\mzz$ for both production processes
in combination with $H\rightarrow ZZ^{(*)}$
for different energies $\sqrt{s}=250,\, 350,\, 500$\,GeV and $1$\,TeV and a fixed polarisation
of the initial state Pol$(e^+,e^-)=(0.3,-0.8)$ are shown in \fig{fig:ZZinvmass}.
For $e^+e^-\rightarrow ZZZ$ we distinguish the pure usage of \eqn{eq:sigmaoff},
$e^+e^-\rightarrow Z_1H\rightarrow Z_1Z_2Z_3$,
from the case with averaging over the three possible invariant mass combinations of $ZZ$ pairs,
$e^+e^-\rightarrow ZH\rightarrow ZZZ$, the latter presented by the 
red, dot-dashed curve. As expected the average over the three $ZZ$ pairs
results in a larger off-shell cross section due to on-shell Higgs events
which contribute also in the region $\mzz>2\mz$.
Similarly, \fig{fig:WWinvmass} shows the differential cross sections
for $H\rightarrow WW^{(*)}$, where for $e^+e^-\rightarrow \nu\bar\nu WW$
the red, dot-dashed curve includes also the $t$-channel Higgs induced contributions.
For $\mvv>2\mv$ we add the contributions from background diagrams leading to the same
final state as blue curve in both \fig{fig:ZZinvmass} and \fig{fig:WWinvmass}.
The full process $e^+e^-\rightarrow ZZZ$ includes all signal and background Feynman diagrams and
thus averaging over the three $ZZ$ pairs is obsolete.
A detailed description of the background is provided in \sct{sec:backgroundVV}.
For $\mvv<2\mv$ only resonant background processes are of relevance. Thus in case of
$H\rightarrow WW^{(*)}$ the $Z$ boson peak at $\mww\approx \mz$ is present, but
not visible in our diagrams which start at $\mww=100$\,GeV.

Given the inclusive cross section for a lower and upper bound of invariant masses $\mvv$
\begin{align}
\sigma_{X}(\mvv^d,\mvv^u)=\int_{\mvv^d}^{\mvv^u} d\mvv\left(\frac{d\sigma_X}{d\mvv}\right)
\label{eq:uplowsigma}
\end{align}
we define the relative importance of the off-shell signal contributions in the form
\begin{align}
\label{eq:deltaoff}
\Delta_{\text{off}}^\zvv=\frac{\sigma^\zvv_{\text{off}}(130\text{GeV},\sqrt{s}-\mz)}{\sigma^\zvv_{\text{off}}}\qquad
\text{and}
\qquad
\Delta_{\text{off}}^\nunuvv=\frac{\sigma^\nunuvv_{\text{off}}(130\text{GeV},\sqrt{s})}{\sigma^\nunuvv_{\text{off}}}\quad,
\end{align}
with $\sigma^\zvv_{\text{off}}=\sigma^\zvv_{\text{off}}(0,\sqrt{s}-\mz)$
and $\sigma^\nunuvv_{\text{off}}=\sigma^\nunuvv_{\text{off}}(0,\sqrt{s})$.
Our discussion is hardly sensitive to the precise numerical 
value of the boundary between on- and off-shell contributions,
which we choose to be at $130$\,GeV.
In contrast to the absolute size of the off-shell contributions, their
relative size $\Delta_{\text{off}}$ is independent of the polarisation
of the initial state electron/positron. As a function of the \cms{} energy the 
latter values are given in \tab{tab:ZZ}.

\begin{table}[htb]
\begin{center}
\begin{tabular}{| c || c | c || c | c |}
\hline
$\sqrt{s}$ & $\sigma^\zzznr_{\text{off}}$ ($\sigma^\zzz_{\text{off}}$) & $\Delta_{\text{off}}^\zzznr$ ($\Delta_{\text{off}}^\zzz$)
& $\sigma^\nunuzz_{\text{off}}$ & $\Delta_{\text{off}}^\nunuzz$
\\\hline\hline
$250$\,GeV & $3.12(3.12)$\,fb   & $0.03(0.03)$\,\% & $0.490$\,fb & $0.12$\,\%\\\hline
$300$\,GeV & $2.36(2.40)$\,fb   & $0.46(1.83)$\,\% & $1.12$\,fb & $0.40$\,\%\\\hline
$350$\,GeV & $1.71(1.82)$\,fb   & $1.82(7.77)$\,\% & $1.91$\,fb & $0.88$\,\%\\\hline
$500$\,GeV & $0.802(0.981)$\,fb & $7.20(24.1)$\,\% & $4.78$\,fb & $2.96$\,\%\\\hline
$1$\,TeV   & $0.242(0.341)$\,fb & $30.9(50.9)$\,\% & $15.0$\,fb & $13.0$\,\%\\\hline
  \hline
$\sqrt{s}$ & $\sigma^\zww_{\text{off}}$ & $\Delta_{\text{off}}^\zww$ 
& $\sigma^\nunuww_{\text{off}}$ & $\Delta_{\text{off}}^\nunuww$
\\\hline\hline
$250$\,GeV & $76.3$\,fb & $0.03$\,\% & $3.98(3.99)$\,fb & $0.13(0.12)$\,\%\\\hline
$300$\,GeV & $57.7$\,fb & $0.42$\,\% & $9.07(9.08)$\,fb & $0.29(0.26)$\,\%\\\hline
$350$\,GeV & $41.4$\,fb & $0.92$\,\% & $15.5(15.5)$\,fb & $0.49(0.43)$\,\%\\\hline
$500$\,GeV & $18.6$\,fb & $2.61$\,\% & $38.2(38.1)$\,fb & $1.21(0.96)$\,\%\\\hline
$1$\,TeV   & $4.58$\,fb & $11.0$\,\% & $110.8(108.9)$\,fb & $4.45(2.78)$\,\%\\\hline
\end{tabular}
\end{center}
\vspace{-5mm}
\caption{Inclusive cross sections $\sigma_{\text{off}}(0,\sqrt{s}-\mz)$
for $e^+e^-\rightarrow ZH\rightarrow ZVV$
and $\sigma_{\text{off}}(0,\sqrt{s})$ for $e^+e^-\rightarrow \nu\bar\nu H \rightarrow \nu\bar\nu VV$
for Pol$(e^+,e^-)=(0.3,-0.8)$ and
relative size of the off-shell contributions $\Delta_{\text{off}}$ in \%.
In brackets we add the results averaging over the $ZZ$ pairs
for $e^+e^-\rightarrow ZZZ$ and taking into account the $t$-channel Higgs contribution for
$e^+e^-\rightarrow \nu\bar\nu WW$. $\Delta_{\text{off}}$ is independent
of the polarisation.}
\label{tab:ZZ}
\end{table}

The off-shell contributions are sizeable and reach $\mathcal{O}(10$\%) for
large enough \cms{} energies.
For $H\rightarrow WW^{(*)}$ the off-shell contributions are generally smaller
than for $H\rightarrow ZZ^{(*)}$,
since the difference between $\Gamma_{H\rightarrow WW}(\mh)$ and $\Gamma_{H\rightarrow WW}(2\mw)$
is not as pronounced as for $H\rightarrow ZZ^{(*)}$.
On the other hand the off-shell contributions are very small for low
\cms{} energies, $\sqrt{s}=250-300$\,GeV. As a consequence, 
the determination of the $HVV$ couplings based on the \zwat{}
in that energy range is to a good approximation
not affected by off-shell contributions.
For the case of $H\rightarrow ZZ^{(*)}$ followed by decays of the two gauge bosons
into leptons or quarks, i.e.\ $Z\rightarrow l^\pm l^\mp/q\bar q$,
on-shell and off-shell contributions can be discriminated by the invariant mass
of the four leptons/quarks. This is not necessarily the case when neutrinos
are involved in the final state like e.g.\
in $H\rightarrow WW^{(*)}$ followed by $W\rightarrow l^\pm \nu(\bar\nu)$,
since the four particle invariant mass is not directly accessible, possibly only indirectly
e.g.\ by the recoil mass in $e^+e^-\rightarrow ZH$.
We discuss the implications for the $Z$ recoil mass measurement and the
extraction of $HVV$ couplings in more detail in \sct{sec:Zrecoil}.

As a final step of this discussion we investigate
the quality of the \zwat{} in the ``on-shell'' region
between $(\mvv^d,\mvv^u)=(120,130)$\,GeV for $H\rightarrow VV^{(*)}$
by comparing $\sigma_{\zwa}(\mvv^d,\mvv^u)$ as defined in
\eqn{eq:ZWA} with $\sigma_{\text{off}}(\mvv^d,\mvv^u)$ defined in
\eqn{eq:sigmaoff}.
Within the specified interval the agreement between
the cross sections is at the permil level, and most of the contribution
stems from the small interval $\mvv=124-126$\,GeV. However, as soon as larger
invariant masses $\mvv>130$\,GeV are taken into account
the difference between both methods becomes visible
in the plots of \fig{fig:ZZinvmass} and \fig{fig:WWinvmass}.
Lower invariant masses in the range $(\mvv^d,\mvv^u)=(100,120)$\,GeV,
on the other hand, are negligible.
The difference between the two inclusive cross sections $\sigma_{\zwa}$
and $\sigma_{\text{off}}$ in the ``on-shell region''
is slightly increasing with the \cms{} energy,
but always stays at the permil level. We conclude that within the ``on-shell region''
the \zwat{} is a valid approximation with an accuracy at the (sub-)permil
level.%
\footnote{Our investigation here has been done for partonic cross
sections. In an experimental simulation the definition of the ``on-shell
region'' may have to be adjusted in order to take into account effects like
detector resolution.}

\subsection{Dependence on the precise numerical value of the Higgs mass}
\label{sec:higgsmass}

Both partial widths $\Gamma_{H\rightarrow ZZ}(\mh)$ and $\Gamma_{H\rightarrow WW}(\mh)$
and accordingly both branching ratios BR$_{H\rightarrow VV}(\mh)$
are strongly sensitive to the precise numerical value of the Higgs mass $\mh$ for $\mh<2\mv$.
As an example, if the Higgs mass $\mh=125$\,GeV is changed
by $\pm 200$\,MeV, both partial widths change by about $\pm 2.5\%$.
Due to the dominance of $H\rightarrow b\bar b$ at $\mh=125$\,GeV
the change in the total width is smaller and amounts to $\pm 0.7\%$.
In order to briefly illustrate these effects we calculate uncertainties
on the cross section using $\mh=124.8-125.2$\,GeV together with $\GaH^{\sm}=4.04-4.10$\,MeV
for $\sqrt{s}=500$\,GeV with a fixed polarisation Pol$(e^+,e^-)=(0.3,-0.8)$
and present the results in \tab{tab:Higgsmass}.

\begin{table}[htb]
\begin{center}
\begin{tabular}{| c || c | c | c | c |}
\hline
$e^+e^-\rightarrow$ & $\sigma_{\text{off}}(\mh)$ & $\sigma_{\text{off}}(\mh\pm 200\,\text{MeV})$ & $\delta \sigma_{\text{off}}$ 
& $\Delta_{\text{off}}$
\\\hline\hline
$Z_1Z_2Z_3$     & $0.802$\,fb & $0.788-0.816$\,fb    & $-1.7,+1.7\%$   & $7.31-7.10$\%\\
($ZZZ$)         & $(0.981$\,fb) & $(0.967-0.995$\,fb)& $(-1.4,+1.4\%)$ & $(24.4-23.8)$\%\\\hline
$ZWW$           & $18.66$\,fb & $18.33-18.90$\,fb     & $-1.7,+1.3\%$   & $2.65-2.58$\%\\\hline
$\nu\bar\nu ZZ$ & $4.78$\,fb & $4.70-4.85$\,fb       & $-1.6,+1.6\%$   & $3.00-2.92$\%\\\hline
$\nu\bar\nu WW$ & $38.16$\,fb & $37.53-38.60$\,fb     & $-1.7,+1.2\%$   & $1.22-1.20$\%\\
                & $(38.09$\,fb) & $(37.46-38.53$\,fb) & $(-1.7,+1.2\%)$ & $(0.957-0.973\%)$\\\hline
\end{tabular}
\end{center}
\vspace{-5mm}
\caption{Cross sections $\sigma_{\text{off}}$ (defined as in \tab{tab:ZZ}) and their dependence
on the Higgs mass $\mh\pm 200$\,MeV for $\sqrt{s}=500$\,GeV with a fixed polarisation Pol$(e^+,e^-)=(0.3,-0.8)$.
$\delta\sigma_{\text{off}}$ shows the variation with respect
to the central value. $\Delta_{\text{off}}$ shows the relevance of the off-shell contributions.
In brackets we add the results averaging over the $ZZ$ pairs
for $e^+e^-\rightarrow ZZZ$ and taking into account the $t$-channel Higgs contribution for
$e^+e^-\rightarrow \nu\bar\nu WW$.}
\label{tab:Higgsmass}
\end{table}

The change is mainly induced in the on-shell region between
$\mvv=120-130$\,GeV, whereas the off-shell contributions $\mvv>130$\,GeV
only change at the sub-permil level.
Thus, the relative fraction of the off-shell contributions, $\Delta_{\text{off}}$,
changes due to the effect in the on-shell region. Since $\Delta_{\text{off}}$
is inversely proportional to the on-shell contributions,
the increase in the central value of the
cross section by e.g.\ $\delta\sigma_{\text{off}}=1$\% with increasing 
Higgs mass $\mh$ translates into a (relative)
decrease of the off-shell contributions $\Delta_{\text{off}}$ 
by about $\mathcal{O}(1\%)$.
The values provided in brackets again average over the three $ZZ$ pairs
for $e^+e^-\rightarrow ZZZ$ and take into account
the $t$-channel Higgs contributions for $e^+e^-\rightarrow \nu\bar\nu WW$.
For practical purposes the relative variation $\delta\sigma_{\text{off}}$ can be 
scaled linearly in order to investigate the case of
higher accuracies in the Higgs mass, e.g.\ for $\mh\pm 50$\,MeV
$\delta\sigma_{\text{off}}(e^+e^-\rightarrow ZWW)$ is given by $-0.4,+0.3$\%.
Due to the main effect in the on-shell region the variations $\delta\sigma_{\text{off}}$
are of similar size for different $\sqrt{s}$.

We conclude that the measurement of the Higgs mass to a precision below $100$\,MeV
is of crucial importance to enable
cross section predictions at the percent level
for processes where the Higgs boson decays to weak gauge bosons.
The strong dependence of the Higgs decay into weak gauge bosons on the Higgs
mass has the consequence that any uncertainty in the Higgs mass
translates into an uncertainty in the extraction of the couplings $g_{\hvv}$ from
the observation of the decay products.
In contrast, the coupling determination $g_{\hvv}$ from production
processes followed by decays $H\rightarrow b\bar b$
or $\gamma\gamma$ is less sensitive to the Higgs mass.
On the other hand the argument can also be turned around: the off-shell 
contributions
in $H\rightarrow VV^{(*)}$ are not very sensitive to the Higgs mass value $\mh$.
The different dependence of on- and off-shell contributions on the
Higgs mass and width can be exploited in various ways, as will be discussed
for some examples below.

\subsection{Background contributions to $e^+e^-\rightarrow ZVV$ and $e^+e^-\rightarrow \nu\bar\nu VV$}
\label{sec:backgroundVV}

\begin{figure}[htp]
\begin{center}
\begin{tabular}{ccccc}
\includegraphics[width=0.17\textwidth]{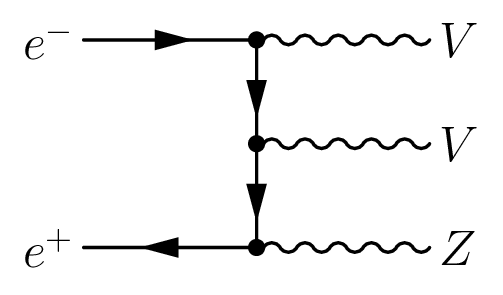} & 
\includegraphics[width=0.17\textwidth]{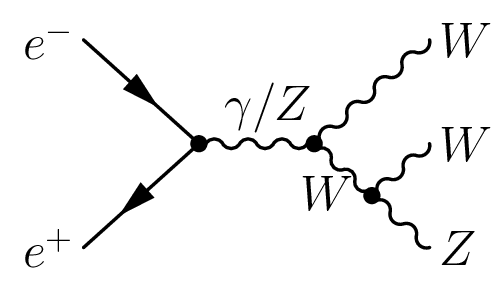} & 
\includegraphics[width=0.17\textwidth]{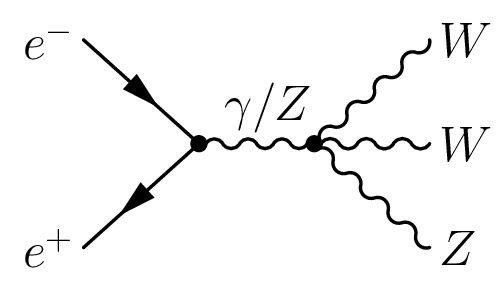} & 
\includegraphics[width=0.17\textwidth]{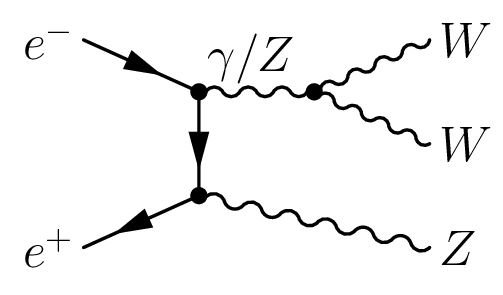} & 
\includegraphics[width=0.17\textwidth]{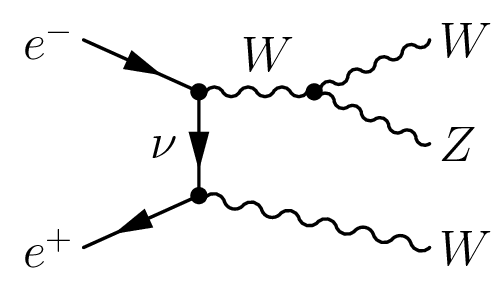} \\[-0.2cm]
 (a) & (b) & (c) & (d) & (e)
\end{tabular}
\end{center}
\vspace{-0.6cm}
\caption{Example background diagrams $e^+e^-\rightarrow ZZZ$
and $e^+e^-\rightarrow ZWW$ with $V=W,Z$.}
\label{fig:feynmanZVV} 
\end{figure}

\begin{figure}[htp]
\begin{center}
\begin{tabular}{cccc}
\includegraphics[width=0.17\textwidth]{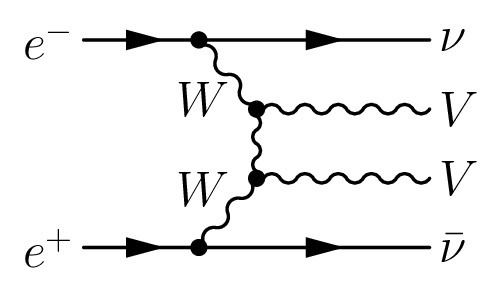} & 
\includegraphics[width=0.17\textwidth]{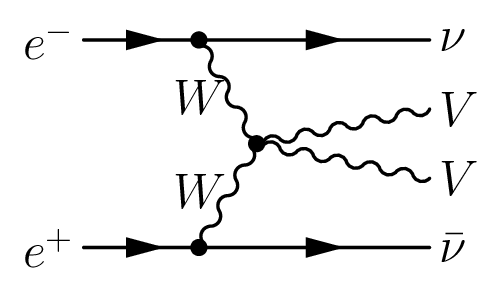} & 
\includegraphics[width=0.17\textwidth]{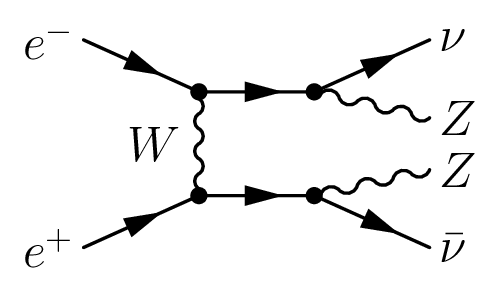} & 
\includegraphics[width=0.17\textwidth]{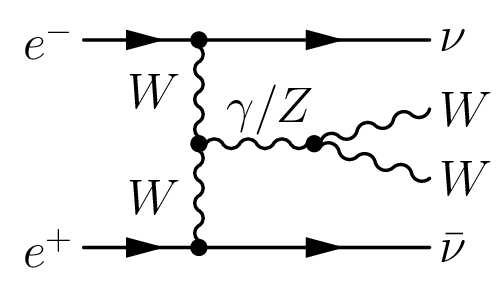}  \\[-0.2cm]
 (a) & (b) & (c) & (d)
\end{tabular}
\end{center}
\vspace{-0.6cm}
\caption{Example background diagrams $e^+e^-\rightarrow \nu\bar\nu ZZ$
and $e^+e^-\rightarrow \nu\bar\nu WW$ with $V=W,Z$.}
\label{fig:feynmannunuVV} 
\end{figure}

In order to estimate the quality of a measurement of the Higgs
induced off-shell contributions (named signal $S$ in the following),
for $\mvv> 2\mv$ a background ($B$) calculation as well as the interference ($I$)
of the signal with the background for large invariant masses $\mvv$ is needed.
For low invariant masses $\mvv<2\mv$ only resonant diagrams are of relevance,
which are either the signal diagrams through the Higgs
at $\mvv\approx \mh$ or possibly a $Z$~boson peak at $\mww\approx \mz$,
the latter for the $WW$ final state only.
The different types of Feynman diagrams for $e^+e^-\rightarrow ZVV$ are depicted in \fig{fig:feynmanZVV}.
For $e^+e^-\rightarrow \nu\bar\nu VV$ we will not list all types of diagrams.
Few examples are shown in \fig{fig:feynmannunuVV}, more diagrams can be directly
constructed from \fig{fig:feynmanZVV} by adding $Z\rightarrow \nu\bar\nu$.
All neutrino flavours have to be taken into account for the background.

\begin{figure}[htp]
\begin{center}
\begin{tabular}{ccc}
\includegraphics[width=0.33\textwidth]{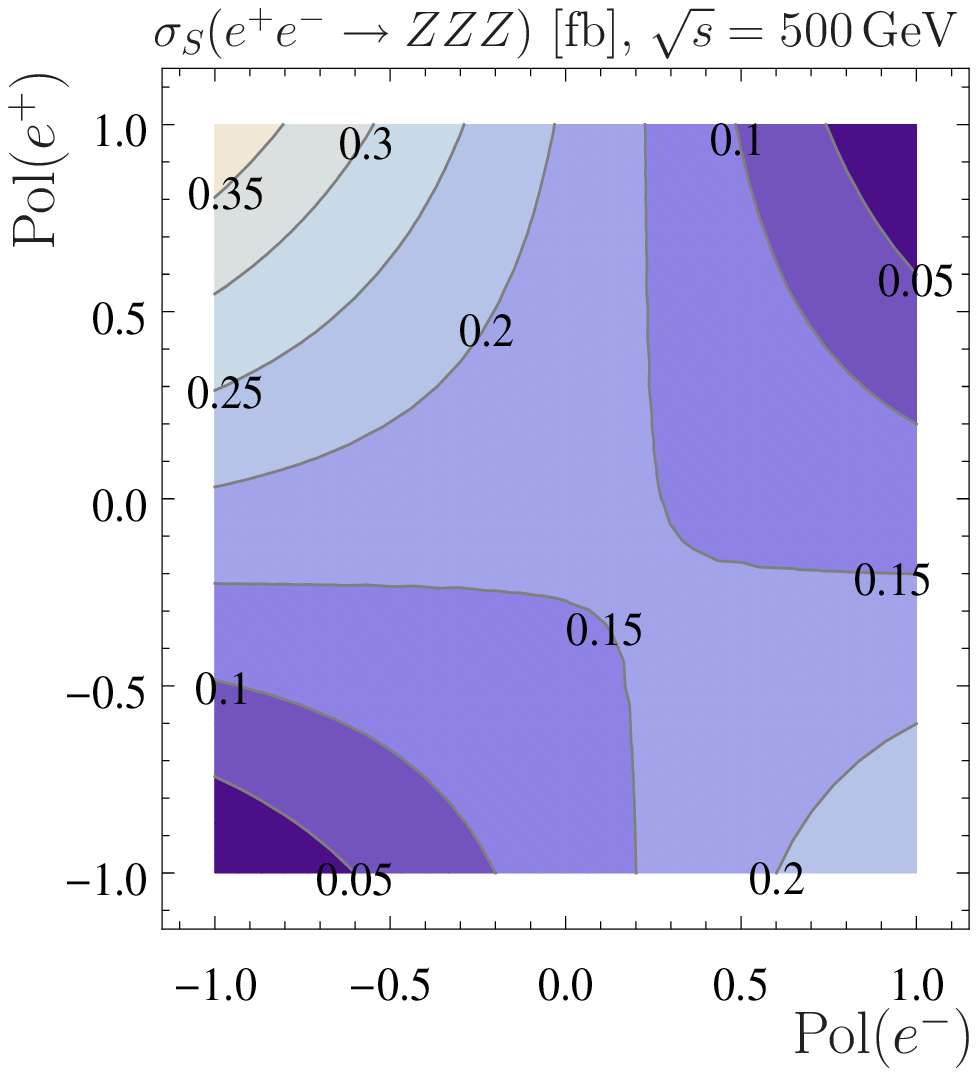} &
\includegraphics[width=0.33\textwidth]{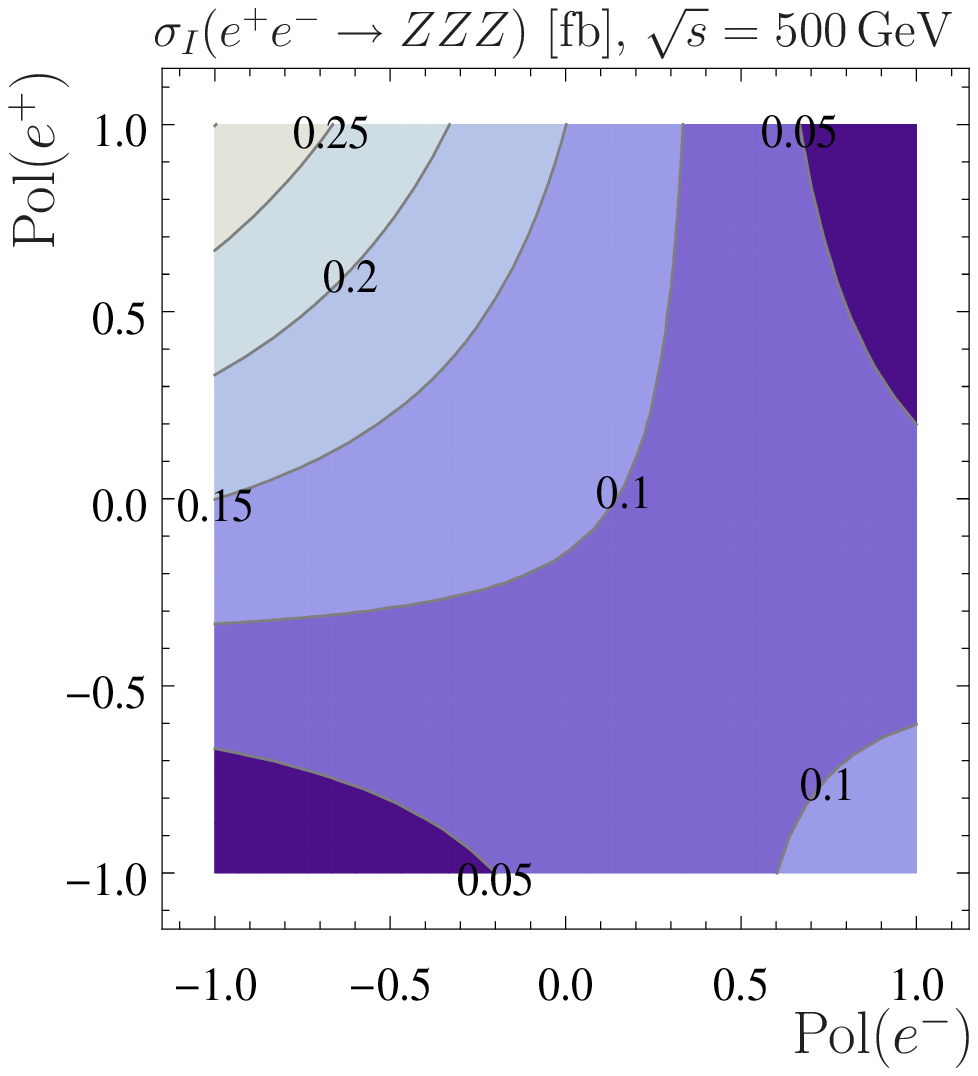} &
\includegraphics[width=0.33\textwidth]{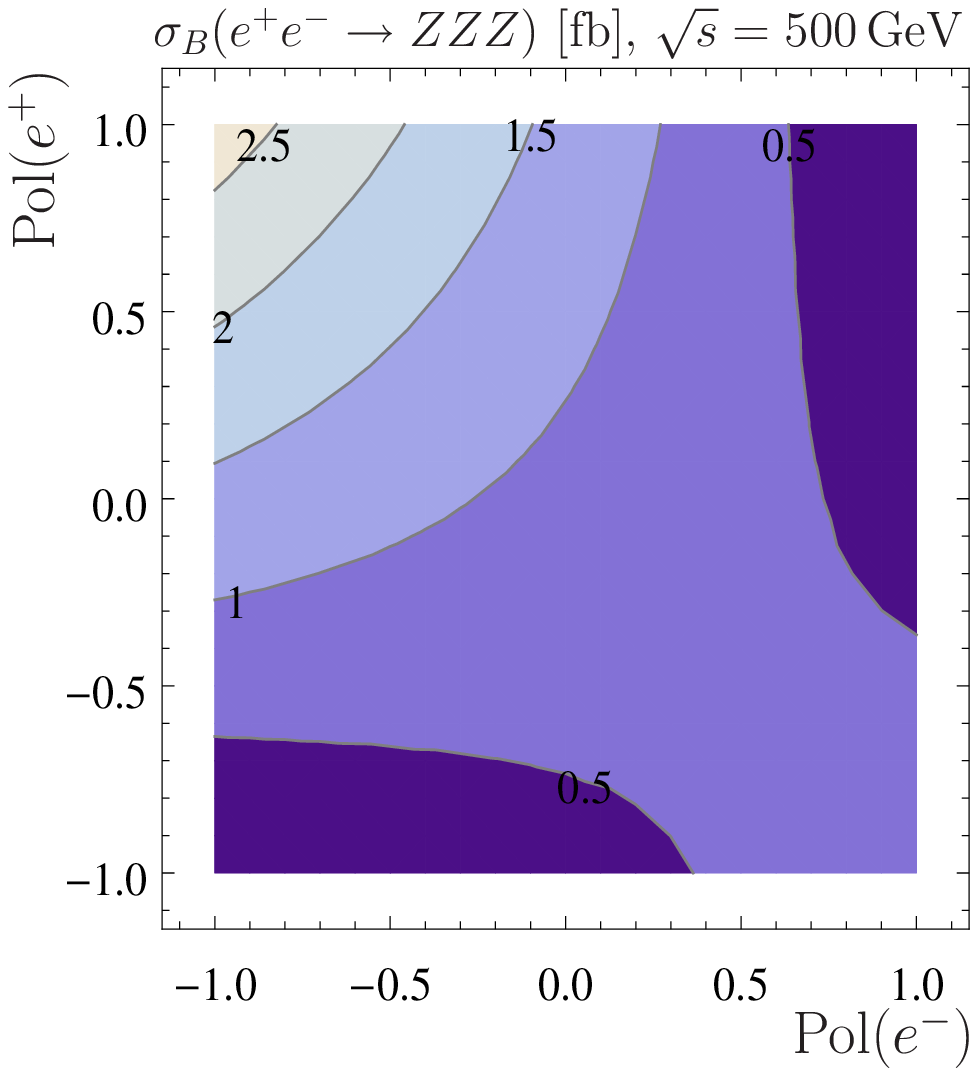} \\[-0.5cm]
 (a) & (b) & (c)
\end{tabular}
\end{center}
\vspace{-0.6cm}
\caption{$\sigma(2\mv,\sqrt{s}-\mv)$ in fb for $e^+e^-\rightarrow ZZZ$ as a function of the polarisation
of the initial state for $\sqrt{s}=500$\,GeV shown separately for
(a) the signal contribution~$S$; (b) the signal-background interference~$I$
and (c) the background contribution~$B$.
}
\label{fig:SBIpol} 
\end{figure}

In \fig{fig:ZZinvmass} above we display apart from the
Higgs induced contributions $S$ in black also 
the sum of all contributions $S+B+I$ (blue) for $e^+e^-\rightarrow ZZZ/\nu\bar\nu ZZ$.
The sum of all contributions is obtained
by {\tt MadGraph5\_aMC@NLO}~\cite{Alwall:2014hca} and extracted as a function of
$\mvv$ with the help of {\tt MadAnalysis}~\cite{Conte:2012fm}.
For the process $e^+e^-\rightarrow ZZZ$ we have also calculated the separate
contributions $S$, $B$ and $I$ with our own code and found agreement at
the permil level with the {\tt MadGraph5\_aMC@NLO} result.
The background prediction for this process is in accordance with \citere{Barger:1988fd}.
\fig{fig:WWinvmass} provides the corresponding result for
$e^+e^-\rightarrow ZWW/\nu\bar\nu WW$.
For $e^+e^-\rightarrow ZWW/\nu\bar\nu WW/\nu\bar\nu ZZ$
the interference term $I$ gives a negative contribution.
The destructive interference is of particular importance for the process
$e^+e^-\rightarrow \nu\bar\nu+4$\,jets, see \sct{sec:example1},
and the sensitivity to the Higgs width in the off-shell region,
see \sct{sec:width}. The destructive interference between signal and
background in this process
is in fact closely related to the preservation
of unitarity in scattering of longitudinal gauge bosons.
For $e^+e^-\rightarrow ZZZ$, however, the interference term provides a positive contribution
for all \cms{} energies and the initial state polarisations as shown in~\fig{fig:SBIpol}.
The relevant background diagram shown in \fig{fig:feynmanZVV}~(a) only includes
couplings of the $Z$ bosons to fermions, but no couplings of weak gauge bosons
among themselves.
Accordingly, the background induced by the diagram~\fig{fig:feynmanZVV}~(a) is by itself
not increasing with the \cms{} energy.
In case of $e^+e^-\rightarrow ZZZ$ the result after averaging over
the $ZZ$ pairs is understood as signal $S$. \fig{fig:SBIpol} additionally shows
that the maximal cross section for $e^+e^-\rightarrow ZZZ$ is obtained for the polarisation
Pol$(e^+,e^-)=(1.0,-1.0)$, however a suppression of the background and thus the signal-background interference
with respect to the signal contribution can be obtained for Pol$(e^+,e^-)=(-1.0,1.0)$.

\setlength{\tabcolsep}{-2pt}
\setlength{\tabcolsep}{6pt}

As depicted in \fig{fig:ZZinvmass} in case of the $H\rightarrow ZZ^{(*)}$ decay
for both production processes the signal $S$ and the signal-background interference $I$
for $\mzz>2\mz$ are about an order of magnitude smaller than the background.
According to \fig{fig:WWinvmass} in case of $H\rightarrow WW^{(*)}$ for both
production processes, however,
more background diagrams lead to a further suppression of the signal to background
ratio $S/B$. The absolute contribution from the interference term~$I$
gains in its relative size with respect to the signal $S$ and easily exceeds it.
We quantify the signal to background ratio below in \tab{tab:VVbackground}.
For $e^+e^-\rightarrow \nu\bar\nu VV$ the interference structures
in \fig{fig:ZZinvmass} and \fig{fig:WWinvmass} at $\mvv=\sqrt{s}-\mz$ are
induced by the process $e^+e^-\rightarrow ZVV$ followed
by $Z\rightarrow \nu\bar\nu$. The latter process is kinematically
strongly suppressed for larger invariant masses, since the $Z$ boson
decaying into a pair of (anti-)neutrinos needs to be off-shell
for $\mvv>\sqrt{s}-\mz$.

The ($s$-channel) Higgs induced contributions
$e^+e^-\rightarrow ZH\rightarrow ZVV\rightarrow \nu\bar\nu VV$,
where all neutrino flavours need to be taken into account,
are treated as a background to $e^+e^-\rightarrow \nu\bar\nu VV$.
However, the relevance
of Higgs induced contributions including $Z\rightarrow \nu\bar\nu$
for the final state $\nu\bar\nu VV$ is small.
Multiplying Higgsstrahlung $e^+e^-\rightarrow ZH$ with the
branching ratio BR$(Z\rightarrow \nu\bar\nu)\approx 20\%$ yields
cross sections smaller than the ones through
vector boson fusion $e^+e^-\rightarrow \nu\bar\nu H$ even
for energies $\sqrt{s}<500$\,GeV.
Additional cuts can moreover further reduce the amount
of Higgsstrahlung in $e^+e^-\rightarrow \nu\bar\nu VV$, since
the two neutrinos in vector boson fusion only induce a relatively small
missing transverse energy/momentum.

We quantify the signal/background ratio by defining
\begin{align}
\Delta_{\text{SB}}^\zvv=\frac{\sigma^\zvv_{\text{off}}(130\text{GeV},\sqrt{s}-\mz)}{\sigma^\zvv_{\text{all}}(130\text{GeV},\sqrt{s}-\mz)}\qquad
\text{and}
\qquad
\Delta_{\text{SB}}^\nunuvv=\frac{\sigma^\nunuvv_{\text{off}}(130\text{GeV},\sqrt{s})}{\sigma^\nunuvv_{\text{all}}(130\text{GeV},\sqrt{s})}\quad,
\end{align}
where $\sigma_{\text{all}}$ includes $S$, $I$ and $B$
contributions, whereas $\sigma_{\text{off}}$ just contains
the signal contributions $e^+e^-\rightarrow ZH\rightarrow ZVV$ and
$e^+e^-\rightarrow \nu\bar\nu H\rightarrow \nu\bar\nu VV$ as before.
The results for different \cms{} energies are presented in
\tab{tab:VVbackground}. Again we distinguish between the pure
usage of \eqn{eq:sigmaoff}, $e^+e^-\rightarrow ZH\rightarrow Z_1Z_2Z_3$,
for the signal contribution and the inclusion of all Higgs induced
diagrams for the process $e^+e^-\rightarrow ZH\rightarrow ZZZ$, where an averaging
over the three $ZZ$ pairs is implied.
Again we would like to stress that for the full process $e^+e^-\rightarrow ZZZ$ the
averaging over the three $ZZ$ pairs is obsolete.
Similarly, for $e^+e^-\rightarrow \nu\bar\nu WW$ we add the $t$-channel Higgs contributions
to capture all Higgs induced diagrams.

\begin{table}[htp]
\begin{center}
\begin{tabular}{| c || c | c || c | c |}
\hline
$\sqrt{s}$ & $\sigma^\zzz_{\text{all}}$ & $\Delta_{\text{SB}}^\zzznr$ ($\Delta_{\text{SB}}^\zzz$ )
& $\sigma^\nunuzz_{\text{all}}$ & $\Delta_{\text{SB}}^\nunuzz$
\\\hline\hline
$250$\,GeV & $---$      & $---$            & $1.51$\,fb & $0.04$\,\%\\\hline
$300$\,GeV & $0.34$\,fb & $3.19(12.9)$\,\% & $1.36$\,fb & $0.33$\,\%\\\hline
$350$\,GeV & $1.19$\,fb & $2.62(11.9)$\,\% & $1.66$\,fb & $1.01$\,\%\\\hline
$500$\,GeV & $2.06$\,fb & $2.83(11.6)$\,\% & $2.85$\,fb & $4.96$\,\%\\\hline
$1$\,TeV   & $1.71$\,fb & $4.40(10.2)$\,\% & $16.7$\,fb & $11.6$\,\%\\\hline\hline
$\sqrt{s}$ & $\sigma^\zww_{\text{all}}$ & $\Delta_{\text{SB}}^\zww$ 
& $\sigma^\nunuww_{\text{all}}$ & $\Delta_{\text{SB}}^\nunuww$\\\hline\hline
$250$\,GeV & $---$       & $---$      & $0.05$\,fb & $9.87(9.87)$\,\% \\\hline
$300$\,GeV & $7.34$\,fb  & $3.27$\,\% & $1.68$\,fb & $1.57(1.42)$\,\% \\\hline
$350$\,GeV & $29.2$\,fb  & $1.30$\,\% & $6.44$\,fb & $1.18(1.03)$\,\% \\\hline
$500$\,GeV & $91.8$\,fb  & $0.53$\,\% & $22.4$\,fb & $2.05(1.63)$\,\% \\\hline
$1$\,TeV   & $136.7$\,fb & $0.37$\,\% & $67.3$\,fb & $7.31(4.49)$\,\% \\\hline
\end{tabular}
\end{center}
\vspace{-0.5cm}
\caption{Inclusive cross sections $\sigma_{\text{all}}(130\text{GeV},\sqrt{s}-\mz)$
for $e^+e^-\rightarrow ZH\rightarrow ZVV$ and $\sigma_{\text{all}}(130\text{GeV},\sqrt{s})$ for
$e^+e^-\rightarrow \nu\bar\nu H \rightarrow \nu\bar\nu VV$
for Pol$(e^+,e^-)=(0.3,-0.8)$ and for different \cms{} energies $\sqrt{s}$ and
relative size of the signal/background ratio $\Delta_{\text{SB}}$ in \%.
In brackets we add the results averaging over the three $ZZ$ pairs
for $e^+e^-\rightarrow ZZZ$ and taking into account the $t$-channel Higgs contribution for
$e^+e^-\rightarrow \nu\bar\nu WW$.}
\label{tab:VVbackground}
\end{table}

As it can be seen from \tab{tab:VVbackground} $\Delta_{\text{SB}}$ is rather
constant for $e^+e^-\rightarrow ZZZ$ and increases
with energy for $e^+e^-\rightarrow \nu\bar\nu VV$, if the very small
cross section for $e^+e^-\rightarrow \nu\bar\nu WW$ for the \cms{}
energy $\sqrt{s}=250$\,GeV is not taken into account.
For $e^+e^-\rightarrow ZWW$ instead the ratio $\Delta_{\text{SB}}$
decreases with the \cms{} energy.
In all cases $\Delta_{\text{SB}}$ is of order $1-10$\,\% in the relevant regions
of the production processes. The influence of the signal-background interference is possibly
even larger and thus the prospects concerning the 
sensitivity to the Higgs contributions look promising.
In contrast to $\Delta_{\text{off}}$ the signal/background ratio $\Delta_{\text{SB}}$
is dependent on the initial polarisation, as it can also be seen from
\fig{fig:SBIpol}.
In \sct{sec:example1} we will perform a simulation with fermionic/hadronic
final states for $e^+e^-\rightarrow \nu\bar\nu +4$\,jets to investigate
the significance in terms of event rates. However, we can only provide
rough estimates and a rather qualitative discussion.
Further studies taking into account higher
order corrections, beamstrahlung, hadronization of jets
and a proper detector simulation would be desirable.

\section{Phenomenological implications of off-shell contributions}
\label{sec:practical}

In this section we want to investigate the consequences of the off-shell Higgs contributions
for the $Z$ recoil method and the extraction of $HVV$ couplings. Moreover, we comment on
their role for unitarity cancellations in gauge boson scattering
and their possible impact on constraining higher-dimensional
operators, which can, for instance, be induced in composite Higgs scenarios.
The connection to the Higgs width is analysed in the subsequent sections.

\subsection{$Z$ recoil method}
\label{sec:Zrecoil}

\begin{figure}[ht]
\begin{center}
\begin{tabular}{ccc}
\includegraphics[width=0.33\textwidth]{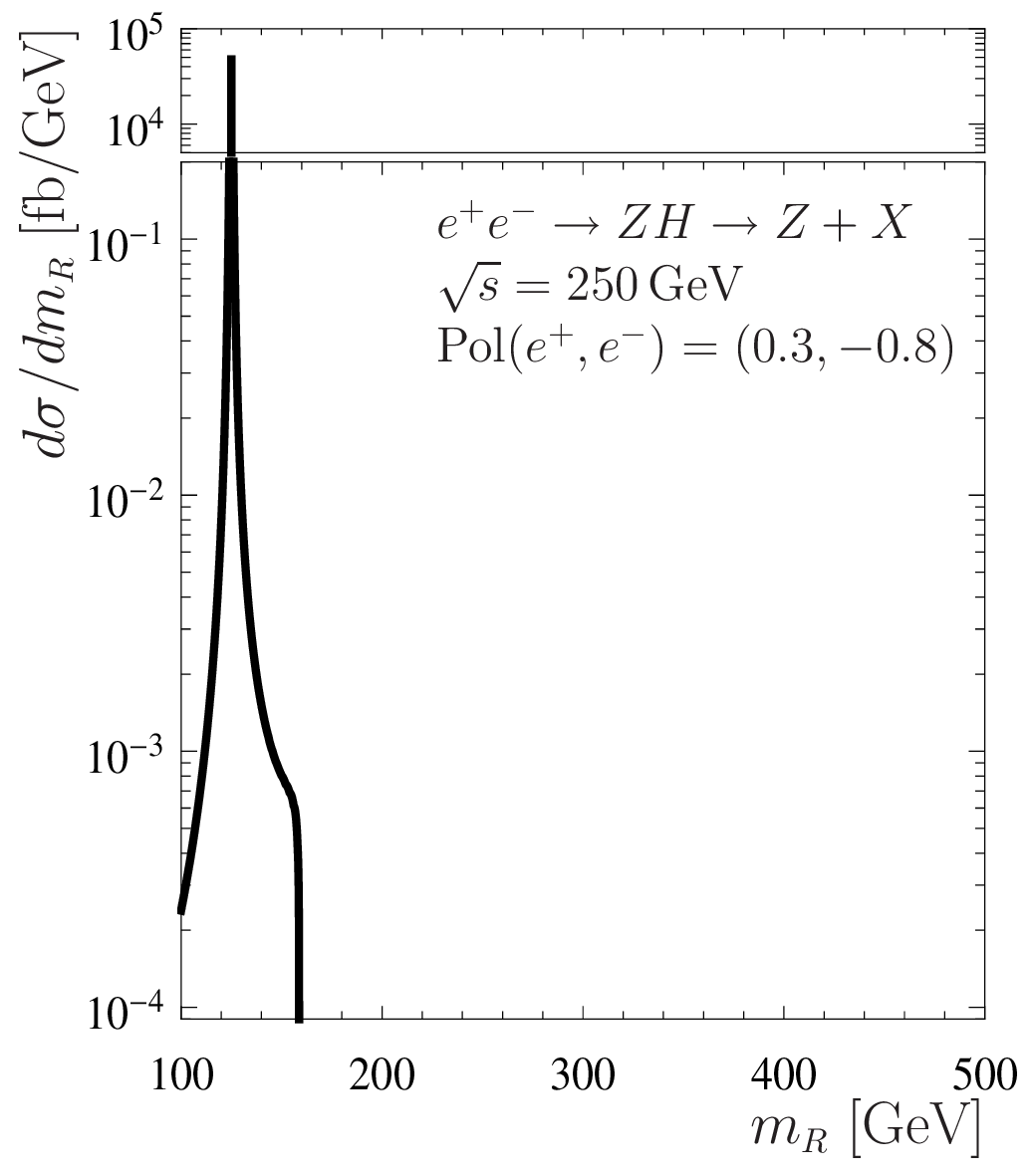} &
\includegraphics[width=0.33\textwidth]{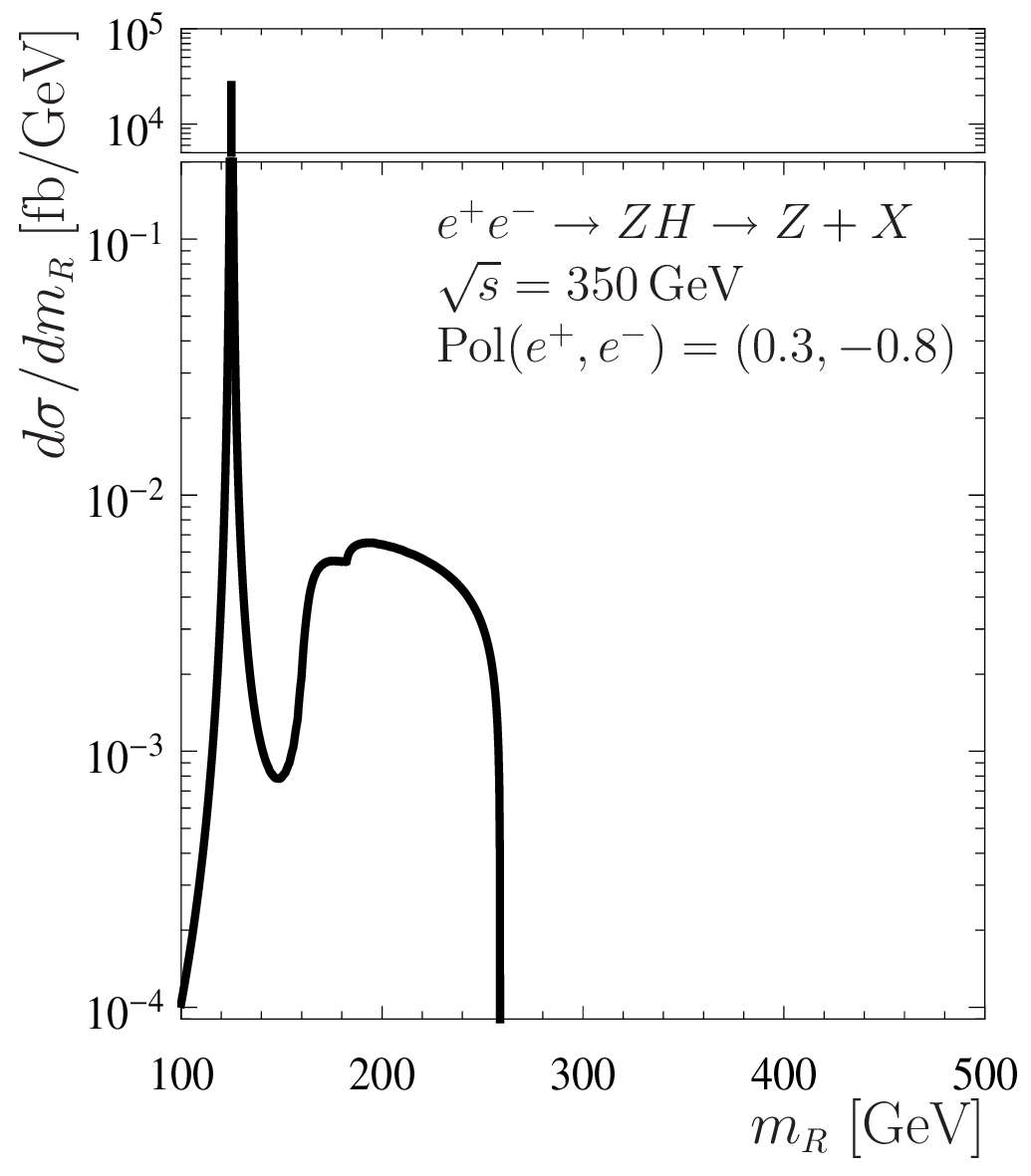} &
\includegraphics[width=0.33\textwidth]{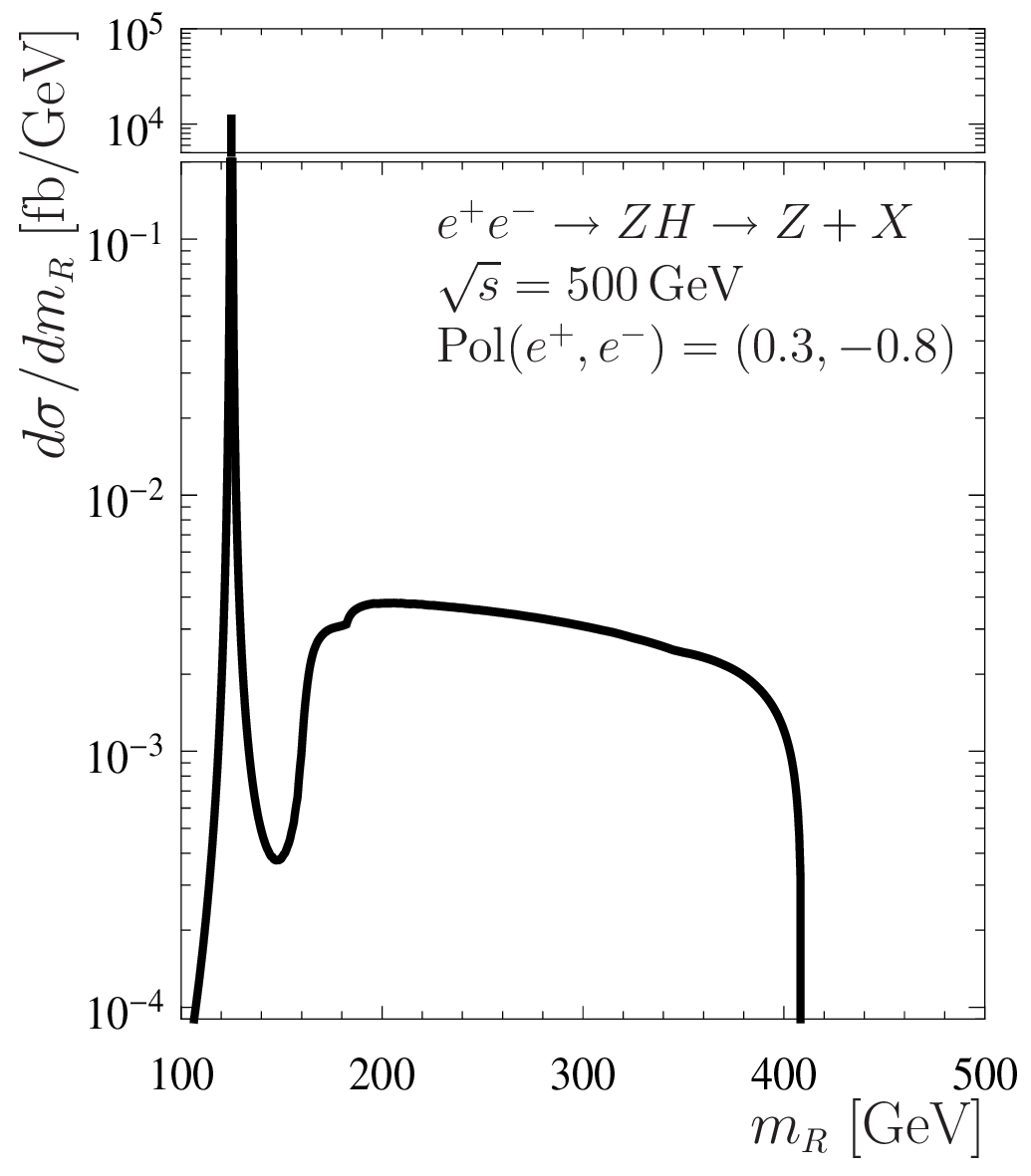} \\[-0.5cm]
 (a) & (b) & (c) 
\end{tabular}
\end{center}
\vspace{-0.6cm}
\caption{$d\sigma/d\mr$ in fb/GeV as a function of $\mr$ in GeV
 for $e^+e^-\rightarrow ZH\rightarrow Z+X$ defined in \eqn{eq:sigmaoff} (with $\mvv$
 replaced by $\mr$)
 combined with the sum over $H\rightarrow X=ZZ^{(*)}, WW^{(*)}, b\bar b, t\bar t, gg, \tau^+\tau^-$
 for a fixed polarisation Pol$(e^+,e^-)=(0.3,-0.8)$ and \cms{} energies
 (a-c) $\sqrt{s}=250,\, 350,\, 500$\,GeV.}
\label{fig:Zrecoil} 
\end{figure}

As pointed out, the $Z$ recoil mass measurement is a key feature
of a linear collider which allows to access the production process
$e^+e^-\rightarrow ZH$ through the decays of the $Z$ boson only, so that
an absolute measurement of the production cross section is possible.
The analysis is primarily based on the decays $Z\rightarrow e^+e^-/\mu^+\mu^-$ \cite{Li:2012taa},
where by the invariant mass and the energy of the $l^+l^-$ system
the reconstructed mass $\hat{m}_\smallz$ and the energy~$E_\smallz$
of the $Z$ boson are obtained. Recently also hadronic final states
were discussed \cite{Miyamoto:2013zva,TalksLCWS}. The recoil mass $\mr$
is computed according to
\begin{align}
 \mr^2 = s + \hat{m}_\smallz^2 - 2E_\smallz\sqrt{s}
\end{align}
and thus equals the invariant mass of the Higgs boson $p_H^2$.
According to our discussion off-shell effects in Higgs boson decays manifest themselves
in the differential cross section $d\sigma/d\mr$, which we demonstrate
in \fig{fig:Zrecoil} for the Higgsstrahlung production process.
The figures show the results obtained by \eqn{eq:sigmaoff}, where the invariant
mass $\mvv$ is replaced by $\mr$, combined with the sum over the partial decays
$H\rightarrow ZZ^{(*)}, WW^{(*)}, b\bar b, t\bar t, gg, \tau^+\tau^-$
as provided by the \lhchxswg{}.
The increase in the differential cross section at the thresholds $\mr=2\mw$
and $\mr=2\mz$ is clearly visible. Moreover at $\mr=2m_t$ additionally
the decay $H\rightarrow t\bar t$ opens kinematically.
In order to quantify the off-shell contributions we use again $\Delta_{\text{off}}$
defined in \eqn{eq:deltaoff} translated to $e^+e^-\rightarrow ZH\rightarrow Z+X$
with $\mr$ instead of $\mvv$ and present the results in \tab{tab:Zrecoil}.

\setlength{\tabcolsep}{-1pt}
\setlength{\tabcolsep}{6pt}

\begin{table}[htb]
\begin{center}
\begin{tabular}{| c || c | c | c | c | c |}
\hline
$\sqrt{s}$            & $250$\,GeV & $300$\,GeV & $350$\,GeV & $500$\,GeV & $1$\,TeV\\\hline
$\Delta_{\text{off}}$ & $0.02$\,\% & $0.12$\,\% & $0.30$\,\% & $0.91$\,\% & $1.84$\,\%\\\hline
\end{tabular}
\end{center}
\vspace{-0.6cm}
\caption{Off-shell contributions for the signal cross section
determined via the $Z$~recoil method.}
\label{tab:Zrecoil}
\end{table}

As expected from the analysis of 
\fig{fig:ZZinvmass} and \fig{fig:WWinvmass}, the
off-shell contributions are unimportant for the case of
$\sqrt{s}=250$\,GeV. Because of the presence of the decay mode 
$H \rightarrow b \bar b$, which dominates for $\mr=120-130$\,GeV, and of the other
relevant decay modes for a \sm{}-like Higgs, the off-shell effects induced by
the $H\rightarrow ZZ^{(*)}$ and $H \rightarrow WW^{(*)}$ modes are less pronounced than in 
\fig{fig:ZZinvmass} and \fig{fig:WWinvmass}, but still clearly visible in
\fig{fig:Zrecoil} for $\sqrt{s}=350$\,GeV and $\sqrt{s}=500$\,GeV.
For $\sqrt{s}=500$\,GeV the off-shell contributions amount to about $1$\% (at 
$\sqrt{s}=1$\,TeV they are close to $2$\%). While these off-shell effects are
relatively small, for $\sqrt{s}=500$\,GeV and above they are nevertheless
relevant for analyses aiming at an accuracy at the percent level.
The potential problem caused by the presence of off-shell
contributions is that the cross section that is determined via the recoil
method actually contains a non-negligible amount of off-shell
contributions, while it is interpreted as an on-shell cross section. The
impact of the off-shell contributions can be reduced by appropriate cuts,
for instance a cut on the recoil mass $\mr\in [115,150]$\,GeV.
Some care is necessary in this case in order to determine the appropriate
efficiencies.
In case of $H\rightarrow ZZ^{(*)}$, where another on-shell $Z$~boson is involved in the process,
a misidentification of the $Z$ boson out of the Higgsstrahlung process can occur.
Again in the most pessimistic approach
an average over the final state $Z$ bosons is performed,
which we included in \fig{fig:Zrecoil}.
We note that this averaging and thus the misidentification of $ZZ$ pairs
lowers the total on-shell cross section by about $1-2$\,\% compared to the
correct discrimination of all $ZZ$ pairs.

While the effects of the off-shell contributions on the determination
of the production cross section via the $Z$ recoil method have turned out
to be relatively small, our analysis nevertheless adds to the motivation
for performing the cross-section determination via the $Z$ recoil method
close to threshold, i.e.\ at about $\sqrt{s}=250-350$\,GeV, rather than at
higher energies where the off-shell effects become relevant.

\subsection{$HVV$ couplings, unitarity and higher-dimensional operators}
\label{sec:unitarity}

Off-shell contributions also play a role for the 
extraction of $HVV$ couplings at an $e^+e^-$ collider. 
While in the studies carried out so far usually the 
validity of the \zwat{} has been assumed, for precision analyses
it will be important to discriminate the on-shell coupling
$g^{\text{on}}_{\hvv}$ from off-shell contributions,
$g_{\hvv}(\mvv)$,
through appropriate cuts on the invariant mass of the 
decay products.
An analysis where this will be relevant
is for example  the determination of the $HWW$ coupling from 
$e^+e^-\rightarrow \nu\bar\nu H\rightarrow \nu\bar\nu WW$
at $\sqrt{s}=500$\,GeV~\cite{Durig:2014lfa,Kumar:2015eea},
where both on- and off-shell Higgs contributions are present.
As mentioned in \sct{sec:higgsmass},
for accurate predictions of processes involving the decay of an 
on-shell Higgs boson into weak bosons and thus for the determination
of $g^{\text{on}}_{\hvv}$ also
a precise knowledge of the Higgs mass $\mh$ will be crucial.

Off-shell Higgs induced contributions in
the scattering of longitudinal gauge bosons are known to be of crucial 
importance for preserving unitarity. The corresponding amplitude involving
contributions from the gauge sector
increases with the square of the \cms{} energy in the high-energy limit.
This bad high-energy
behaviour is cancelled by Higgs-exchange contributions in 
models where a Higgs sector with at least one fundamental scalar
particle gives rise to electroweak symmetry breaking.
Accordingly, the interference term
between the Higgs-exchange and the background contributions
is large and negative, as discussed in \sct{sec:backgroundVV}.
Detailed studies of high-energy vector boson scattering
are an essential test for electroweak symmetry breaking.
The investigation of deviations from the \sm{} prediction in form of effective field
theories requires the application of a unitarization prescription \cite{Kilian:2014zja}.

The off-shell Higgs contributions are known to be a sensitive probe 
in particular of compositeness \cite{Giudice:2007fh} described through higher-dimensional
operators \cite{Barger:2003rs}, which affect $WW\rightarrow WW/HH$.
A detailed study on the sensitivity of 
$e^+e^-\rightarrow \nu\bar\nu VV$ with subsequent $VV\rightarrow 4$\,jets
can be found in \citere{Contino:2013gna}. We will also discuss the Higgs
induced contributions to this process and the process 
$e^+e^-\rightarrow \mu^+\mu^-+4$\,jets in the following. For the case
of the \lhc{} and collider-independent statements we refer to the studies presented in
\citeres{Gainer:2014hha,Ghezzi:2014qpa,Azatov:2014jga,Cacciapaglia:2014rla,Buschmann:2014sia,Barducci:2015oza}.

\section{Processes $e^+e^-\rightarrow \nu\bar\nu +4$\,jets and $e^+e^-\rightarrow \mu^+\mu^-+4$\,jets}
\label{sec:example1}

In order to work out the practical consequences of our discussion
and to investigate the sensitivity to off-shell Higgs contributions,
we have performed a Monte Carlo
simulation of the partonic process $e^+e^-\rightarrow \nu\bar\nu +4$\,jets
using {\tt MadGraph5\_aMC@NLO} at leading order.
We present results for an integrated luminosity of $\lumi=500$\,fb$^{-1}$
at energies $\sqrt{s}=350, 500$\,GeV and $1$\,TeV for a polarisation
of the initial state of Pol$(e^+,e^-)=(0.3,-0.8)$.
In \sct{sec:ISRNLO} we shortly discuss the changes due
to the inclusion of initial state radiation, which also allows one
to estimate effects of beamstrahlung.
Moreover the inclusion of higher-order contributions
is described within this section.

The first process that we consider is 
$e^+e^-\rightarrow \nu\bar\nu +4$\,jets. This process is suitable for
the investigation of off-shell contributions for several reasons: 
the $H\rightarrow ZZ^{(*)}$ and $H\rightarrow WW^{(*)}$ decays both give
rise to $4$\,jet final states with a relatively high event rate, and the
invariant mass of the intermediate Higgs can be reconstructed from the
decay products (without the need
for averaging between different final states).
For the theoretical prediction we include all three neutrino
flavours in the final state (i.e., not only the electron neutrinos
produced in the diagram of \fig{fig:feynmanZZ}b).
A jet is understood as being either a gluon or one of the (anti-)quarks
$u,d,s,c$.
In contrast a $b$-(anti-)quark in the final state would change the picture
due to the decay $H\rightarrow b\bar b$.
It should be noted that we do not employ parton showering/hadronization,
but denote (anti-)quarks and gluons as final state jets.
For a jet we demand a minimum transverse momentum of $p_T>20$\,GeV,
a maximal rapidity of $|y|<5$ and 
$\Delta_R \equiv \sqrt{(\Delta \phi)^2+(\Delta y)^2}>0.4$
between the various jets to allow for jet separation. $\Delta \phi$ and $\Delta y$
denote the azimuthal angular and rapidity differences between two jets.
For a massless particle the rapidity $y=5$ corresponds to an opening angle of
$0.77^\circ$ between the particle three-momentum and the beam axis.
As explained below the transverse momentum of all four jets $\ptj$ is 
required to be larger than $75$\,GeV to reduce background from the
process $e^+e^-+4$\,jets.

\setlength{\tabcolsep}{-2pt}
\begin{figure}[tb]
\begin{center}
\begin{tabular}{ccc}
\includegraphics[width=0.34\textwidth]{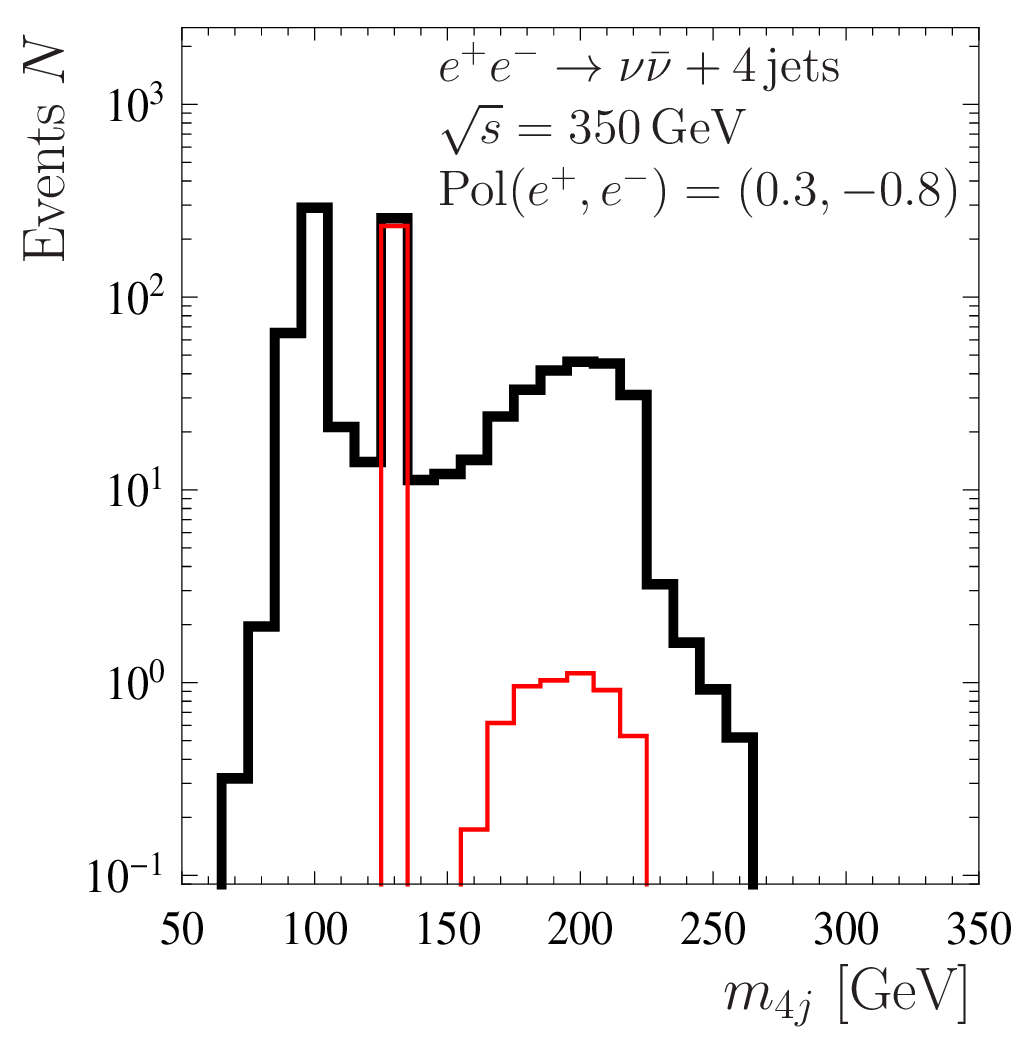} &
\includegraphics[width=0.34\textwidth]{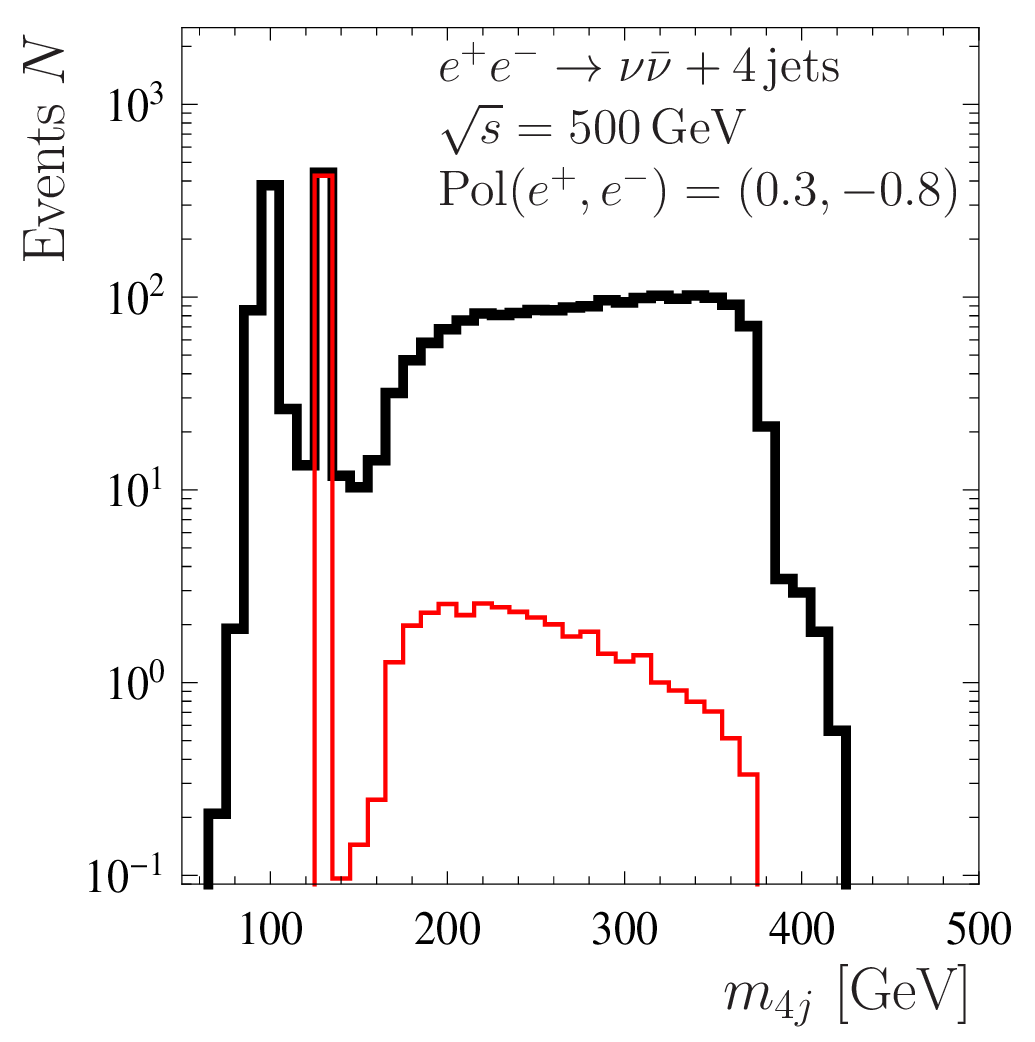} &
\includegraphics[width=0.34\textwidth]{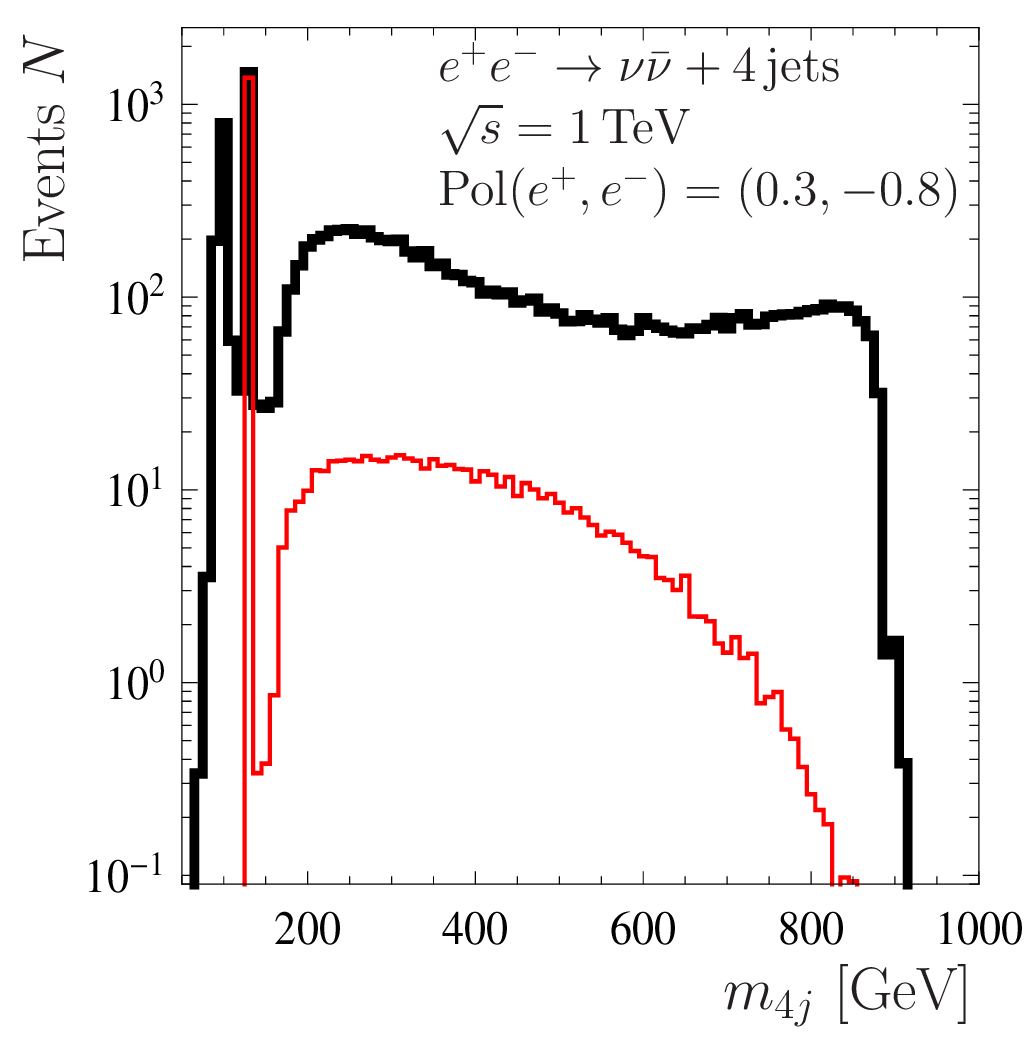} \\[-0.5cm]
 (a) & (b) & (c) 
\end{tabular}
\end{center}
\vspace{-0.6cm}
\caption{Event rates for $e^+e^-\rightarrow \nu\bar\nu +4$jets for $\lumi=500$\,fb$^{-1}$
as a function of the invariant mass of the $4$\,jets $m_{4j}$ in bins
of $10$\,GeV width after the cut $\ptj>75$\,GeV
for (a) $\sqrt{s}=350$\,GeV, (b) $\sqrt{s}=500$\,GeV
and (c) $\sqrt{s}=1$\,TeV.
The Higgs-induced contributions are shown in red.
}
\label{fig:vvjjjjstudy} 
\end{figure}
\setlength{\tabcolsep}{6pt}

We show our results in \fig{fig:vvjjjjstudy}.
The red contribution includes only ($s$-channel) Higgs induced
contributions, as discussed
in previous sections, and corresponds to an estimate of signal and
interference contributions. 
In those we exclude contributions where the $4$\,jets do not
stem from a Higgs. The latter case can occur in case of $e^+e^-\rightarrow ZH$
followed by $Z\rightarrow jj$ and $H\rightarrow \nu\bar\nu jj$.
The treatment of interference terms between
$e^+e^-\rightarrow ZH, Z\rightarrow jj, H\rightarrow \nu\bar\nu jj$
and the signal contributions 
is not straightforward, and we count those interference
contributions as part of the signal (red).
However, since the Higgsstrahlung production process
$e^+e^-\rightarrow ZH$ multiplied with
an additional branching ratio is essentially
irrelevant for $\sqrt{s}>500$\,GeV,
the net effect of this kind of contributions
is negligible. Also for $\sqrt{s}=350$\,GeV the effect
is rather small.
We note that the indicated signal contribution
is --- of course --- not a physical observable. Our aim here is to 
merely illustrate the relevance of Higgs-induced contributions
(assuming \sm{}-like couplings) in comparison with the full cross
section, i.e.\ the total number of events.
Even more important than the signal $S$ 
indicated in red is the signal--background interference $I$, which
for the process under consideration is negative.
Within the \sm{} at high \cms{} energies, where \vbf{} dominates,
its absolute size is larger than $S$, such that the inclusion of
the Higgs-induced contributions lowers the total number of events.

We quantify the number of Higgs-induced events, indicated as signal
in \fig{fig:vvjjjjstudy}, as $N_H$ in \tab{tab:nrevents}
and add the number of events without Higgs contribution
in any Feynman diagram $N_{\rm woH}$, which allows to 
read off the impact of the interference term $I$ when
comparing to the total number of events $N$.
In particular at $\sqrt{s}=1$\,TeV the off-shell contributions 
give rise to a sizable fraction of the total number of events.
At low energies, on the other hand, the sensitivity to off-shell
contributions is statistically
limited. The relevance of Higgs off-shell events 
in $e^+e^-\rightarrow \nu\bar\nu+4$\,jets can be enhanced by
enforcing \vbf{} induced final states by e.g. setting an upper limit
on the missing transverse energy (\met{}) in energy regions where \vbf{}
dominates over Higgsstrahlung.

\begin{figure}[ht]
\begin{center}
\begin{tabular}{cc}
\includegraphics[width=0.45\textwidth]{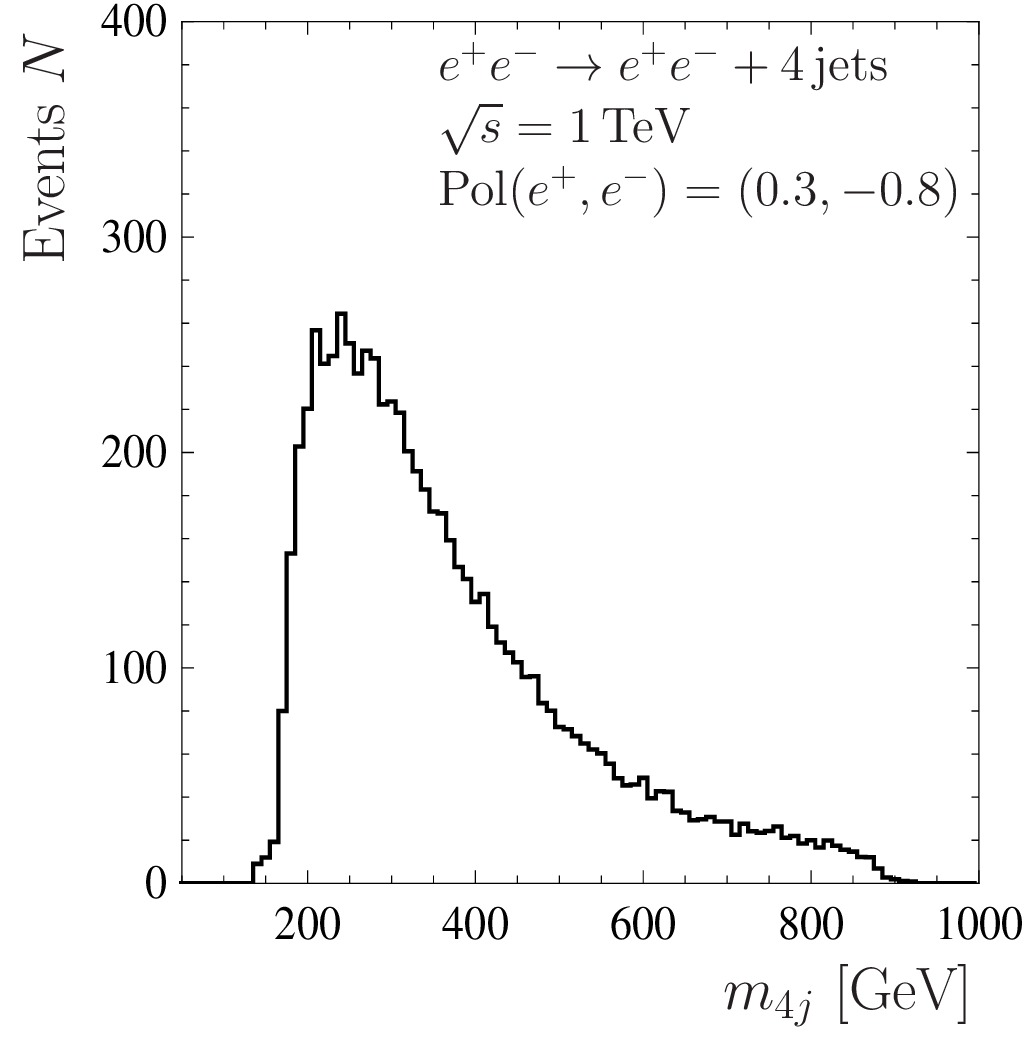} &
\includegraphics[width=0.45\textwidth]{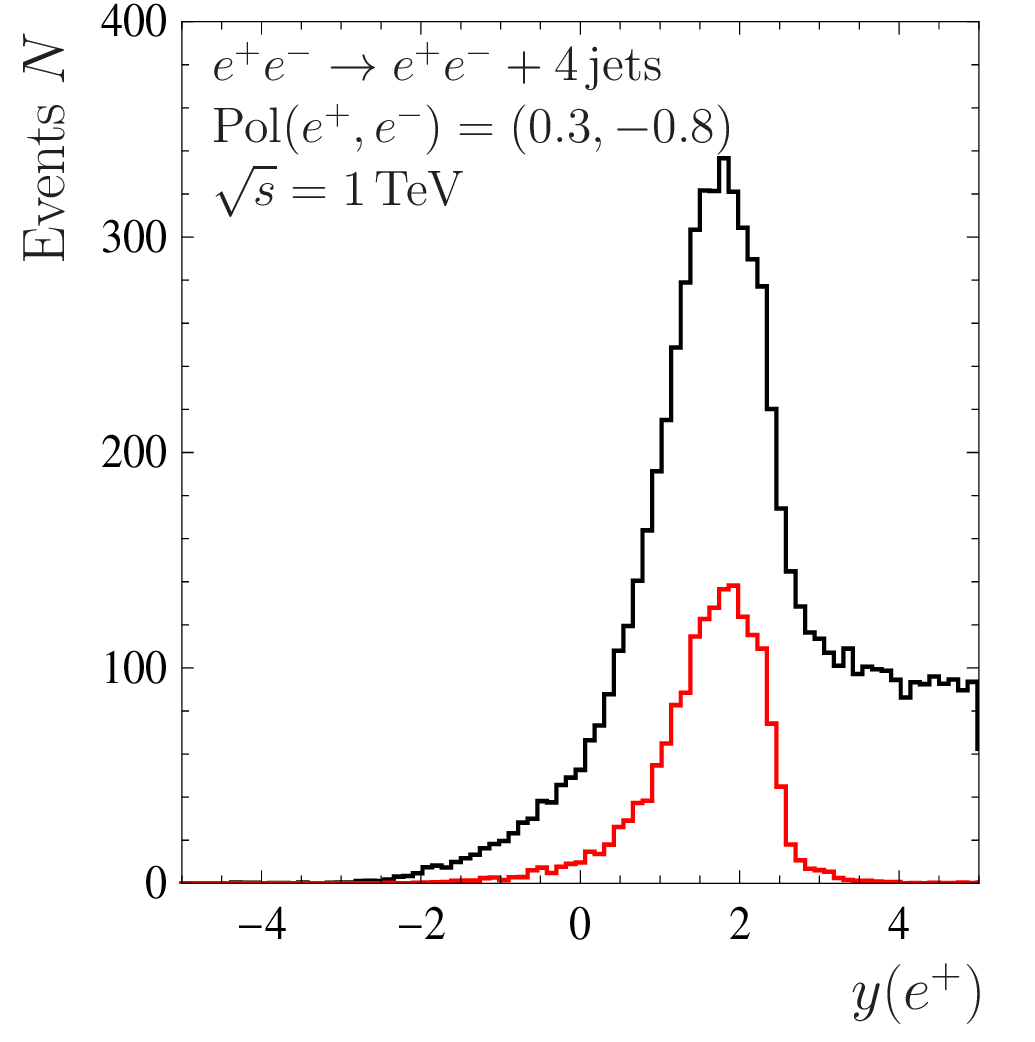} \\[-0.5cm]
 (a) & (b)
\end{tabular}
\end{center}
\vspace{-0.6cm}
\caption{Event rates for $e^+e^-\rightarrow e^+e^- +4$jets 
for $\sqrt{s}=1$\,TeV 
and $\lumi=500$\,fb$^{-1}$
after the cuts $m_{4j}>130$\,GeV and $\ptj>75$\,GeV
as a function of (a) the invariant mass of the $4$\,jets $m_{4j}$;
(b) the rapidity of the positron $y(e^+)$. 
In the right plot we show in red events with $|y(e^-)|>2.5$, 
out of which
events with $|y(e^+)|>2.5$ correspond to the background $N_B$ (see
text).
}
\label{fig:eejjjjstudy} 
\end{figure}

A potentially large background to $e^+e^-\rightarrow \nu\bar\nu+4$\,jets is induced by the
final state $e^+e^-+4$\,jets, where the two leptons remain undetected, i.e.\
stay close to the beam pipe. 
We assume here that leptons can be reconstructed up to rapidities
of $|y|<2.5$, which for a massless particle corresponds to
an angle of $9.38^\circ$ between particle three-momentum and beam axis.
Thus, the background can be strongly suppressed by a lower
cut on the transverse momentum of the four jets $\ptj>75$\,GeV,
which forces the electron and positron to have a combined $p_T$
of more than $75$\,GeV and thus rather small rapidities.
Accordingly, this cut was introduced already
in our investigation of the process
$e^+e^-\rightarrow \nu\bar\nu +4$\,jets (see the 
results shown in \fig{fig:vvjjjjstudy} above),
where less events are lost by the requirement $\ptj>75$\,GeV,
even for \vbf{}.
The remaining number of background events 
from the process $e^+e^-+4$\,jets is 
denoted by $N_B$ in \tab{tab:nrevents}.
In \fig{fig:eejjjjstudy} we show distributions for the process
$e^+e^-\rightarrow e^+e^-+4$\,jets applying the cuts
$m_{4j}>130$\,GeV and $\ptj>75$\,GeV, but at first
allowing for arbitrary rapidities of both leptons.
The relevant background events $N_B$ are those with rapidities $|y(e^\pm)|>2.5$,
which can be deduced from \fig{fig:eejjjjstudy} (b).

It should be noted that the process
$e^+e^-\rightarrow e^+e^-+4$\,jets of course includes Higgs-induced
events. However, the cross section of $Z$ boson fusion is considerably smaller 
than the one of $W$ boson fusion,
and for Higgsstrahlung including $Z\rightarrow e^+e^-$ the probability that both
leptons escape undetected is small. In that manner 
$e^+e^-\rightarrow e^+e^-+4$\,jets with undetected leptons
can be considered as a pure background contribution.

\begin{table}[htp]
\begin{center}
\begin{tabular}{| c || c | c | c |}
\hline
$\sqrt{s}$      & $350$\,GeV & $500$\,GeV & $1$\,TeV \\\hline\hline
$N$             & $265$      & $1793$     & $7994$ \\\hline
$N_H$           & $6$        & $34$       & $510$   \\\hline
$N_{\rm woH}$       & $256$      & $1771$     & $8298$ \\\hline\hline
$N_B$           & $0$        & $0$        & $65$ \\\hline
\end{tabular}
\end{center}
\vspace{-0.5cm}
\caption{Number of events $N$ with $\mj>130$\,GeV and $\ptj>75$\,GeV
for $e^+e^-\rightarrow \nu\bar\nu +4$jets for $\lumi=500$\,fb$^{-1}$.
$N_H$ corresponds to the ($s$-channel) Higgs induced events. $N_{\rm woH}$ is the number of
events without any Higgs contribution allowing to estimate the interference $I$.
$N_B$ are background events due to $e^+e^-\rightarrow e^+e^-+4$\,jets
(see text).}
\label{tab:nrevents}
\end{table}

As a second example we consider the process $e^+e^-\rightarrow \mu^+\mu^-+4$\,jets, where
we again demand a minimum transverse momentum of $p_T>20$\,GeV for the jets,
a maximal rapidity of $|y|<5$ and $\Delta_R>0.4$
between the various jets to ensure jet separation.
Concerning the detection of the two final state leptons,
we again assume that a rapidity of $|y|<2.5$ is required.
In contrast to the previous process, the final state $\mu^+\mu^-+4$\,jets is not
induced by \vbf{} and therefore shows a generally smaller interference term,
but necessarily also overall smaller event rates.
The sensitivity of both processes to effects of the Higgs-boson width
will be investigated in \sct{sec:width}.

\section{Initial state radiation and higher-order effects}
\label{sec:ISRNLO}

\begin{figure}[ht]
\begin{center}
\begin{tabular}{ccc}
\includegraphics[width=0.34\textwidth]{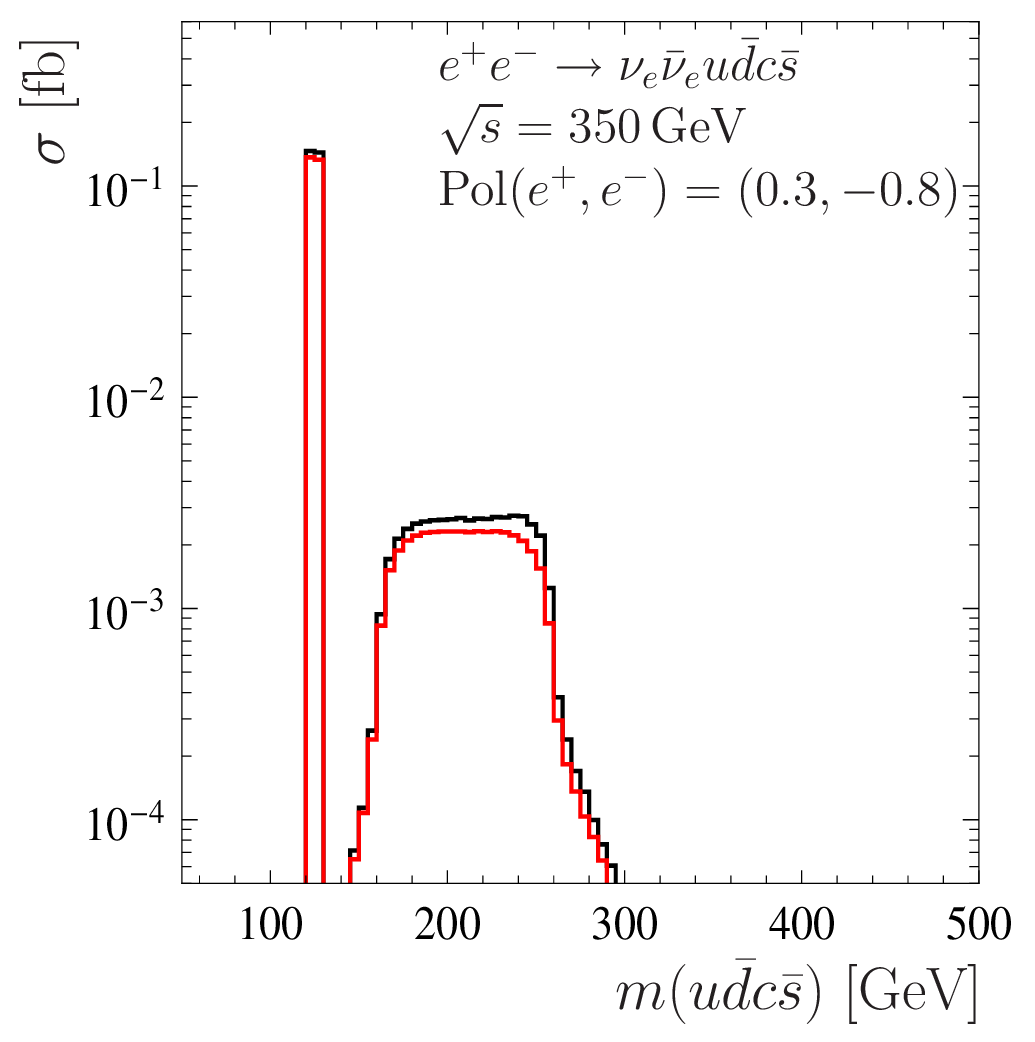} &
\includegraphics[width=0.34\textwidth]{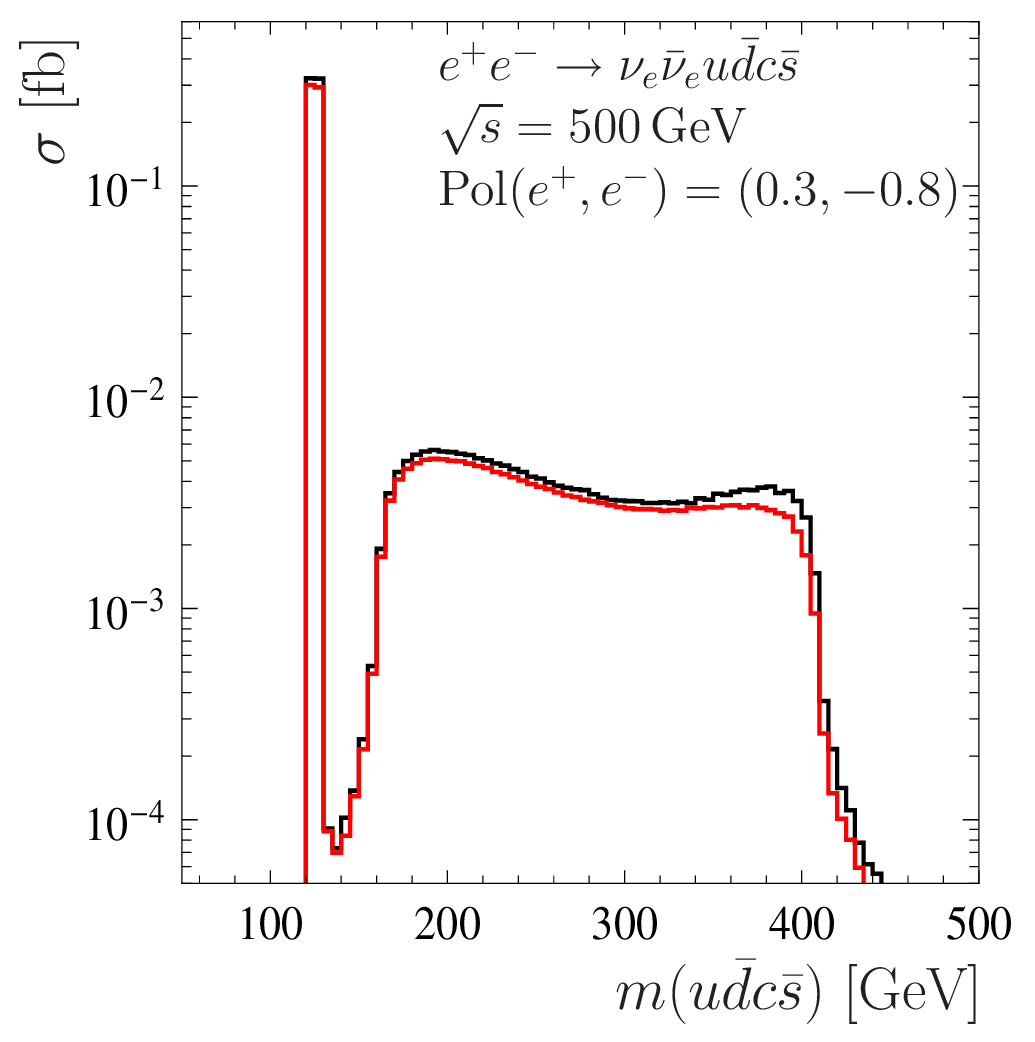} &
\includegraphics[width=0.34\textwidth]{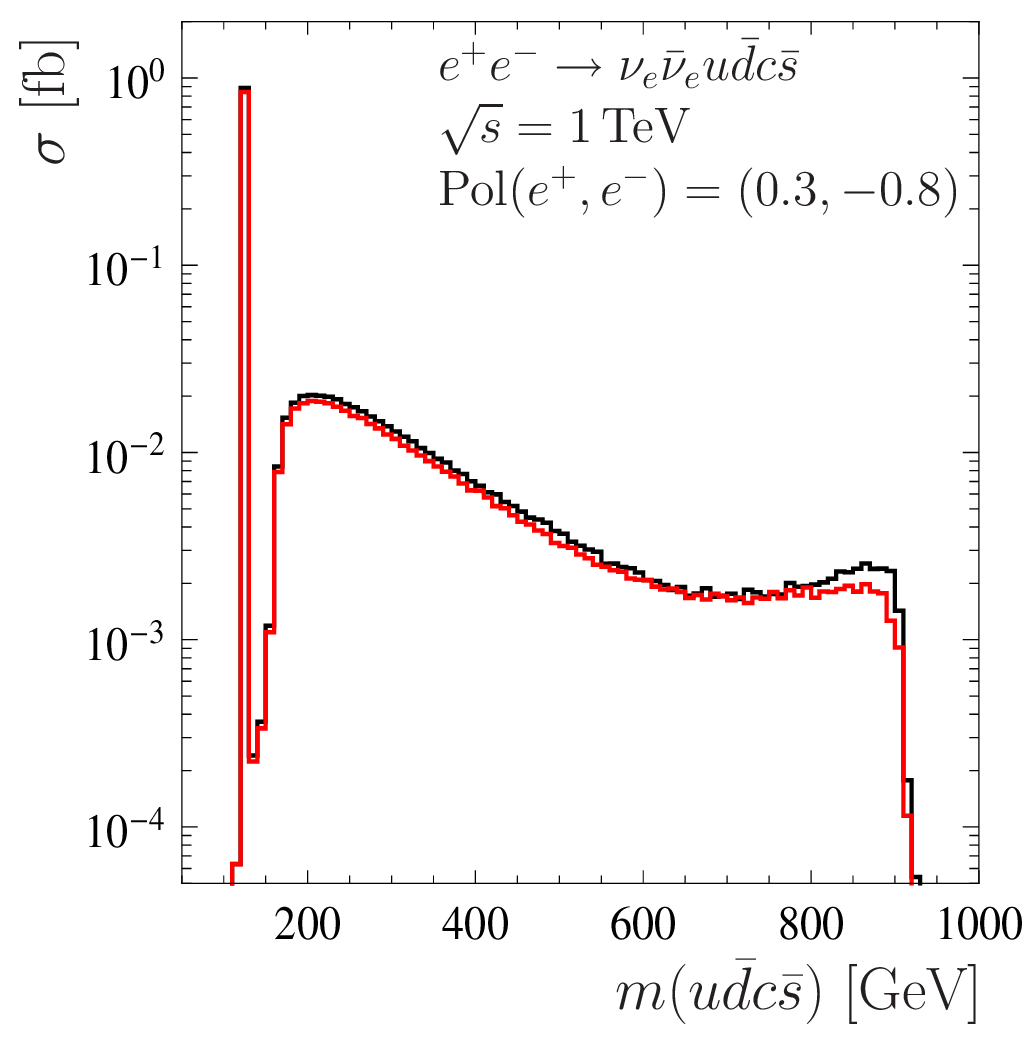} \\[-0.5cm]
 (a) & (b) & (c) 
\end{tabular}
\end{center}
\vspace{-0.6cm}
\caption{Cross section for $e^+e^-\rightarrow \nu_e\bar\nu_e u\bar d c\bar s$
in fb in bins as a function of $m(u\bar d c\bar s)$ in GeV for
(a) $\sqrt{s}=350$\,GeV, (b) $\sqrt{s}=500$\,GeV and (c) $\sqrt{s}=1$\,TeV. 
The black curves show the results without and the red curves the
results including initial state radiation.}
\label{fig:lusifer} 
\end{figure}

In this section we want to investigate the impact of the inclusion of 
initial state radiation and other higher-order effects.
For this purpose we have repeated our study for the process
$e^+e^-\rightarrow \nu_e\bar\nu_e u\bar d s\bar c$,
being a subprocess of $e^+e^-\rightarrow \nu\bar\nu+4$\,jets,
with the code {\tt LUSIFER} \cite{Dittmaier:2002ap}, which allows 
the user to apply different schemes for the treatment of finite widths
and to include initial state radiation. We choose $\sqrt{s}=500$\,GeV, Pol$(e^+,e^-)=(0.3,-0.8)$
and require for a minimum jet energy of $10$\,GeV. 
Since $\ptj$ is only nontrivially accessible in {\tt LUSIFER}
the corresponding cut is not applied, however for each outgoing
(anti-)quark a minimal energy of $E_q>10$\,GeV is required.
The inclusive cross sections obtained
with {\tt LUSIFER} and {\tt MadGraph5\_aMC@NLO} agree at lowest order without
the inclusion of initial state radiation within their numerical errors
applying the cut of $E_q>10$\,GeV for each outgoing (anti-)quark.
We have also verified that the predictions of the two codes 
for the differential cross section as function of the invariant mass of the 
four quarks agree with each other very well.
We choose input-parameter scheme $2$ of {\tt LUSIFER} and
include gluonic contributions, which are however tiny and
also part of the {\tt MadGraph5\_aMC@NLO} result.
In order to test the reliability of the 
prediction in the high invariant mass region, we choose different
width schemes for the gauge bosons and find negligible differences. In particular 
the usage
of a fixed width leads to results that just differ at the permil level
from the ones obtained using a complex-pole scheme, even for large invariant masses
of the four quarks final state.
The effect of initial state radiation turns out to be more relevant. It
reduces the overall cross section by a few percent and affects
mostly the region of high invariant masses, see \fig{fig:lusifer}.

\setlength{\tabcolsep}{-2pt}
\setlength{\tabcolsep}{6pt}

A similar effect can be expected from the inclusion of beamstrahlung.
Thus, both effects should be taken into account for precision analyses, 
but they do not change the qualitative features of our results.
The inclusion of higher-order corrections is expected to have a bigger
impact for $e^+e^-\rightarrow \nu_e\bar\nu_e W^+W^-$,
as reported in \citere{Accomando:2006hq}.
In \citere{Accomando:2006hq} 
the corrections have been calculated 
in the equivalent vector-boson approximation. This
approximation only includes diagrams involving the subprocess $WW\rightarrow
WW$, and the quality of this approximation is expected to improve with increasing
\cms{} energy. However, 
even in the high mass region this approximation has been found to depart 
by up to $10$\% from the exact matrix element calculation at tree-level.
Nevertheless, this method can serve as an estimate of the impact of 
higher-order effects: the 
(logarithmic) electroweak higher-order corrections
for $\sqrt{s}=1$\,TeV reach $-6.7$\%, where several cuts
like e.g.\ $\mww>400$\,GeV have been employed. However, the correction
strongly increases with the invariant mass of the gauge bosons
and amounts up to $-20$\,\% in the region $\mvv=700-800$\,GeV.
Thus, higher-order effects are relevant in this context and should be taken
into account for a precise analysis of off-shell effects. As one can see 
from the dependence of the corrections on the invariant mass, this kind of
corrections can in general not be taken into account by a simple global 
``K-factor''.

\section{Constraints on the Higgs width}
\label{sec:width}

For the reconstruction of the width of the Higgs boson, $\GaH$, the linear
collider offers a unique method through the measurement of
the recoil against the $Z$ boson in $e^+e^-\rightarrow ZH$ in combination with
the measurement of branching ratios.
The method was already discussed in \sct{sec:Zrecoil}. It is based on
the validity of the \zwat{}
\begin{align}
\sigma_\zwa=
\frac{\sigma_{A\rightarrow H}\Gamma_{H\rightarrow B}}{\GaH}\propto 
\frac{(g_A^{\text{on}}g_B^{\text{on}})^2}{\GaH}\quad,
\label{eq:sigZWAon}
\end{align}
where the index ``$\text{on}$'' refers to on-shell couplings.
Running an $e^+e^-$ collider at rather low \cms{} energies of $\sqrt{s}=250-350$\,GeV,
the $Z$ recoil method allows to determine $\sigma_{e^+e^-\rightarrow ZH}$ and 
thus $\left(g^{\text{on}}_{HZZ}\right)^2$
by just observing the decay products of the $Z$ boson into leptons.
Recently also $Z$ boson decays into hadrons were discussed, which provide a
higher sensitivity, but on the other hand are more difficult
to distinguish from Higgs decays~\cite{Miyamoto:2013zva,TalksLCWS}.
By a combination of various final states of the Higgs decays
(see e.g.\ \citere{Han:2013kya}) the Higgs width $\GaH$ can be extracted,
in case the \zwat{} is valid or at least off-shell contributions can be reduced
by reasonable cuts.

\subsection{Combination of on- and off-shell contributions}

On the other hand, similar to proposed methods at the \lhc{}
\cite{Caola:2013yja,Kauer:2012hd,Kauer:2013cga,Kauer:2013qba,Campbell:2013una,Campbell:2013wga,Chen:2013waa},
the combination of on- and off-shell contributions can be used to
obtain constraints on the Higgs width. We will discuss this method
and the underlying assumptions in the following.

Changing the product of the squared couplings entering 
the production cross section and the partial decay width by a common factor
leaves the inclusive on-shell cross section $\sigma_\zwa$ in \eqn{eq:sigZWAon}
unchanged, if at the same time the total width appearing in the denominator of
\eqn{eq:sigZWAon} is rescaled accordingly.
For instance, for the process $e^+e^-\rightarrow ZH\rightarrow ZZZ$
the inclusive on-shell cross section can be expressed by scale
factors relative to the \sm{} cross
section~\cite{LHCHiggsCrossSectionWorkingGroup:2012nn,Heinemeyer:2013tqa},
\begin{align}
 \sigma^{\zzz}_{\zwa}
 = \frac{(\kappa_\smallz^{\text{on}})^4}{r}\sigma_{\zwa}^{\zzz,\sm}
 =: \mu_{\text{on}}^{\zzz} \sigma_{\zwa}^{\zzz,\sm}\quad,
\end{align}
where $\kappa^{\text{on}}_\smallv
=g^{\text{on}}_{\hvv}/g_{\hvv}^{\text{on},\sm}$, $r=\GaH/\GaH^{\sm}$, and $\mu_{\text{on}}$ is
the signal strength obtained from on-shell measurements for the process under consideration.
Similarly, for the Higgs induced processes $e^+e^-\rightarrow ZWW/\nu\bar\nu ZZ/\nu\bar\nu WW$
the scale factors in the numerator read
$\kappa_\smallz^2\kappa_\smallw^2$, $\kappa_\smallz^2\kappa_\smallw^2$
and $\kappa_\smallw^4$, respectively.
The off-shell contributions are not proportional to $1/\GaH$, but depend on the off-shell
propagator, see e.g.\ \eqn{eq:sigmaoff}.
In the approximation where the dependence of the off-shell propagator on
the total width is neglected, the off-shell cross section can formally be
expressed in terms of off-shell scale factors%
\footnote{The extension of the concept of tree-level inspired scale
factors $\kappa_i$ to off-shell quantities is in general questionable,
see the discussion in \citere{GWtalk}. We merely use off-shell scale
factors in a formal sense here as a shorthand for the parametrisation of
deviations from the \sm{}.}
as
\begin{align}
 \frac{d\sigma^{\zzz}_{\text{off}}}{d\mzz} = 
(\kappa^{\text{off}}_\smallz (\mzz))^4
 \frac{d\sigma_{\text{off}}^{\zzz,\sm}}{d\mzz} = 
\mu_{\text{off}}^{\zzz}(\mzz) \frac{d\sigma_{\text{off}}^{\zzz,\sm}}{d\mzz}
 \qquad\text{for}\qquad \mzz > \mh \quad.
\end{align}
The off-shell scale factors $\kappa^{\text{off}}(\mvv)$ and signal strengths 
$\mu_{\text{off}}(\mvv)$ depend in general on the invariant mass $\mvv$.
The change in the scale factors for different values of $\mvv$ arises
from the running of the couplings induced by loop contributions and in
particular from threshold effects associated with additional particles
beyond the \sm{}. If effects of this kind are disregarded and it is assumed
that the off-shell scale factors $\kappa^{\text{off}}(\mvv)$ can be set
equal to their on-shell values $\kappa^{\text{on}}$ for the whole
considered range of $\mvv$ values, the ratio of the off-shell and
on-shell signal strengths provides information about the total width,
\begin{align}
\frac{\mu_{\text{off}}(\mvv)}{\mu_{\text{on}}} = r \quad \text{for} \quad 
\kappa^{\text{off}}(\mvv) = \kappa^{\text{on}} \quad.
\end{align}
In particular, an upper bound on the total width can be obtained under
those assumptions from the
measurement of $\mu_{\text{on}}$
and the determination of an upper
bound on $\mu_{\text{off}}(\mvv)$. This procedure can be repeated
for all final states independently.

The question how well the off-shell contribution of the signal can
be discriminated against the background clearly plays an important role
regarding the sensitivity that can be achieved with this method. 
In this context the signal-background interference~$I$
for $\mvv>2\mv$ is of large relevance, see \sct{sec:backgroundVV}.
Assuming that the background behaves \sm{}-like, the interference term
$I$ is expected to scale like 
$\sqrt{\mu_{\text{off}}(\mvv)}=\sqrt{\mu_{\text{on}} r}$.
The interference term yields a mostly negative contribution
and thus lowers the sensitivity to the Higgs width.
In our numerical analysis below we assume that the measured value
for the on-shell cross section agrees with the \sm{} value, i.e.\
$\mu_{\text{on}}=1$, and we furthermore assume
$\kappa_\smallv \equiv \kappa_\smallz = \kappa_\smallw$ 
for simplicity.

\subsection{Impact of \bsm{} contributions}

The method described above relies on the strong theoretical
assumption that the effective couplings far off-shell are the same as
their on-shell counterparts. The relation between
$\kappa^{\text{off}}(\mvv)$ and $\kappa^{\text{on}}$ can however be
severely affected by contributions from physics beyond the 
\sm{} (\bsm{}), in particular via threshold effects. In fact,
\bsm{} effects of this kind may actually be needed to give rise
to a Higgs-boson width that differs from the one of the \sm{} by the amount
that is currently probed in the analyses at the \lhc{}. 
Examples for the possible impact of \bsm{} effects on the \lhc{}
analyses have recently been investigated in
\citeres{Englert:2014aca,Ghezzi:2014qpa,Englert:2014ffa,Logan:2014ppa}.
In particular, in \citere{Englert:2014ffa}
the validity of the Higgs width bound has been discussed in different
\bsm{} models, and non-resonant and resonant
contributions in the off-shell region have been classified. 
As an example, squark contributions in the minimal supersymmetric standard model (\mssm{})
or it simplest extension by adding a singlet (\nmssm{}) can alter the off-shell
region by non-resonant contributions affecting the gluon fusion
production cross section at the \lhc{}.
It should be noted that even the pure \sm{} loop contribution from
top quarks to the decay of an off-shell Higgs boson,
$H\rightarrow VV$, show a sensitive dependence on $\mvv$, in 
particular across the threshold where $\mvv=2m_t$. The description of
those loop contributions in terms of a universal, i.e.\
$\mvv$-independent, scale factor is therefore a rather poor approximation in
the off-shell region.

In contrast to the loop-induced gluon fusion production process
at the \lhc{}, the corresponding processes at an $e^+e^-$ collider are 
a priori less affected by loop contributions of \bsm{} particles,
since those loop contributions
have to compete with the leading tree-level type contributions to
Higgs production at an $e^+e^-$ collider.
Nevertheless, sizeable effects in the off-shell contributions could also arise
from the presence of additional Higgs bosons at tree level, see e.g.\
\citere{Logan:2014ppa}.

\subsection{Application to the linear collider and \lhc{} implications}

In the following we want to investigate which sensitivity one obtains
at a linear collider for constraining the Higgs width from the comparison of
on-shell and off-shell contributions, if one uses the same assumption about
the equality of the on-shell and off-shell effective couplings as in the \lhc{}
analyses. Because of the discussed problems of this assumption regarding the
possible presence of sizeable \bsm{} contributions, we focus our
attention to the region where the Higgs width differs from the \sm{} case
only by a rather small amount.
We consider again
the process $e^+e^-\rightarrow\nu\bar\nu +4$\,jets simulated with {\tt MadGraph5\_aMC@NLO}.
We apply the same cuts as described in \sct{sec:example1}.
Assuming an on-shell signal strength of $\mu_{\text{on}}=1$, the 
dependence of the number of events on $r$ can be written in the form
\begin{align}
 N(r)=N_0(1+R_1\sqrt{r}+R_2 r)+N_B\quad.
 \label{eq:nrevents}
\end{align}
$N_0$ differs from $N_{\rm woH}$ by on-shell Higgs events. $N_B$
are background events $e^+e^-\rightarrow e^+e^- +4$\,jets with undetected leptons
and can be taken from \tab{tab:nrevents}. Their dependence on $r$ is negligible for $r<10$. 

\begin{table}[htp]
\begin{center}
\begin{tabular}{| c || c | c | c |}
\hline
$\sqrt{s}$                            & $350$\,GeV & $500$\,GeV & $1$\,TeV \\\hline\hline
$N_0$ ($\lumi=500$\,fb$^{-1}$)        & $263$      & $1775$     & $8420$ \\\hline
$R_1$                                 & $-0.017$   & $-0.010$   & $-0.098$   \\\hline
$R_2$                                 & $0.026$    & $0.019$    & $0.048$ \\\hline
Limit on $r$ ($\lumi=500$\,fb$^{-1}$) & $7.0$     & $3.8$     & $2.8$ \\\hline\hline
Limit on $r$ ($\lumi=1$\,ab$^{-1}$)   & $5.1$     & $3.1$     & $2.5$ \\\hline
\end{tabular}
\end{center}
\vspace{-0.5cm}
\caption{$N_0$, $R_1$ and $R_2$ as a function of the \cms{} energy for $e^+e^-\rightarrow\nu\bar\nu +4$\,jets
with $\mj>130$\,GeV and $\ptj>75$\,GeV. The upper limits on $r$ at $95$\% have been obtained
according to our simplistic Bayesian approach, 
using the assumptions specified in the text.}
\label{tab:r1r2nunubar}
\end{table}

We have performed a simulation for three values of $\sqrt{s}$ 
corresponding to an integrated luminosity of $\lumi=500$\,fb$^{-1}$ at each 
\cms{} energy. The results for the parameters $N_0$, $R_1$ and $R_2$, which
we have obtained from a fit, are given in \tab{tab:r1r2nunubar}.
As expected the interference term, reflected in $R_1$, is large
and negative and thus lowers the sensitivity around $r\sim 1$.
The interference term is largest for the largest \cms{} energy, since
there the \vbf{} channel dominates. For $\sqrt{s}= 350$\,GeV and $500$\,GeV,
on the other hand, the relative importance of the Higgsstrahlung process is
higher, reducing the impact of the interference term.

\begin{figure}[ht]
\begin{center}
\includegraphics[width=0.5\textwidth]{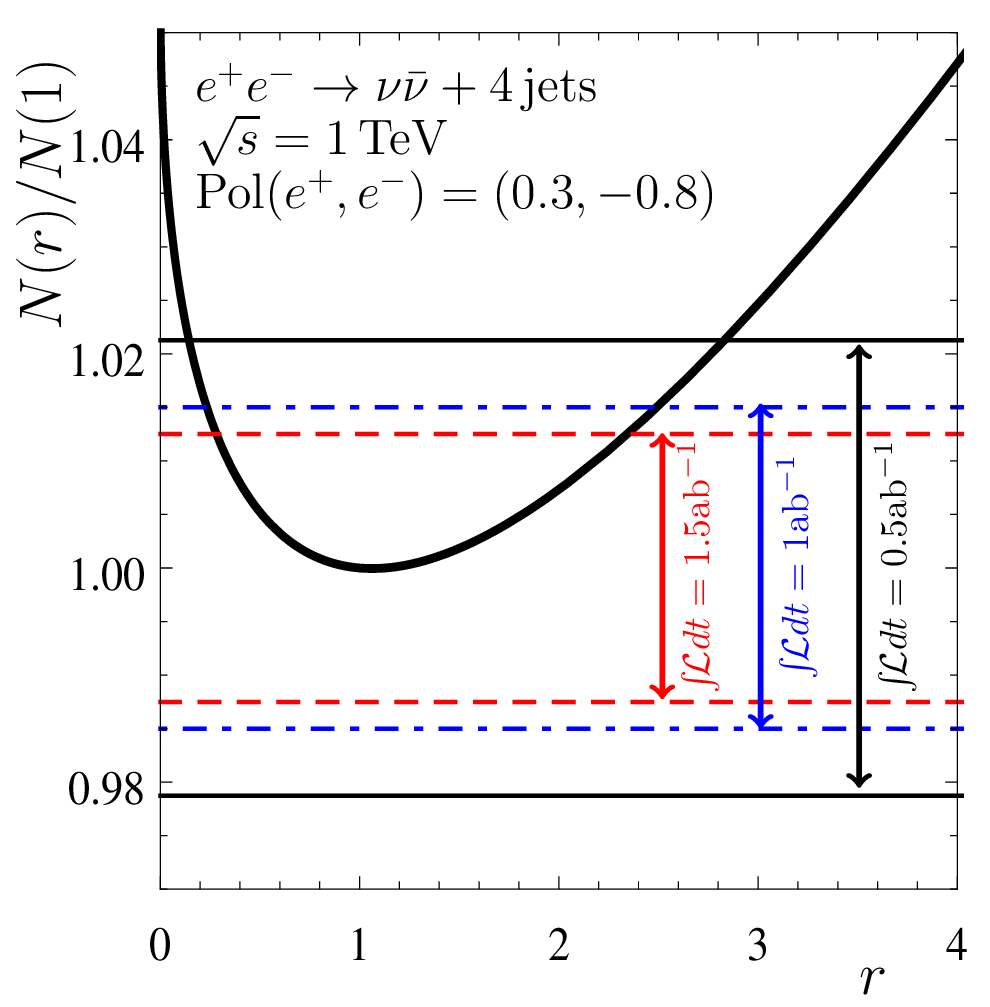} 
\end{center}
\vspace{-0.7cm}
\caption{Normalised event rates $N(r)/N(1)$ as a function of $r$
for the process $e^+e^-\rightarrow \nu\bar\nu +4$jets for $\sqrt{s}=1$\,TeV
and a fixed polarisation with $95$\% uncertainty bands for different
integrated luminosities.}
\label{fig:vvjjjjwidth} 
\end{figure}

In order to investigate the sensitivity to set bounds on $r$ with this
method, we perform a simplistic Bayesian approach: the probability $P(N(r)|N_{\rm obs})$
with $N(r)$ being the expected number of events and $N_{\rm obs}$ the observed number of events
is related to $\mathcal{P}(N_{\rm obs}|N(r))$ through a prior $\pi(N(r))$, which we
assume to be constant in the region of $r$ in the vicinity of $r=1$
which we are considering here. We furthermore assume that the events are
distributed according to a Poisson distribution
\begin{align}
 \mathcal{P}(N_{\rm obs}|N(r))=\frac{e^{-N(r)}(N(r))^{N_{\rm obs}}}{N_{\rm obs}!} 
\end{align}
and that the observed rate equals the \sm{} rate, i.e.\ $N_{\rm obs}=N(1)$. 
Accordingly, values of $r$ are excluded in this way if $N_{\rm obs}(r)$ lies outside of the 
$95$\% band of the Poisson distribution $\mathcal{P}(N_{\rm obs}|N(r))$. The
corresponding exclusion limits for $r$ are also shown in
\tab{tab:r1r2nunubar}. The interference term~$I$ lowers the sensitivity to $r$
even for quite high statistics as it can be seen from \fig{fig:vvjjjjwidth}, 
where the exclusion limits on $r$ are shown for three values of the
integrated luminosity at $\sqrt{s} = 1$\,TeV.
The minimum of $N(r)$ is in the vicinity of $r=1$,
so that a measurement of $N(r)$ in this region has the least sensitivity to
$r$. If $N(r)$ differs sufficiently from the minimum value, a high-precision
measurement of $N(r)$ could result in a two-fold ambiguity in $r$.
The latter might only be resolved within this method
by taking into account different final states.

\begin{table}[htp]
\begin{center}
\begin{tabular}{| c || c | c | c |}
\hline
$\sqrt{s}$      & $350$\,GeV & $500$\,GeV \\\hline\hline
$N_0$ ($\lumi=1$\,ab$^{-1}$)          & $430$    & $1024$  \\\hline
$R_1$           & $0.026$      & $0.006$    \\\hline
$R_2$           & $0.005$      & $0.006$    \\\hline
Limit on $r$ ($\lumi=1$\,ab$^{-1}$)   & $9.5$     & $15$  \\\hline\hline
Limit on $r$ ($\lumi=1.5$\,ab$^{-1}$) & $5.4$     & $8.2$  \\\hline
\end{tabular}
\end{center}
\vspace{-0.5cm}
\caption{$N_0$, $R_1$ and $R_2$ as a function of the \cms{} energy
for $e^+e^-\rightarrow\mu^+\mu^- +4$\,jets with $\mj>130$\,GeV.
The upper limits on $r$ at $95$\% have been obtained according to our simplistic Bayesian approach, 
using the assumptions specified in the text.}
\label{tab:r1r2ee}
\end{table}

For the process $e^+e^-\rightarrow \mu^+\mu^-+4$\,jets the situation is
different, since for this process the interference
term is positive and also no background events of the type $N_B$ as
specified in \eqn{eq:nrevents} need to be considered.
The corresponding results are shown in \tab{tab:r1r2ee}.
However, for this process the achievable 
statistics limits the sensitivity to the Higgs width via this method.

In total we conclude from this investigation that the 
comparison of on-shell and off-shell contributions at a linear collider 
provides constraints on the Higgs width that are complementary to the ones
that can be obtained from the $Z$ recoil method yielding absolute
branching-ratio measurements in combination with the determination of a
partial width.
The numbers obtained from our simplistic approach
can certainly be improved by a better suited analysis identifying
intermediate states, choosing different polarisations of the initial
state and applying more sophisticated cuts.
It can however be inferred that this method, besides relying heavily on 
theoretical assumptions, requires very high statistics and is limited by the negative interference term.
Thus, we find that the approach based on the 
absolute branching-ratio measurements from the $Z$ recoil
remains the by far superior method for determining the Higgs
width at a linear collider, both because of its model-independence and the
much higher achievable precision.

\begin{figure}[ht]
\begin{center}
\begin{tabular}{cc}
\includegraphics[width=0.4\textwidth]{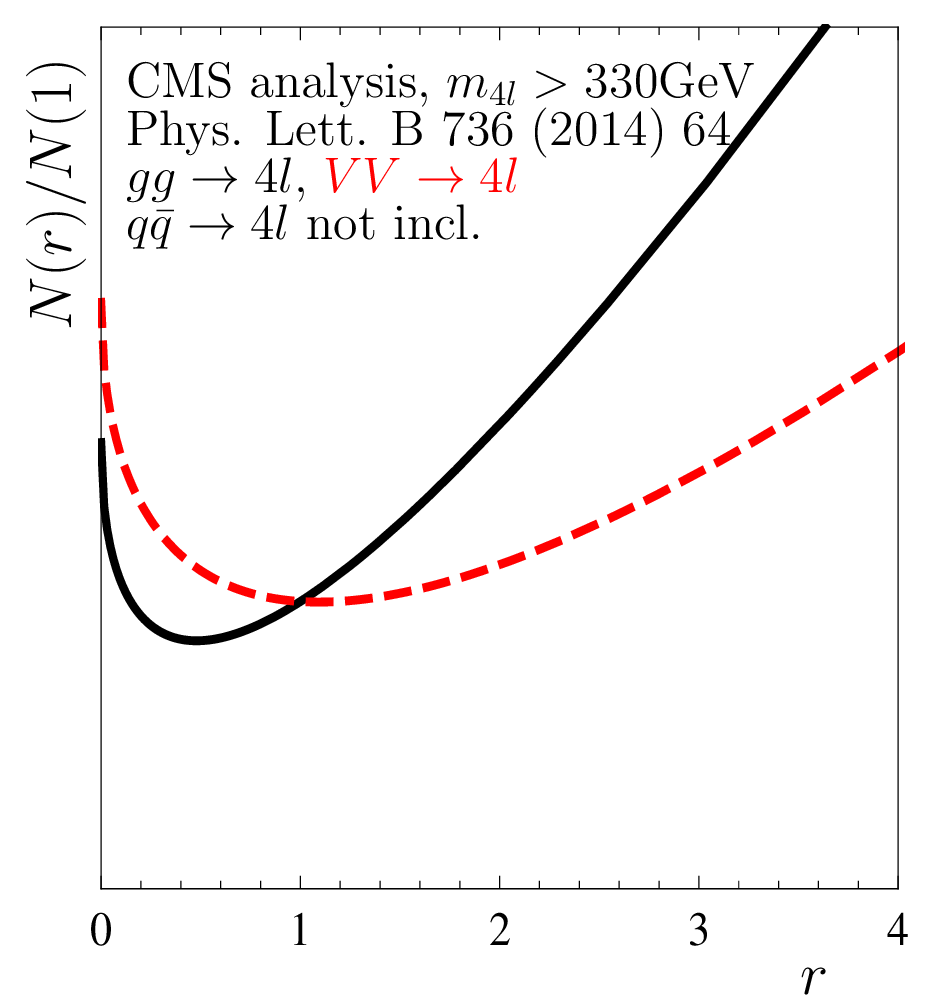} &
\includegraphics[width=0.4\textwidth]{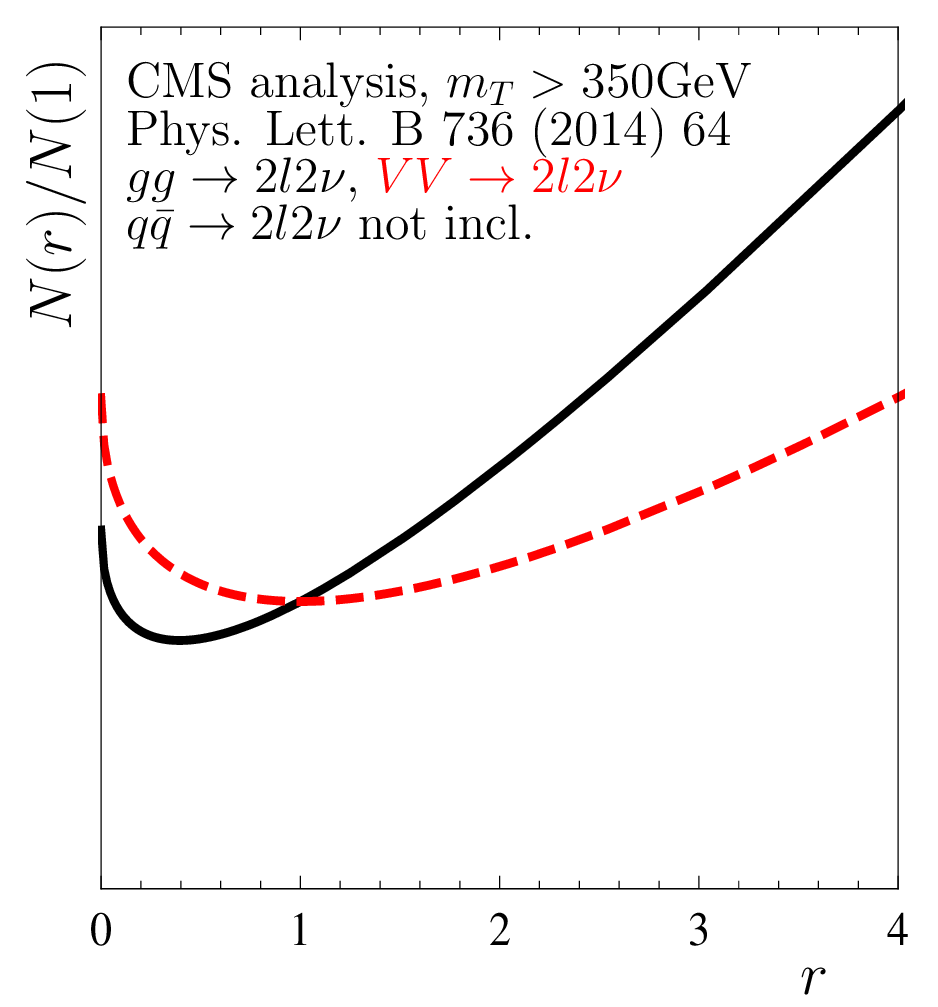} \\[-0.5cm]
 (a) & (b)
\end{tabular}
\end{center}
\vspace{-0.6cm}
\caption{Normalised event rates $N(r)/N(1)$ as a function of $r$
from the CMS analysis presented in \citere{Khachatryan:2014iha} 
for the final states (a) $4l$ and (b) $2l2\nu$. The black solid curves
show the gluon fusion production process, while the red dashed
curves indicate the vector-boson fusion production process.
In case (a) a likelihood
discriminant characterizing the event topology and the cut on the invariant
mass of the four leptons $m_{4l}>330$\,GeV was applied, in case (b) 
a transverse mass of $m_T>350$\,GeV and a missing energy of 
$E_T^{\rm miss}>100$\,GeV was required.
Since background events have been omitted in those plots, no scale is
indicated on the $y$-axis.
More details can be found in \citere{Khachatryan:2014iha}.}
\label{fig:CMS} 
\end{figure}

The qualitative features of our analysis for a linear collider can also
be applied to the case of the \lhc.
In \fig{fig:CMS} we show the normalised event rates $N(r)/N(1)$ 
from the CMS analysis presented in \citere{Khachatryan:2014iha} 
for the four lepton final state ($4l$) as well as for the
two lepton and two neutrino final state ($2l2\nu$) in dependence of 
the production mechanism after applying suitable cuts. 
Since background events have been omitted in those plots, no scale is
indicated for $N(r)/N(1)$.
The $4l$ final state just includes $H\rightarrow ZZ^{(*)}$ contributions,
whereas the $2l2\nu$ final state also contains
$H\rightarrow WW^{(*)}$ contributions.
All curves show a behaviour that is very similar to what we found in
our analysis for a linear collider.
With increasing statistics the sensitivity to $r$ in the vicinity
of $r = 1$ is also considerably reduced at the \lhc{}
due to the negative interference term.
Similar conclusions are obtained from the ATLAS result~\cite{ATLASwidth}
as well as from the various theoretical
works~\cite{Caola:2013yja,Campbell:2013una,Campbell:2013wga,Campbell:2014gha}.

\section{Effects of the heavy Higgs in a \thdm{}}
\label{sec:2hdm}

In this section we address interference effects of an on-shell heavy Higgs
with the off-shell contributions of a \sm-like light Higgs in the context of
a 2-Higgs-Doublet model (\thdm{}). Studies for a singlet extension of
the \sm{} in gluon fusion at proton colliders were carried out in
\citeres{Maina:2015ela,Kauer:2015hia}, and significant interference
effects dependent on the admixture of the singlet and the \sm{} Higgs
doublet were found.
The authors of {\tt VBFNLO}~\cite{Baglio:2014uba} studied the
interference of an off-shell Higgs with a second Higgs
in vector boson fusion at proton colliders. By an appropriate
choice of the couplings the generic two Higgs model of
{\tt VBFNLO} can be used for the description of the light and heavy Higgs of a \thdm{}.
The scheme-dependence of parametrising the width for a heavy Higgs 
in the context of a Higgs portal scenario was discussed in \citere{Englert:2015zra}.

The introduction of a second Higgs doublet is an obvious possibility
for extending the \sm{} Higgs sector, for reviews we
refer to \citeres{Gunion:1989we,Akeroyd:1996he,Akeroyd:1998ui,Aoki:2009ha,Branco:2011iw,Craig:2012vn}.
In our analysis we assume \cp{} conservation in the Higgs sector and
the absence of tree-level flavour-changing neutral currents. In this case
the neutral Higgs sector consists of two \cp{}-even Higgs bosons~$h$ and $H$
with $m_h<m_{H}$ and one \cp{}-odd Higgs boson $A$.
One distinguishes four types of \thdm{}s
according to the structure of the Yukawa couplings.
Before presenting results in the context of \thdm{}s we shortly state
our notation:
The two Higgs doublets $H_1$ and $H_2$ acquire vacuum expectation
values $v_1$ and $v_2$, whose ratio is defined as $\tan\beta\equiv v_2/v_1$.
The mixing angle rotating the neutral components $H_1^0$ and $H_2^0$
to mass eigenstates $h$ and $H$ is called $\alpha$. Then the couplings
of $h$ and $H$ to the gauge bosons are given by
\begin{align}
 g_{hVV}=\sin(\beta-\alpha)g_{HVV}^{\text{SM}}\quad \text{and}\quad
 g_{HVV}=\cos(\beta-\alpha)g_{HVV}^{\text{SM}}\quad,
\end{align}
where $g_{\hvv}^{\text{SM}}$ denotes the coupling to the \sm{} Higgs.
For the light Higgs boson to be \sm{}-like, $|\sin(\beta-\alpha)|$
has to be in the vicinity of one. As a consequence, the heavy Higgs $H$
has heavily suppressed couplings to the gauge bosons in this case. 
In the following we study the
deviations $\sin(\beta-\alpha)=\lbrace 0.95,0.98,0.99\rbrace$ from one, the latter
two providing a light Higgs boson~$h$ which is
hardly discriminable from a \sm{} Higgs boson at the \lhc{}.

Since no Yukawa couplings are involved in our two production processes
followed by $\lbrace h,H\rbrace\rightarrow VV^{(*)}$, the \thdm{} type is only of 
relevance for the total widths $\Gamma_h$ and $\Gamma_{H}$, which
enter the Breit-Wigner propagators. The \cp-odd state $A$ does not couple
to gauge bosons and therefore does not enter our calculation at lowest order
in perturbation theory, when setting external fermion masses to zero.
Our examples are all based on a type II \thdm{} (Yukawa couplings as
in the minimal supersymmetric standard model (\mssm{})) with $\tan\beta=1$.
This guarantees stability of the Higgs potential
and unitarity in gauge boson scattering, which we have checked with the help
of {\tt 2HDMC} \cite{Eriksson:2009ws}. We use the Higgs basis output
of {\tt 2HDMC} for the input file of {\tt MadGraph5\_aMC@NLO}.
For the \mssm{} a rather heavy Higgs~$H$ and a \cp-odd state~$A$ of
similar mass imply that $\cos(\beta-\alpha)$ is close to zero, which reduces the
sensitivity at a linear collider accordingly. In the \mssm{}
values of $\ma\gtrsim 300$\,GeV imply $\sin(\beta-\alpha)>0.99$ for most
of the parameter space.
A heavy Higgs can then be observed through the process
$e^+e^-\rightarrow Z^{*}\rightarrow AH$, which limits the detection
to about $m_{H}\lesssim \sqrt{s}/2$. In \citere{Hahn:2002gm} 
higher order corrections to the process $e^+e^-\rightarrow \nu\bar\nu H$
have been discussed, which can give rise to an upward shift of the
detection limit in favourable regions of the parameter space.

\begin{figure}[htp]
\begin{center}
\begin{tabular}{cc}
\includegraphics[width=0.4\textwidth]{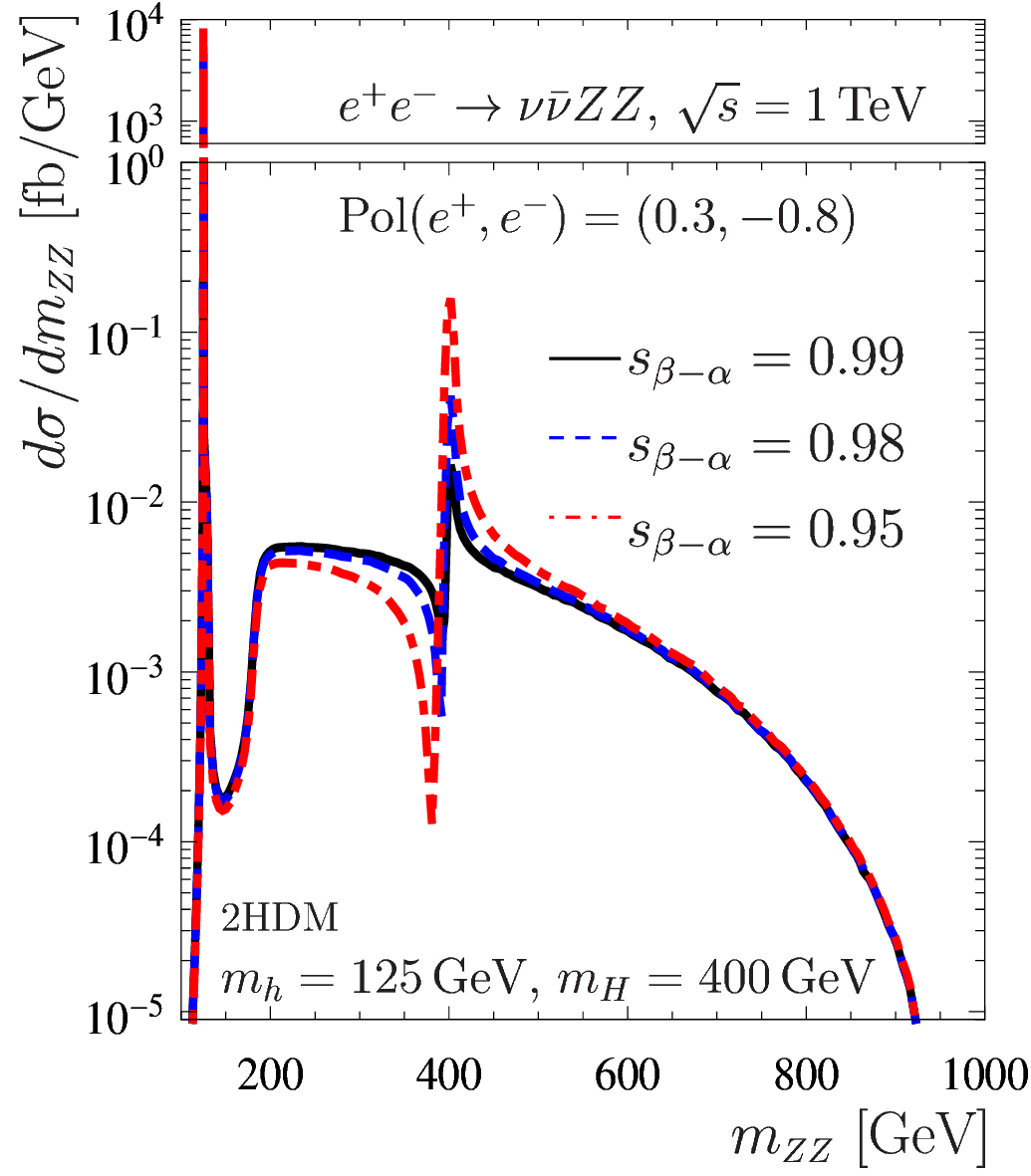} &
\includegraphics[width=0.4\textwidth]{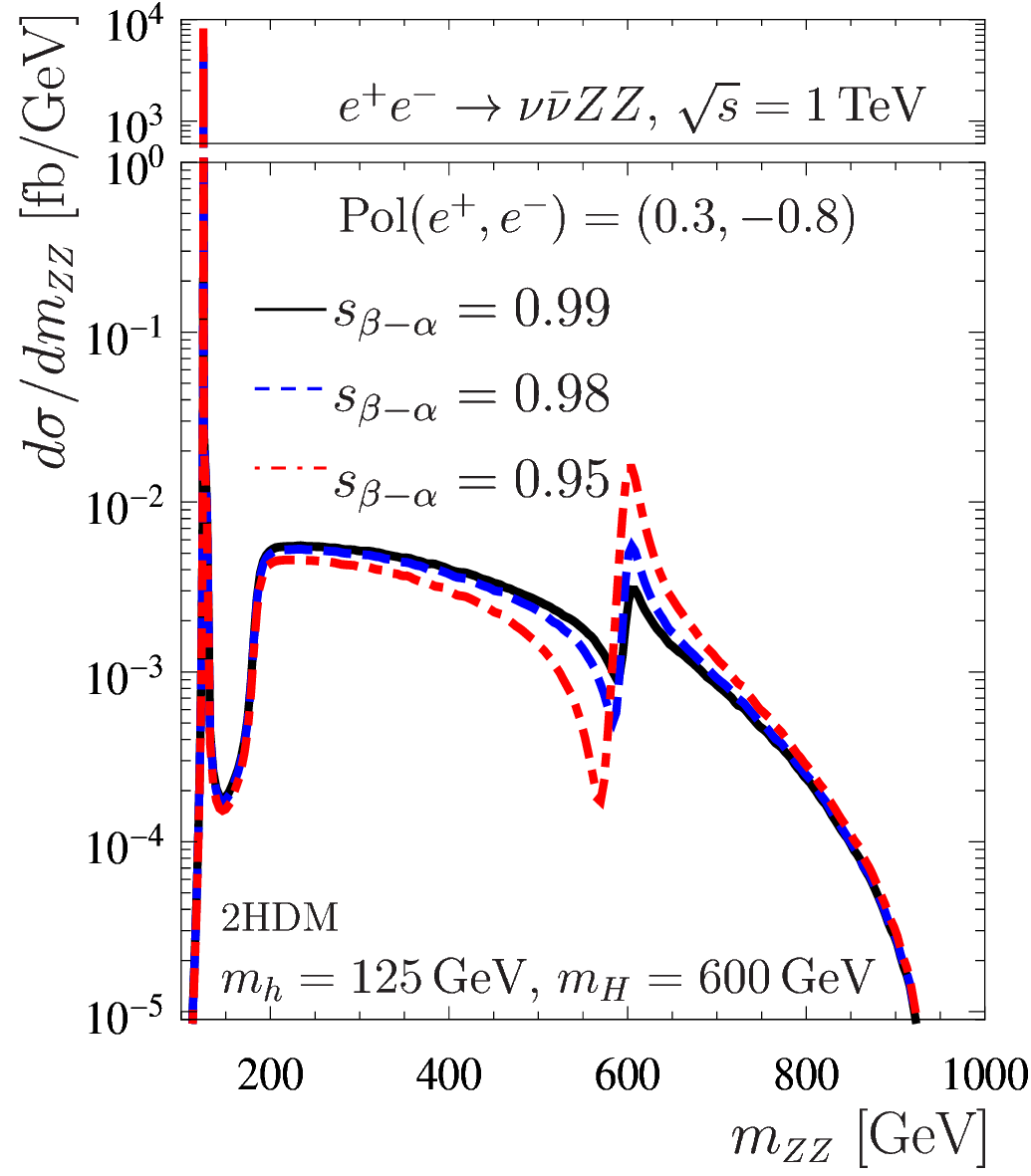} \\[-0.5cm]
 (a) & (b) \\
\end{tabular}
\end{center}
\vspace{-0.6cm}
\caption{
$d\sigma/d\mzz$ in fb/GeV as a function of $\mzz$ in GeV for $\sqrt{s}=1$\,TeV
and a fixed polarisation Pol$(e^+,e^-)=(0.3,-0.8)$ for the process
$e^+e^-\rightarrow\nu\bar\nu \lbrace h,H\rbrace\rightarrow \nu\bar\nu ZZ$
in the context of a type II \thdm{} with $\tan\beta=1$ and three different
values of $s_{\beta-\alpha}:=\sin(\beta-\alpha)$ for
the two mass scenarios (a) $m_{H}=400$\,GeV and (b) $m_{H}=600$\,GeV.
}
\label{fig:2HDM1} 
\end{figure}

In the following we consider the process
$e^+e^-\rightarrow\nu\bar\nu \lbrace h,H\rbrace\rightarrow \nu\bar\nu ZZ$
in the \thdm{} and investigate the impact of the interference between
the contributions of a heavy Higgs $H$ and a \sm{}-like light Higgs $h$ (with
$m_h = 125$\,GeV) on the sensitivity for detecting the heavy Higgs at a
linear collider. We use $\sqrt{s} =  1$\,TeV with a polarisation of 
Pol$(e^+,e^-)=(0.3,-0.8)$ and choose the two scenarios
(a) $m_{H}=400$\,GeV, $m_A=360$\,GeV, $m_{H^\pm}=440$\,GeV and
(b) $m_{H}=600$\,GeV, $m_A=560$\,GeV, $m_{H^\pm}=640$\,GeV,
where for the latter case $m_{H}$ lies beyond the kinematic
reach of the $HA$ pair production process. For 
$\sin(\beta-\alpha)$ we consider the three scenarios
$\sin(\beta-\alpha)=\lbrace 0.95,0.98,0.99\rbrace$.
\fig{fig:2HDM1} shows the $\mzz$ invariant mass distribution arising from
this process. Below the threshold for on-shell production of the heavy Higgs~$H$
the distribution closely resembles the case of a \sm{} Higgs at $125$\,GeV.
The peak for the on-shell production of the light Higgs $h$ and a continuum
of off-shell contributions are clearly visible. At $\mzz = m_H$ the
distribution shows a resonance-type behaviour with a significant
interference contribution from the light Higgs.
Since the heavy Higgs is only observable in $H\rightarrow VV$ for 
non-vanishing $\cos(\beta-\alpha)$, the effect on the $\mzz$ distribution in
the plot is largest for $\sin(\beta-\alpha) = 0.95$ and gets reduced as 
$\sin(\beta-\alpha)$ approaches unity. The shape of the $\mzz$ distribution
is furthermore affected by the total width $\Gamma_{H}$. 
The values for the heavy Higgs width $\Gamma_{H}$ as obtained
by {\tt 2HDMC} are given in \tab{tab:H0width} for the scenarios considered here.
For the heavy Higgs with mass $m_{H}=600$\,GeV the width exceeds $10$\,GeV,
which results in a rather broad peak and $h-H$ interference structure. 
For invariant masses $\mvv>m_{H}$ the distribution receives off-shell
contributions from both Higgs bosons. For high invariant masses the
contributions of the two Higgs bosons add in such a way that they unitarize
this process in the same way as the single contribution from a \sm{} Higgs.
Similar effects as the ones discussed here can also be expected 
at the \lhc{} if the \sm-like Higgs at $125$\,GeV is supplemented with
a heavier neutral Higgs with suppressed couplings to gauge bosons (for a
discussion of the vector boson fusion process
at proton colliders with the help of {\tt VBFNLO} see
\citere{Baglio:2014uba}).

\begin{table}[htp]
\begin{center}
\begin{tabular}{| c || c | c | c |}
\hline
$\Gamma_{H}$       & $s_{\beta-\alpha}=0.95$ & $s_{\beta-\alpha}=0.98$ & $s_{\beta-\alpha}=0.99$ \\\hline\hline
$m_{H}=400$\,GeV   & $4.30$\,GeV  & $3.21$\,GeV  & $2.90$\,GeV \\
$m_{H}=600$\,GeV   & $19.1$\,GeV  & $16.4$\,GeV  & $16.1$\,GeV \\\hline
\end{tabular}
\end{center}
\vspace{-5mm}
\caption{Heavy Higgs width $\Gamma_{H}$ for the scenarios under consideration.}
\label{tab:H0width}
\end{table}

\begin{figure}[htp]
\begin{center}
\begin{tabular}{cc}
\includegraphics[width=0.4\textwidth]{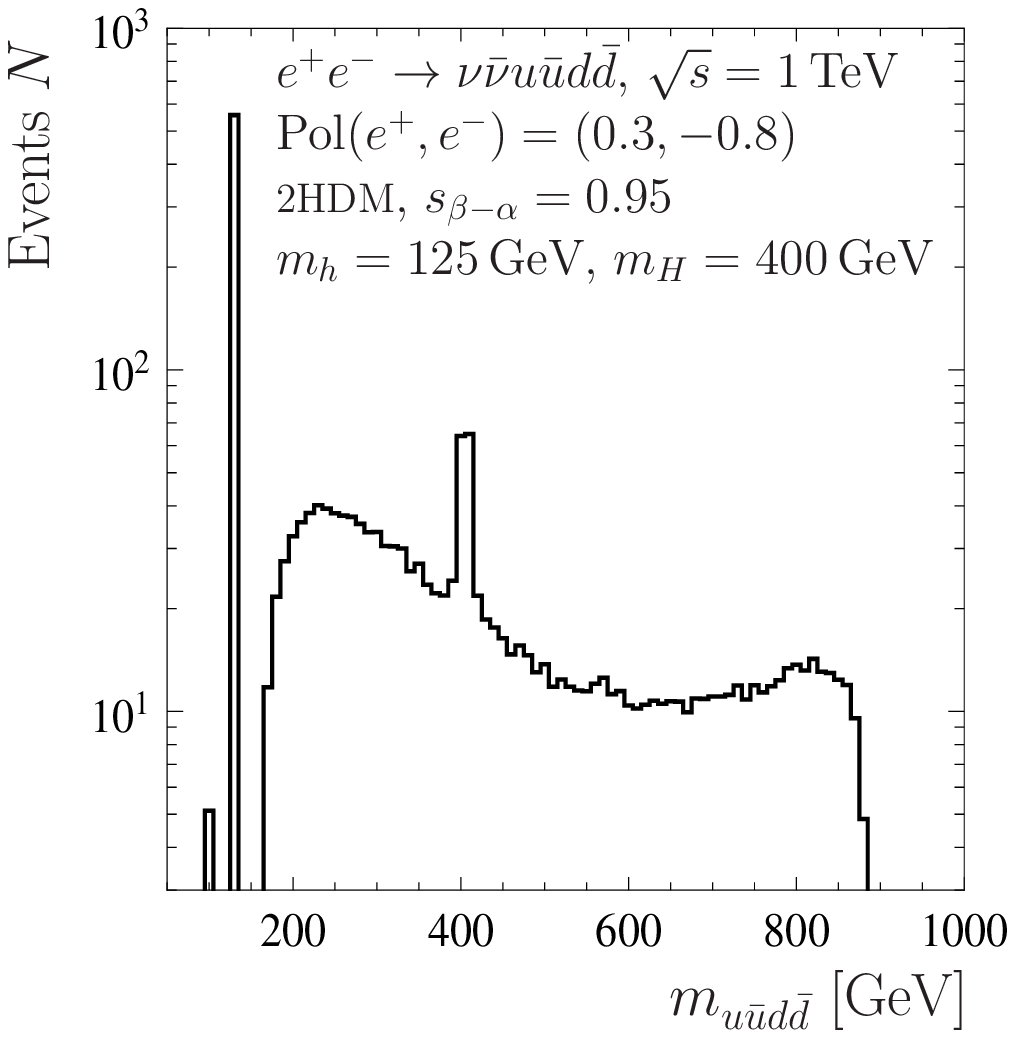} &
\includegraphics[width=0.4\textwidth]{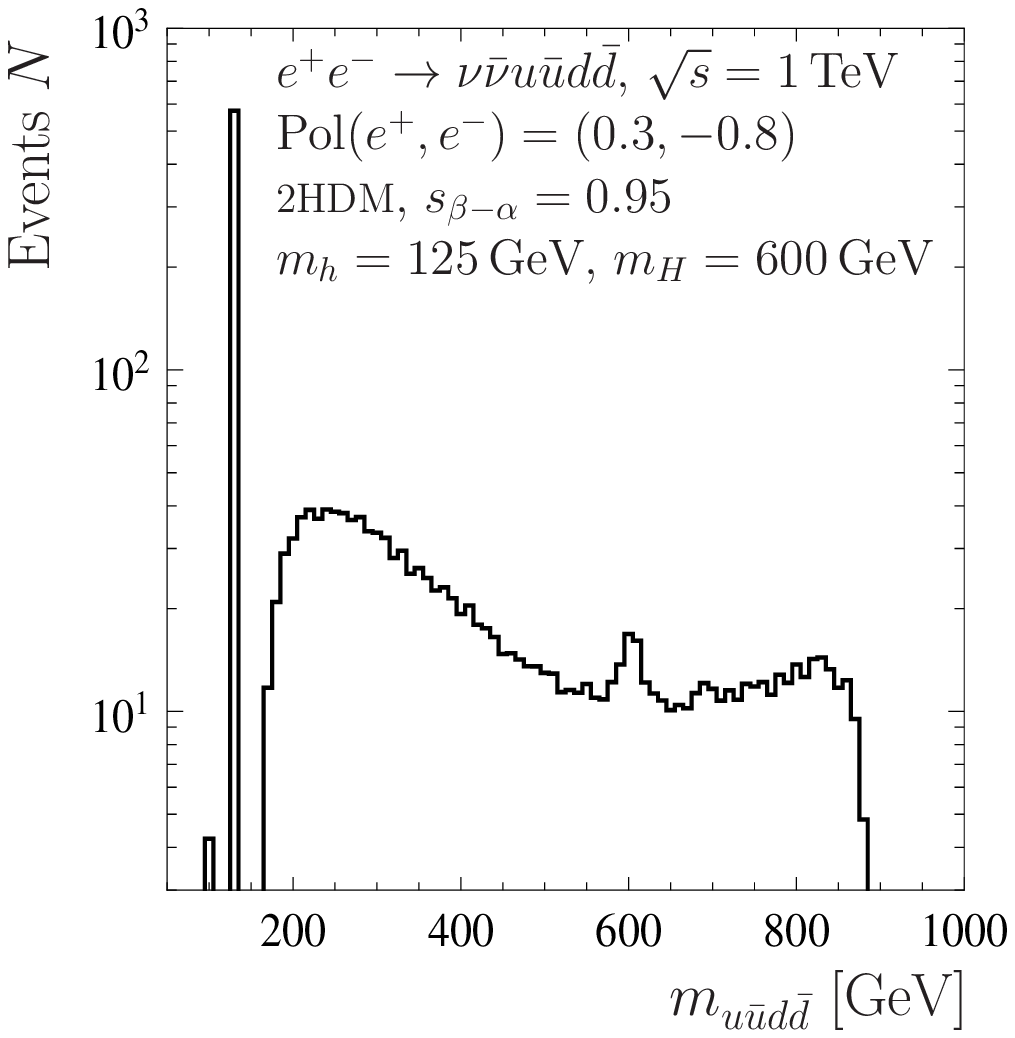} \\[-0.5cm]
 (a) & (b) \\
\includegraphics[width=0.4\textwidth]{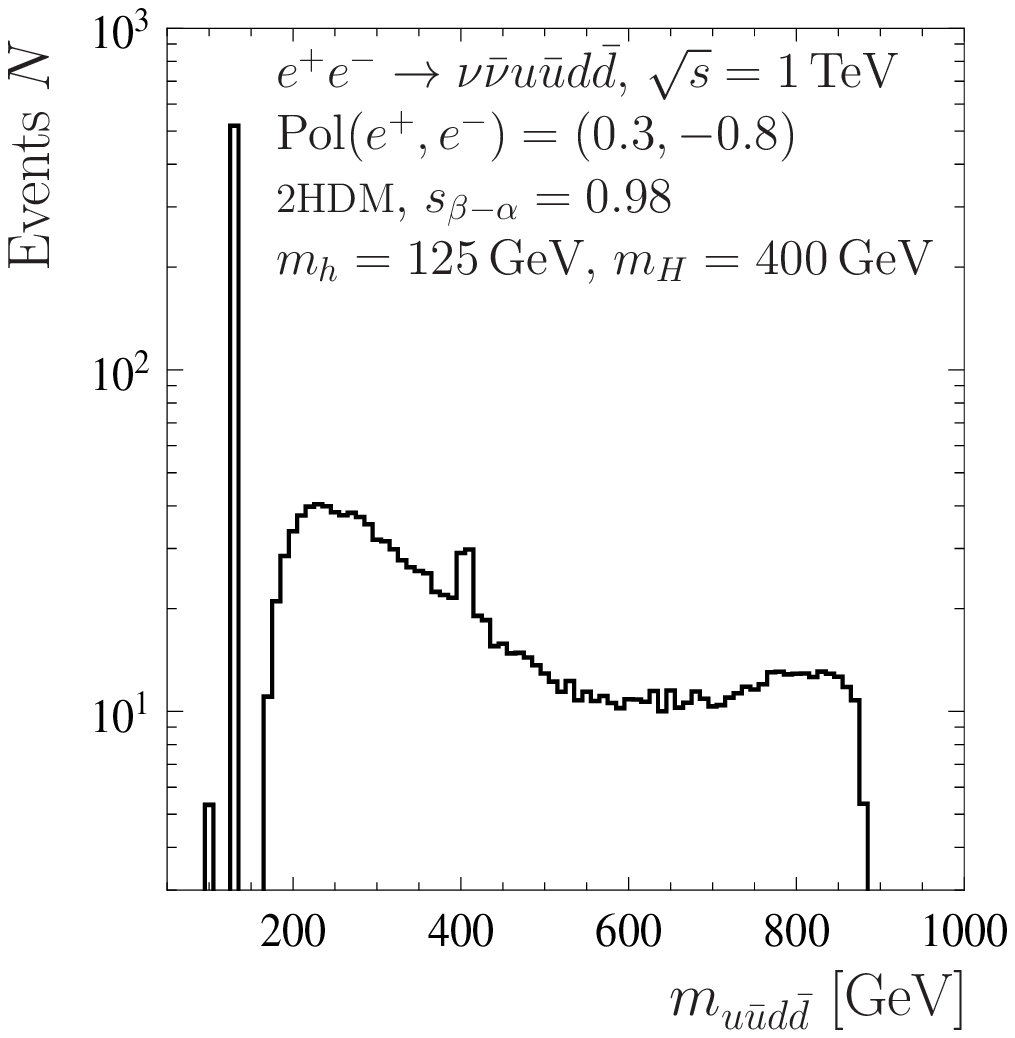} &
\includegraphics[width=0.4\textwidth]{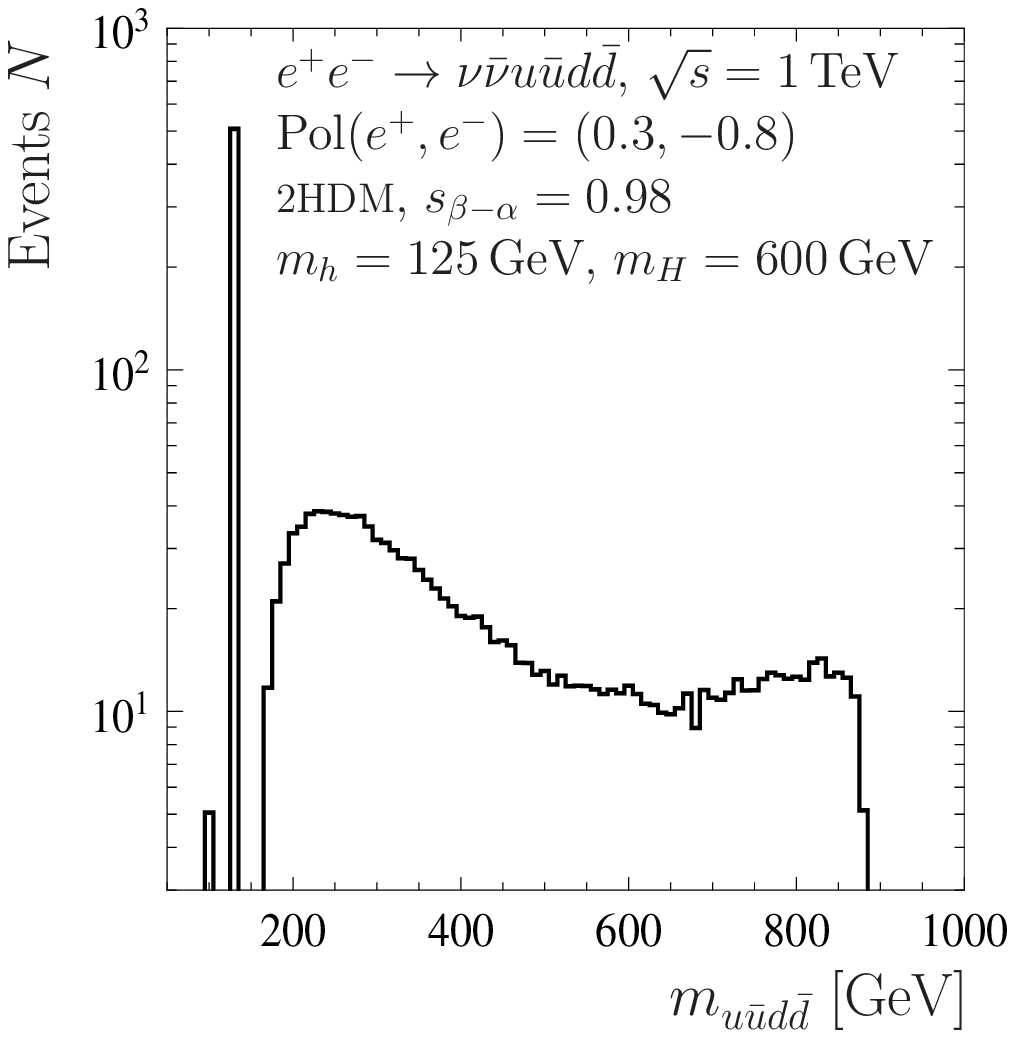} \\[-0.5cm]
 (c) & (d) \\
\includegraphics[width=0.4\textwidth]{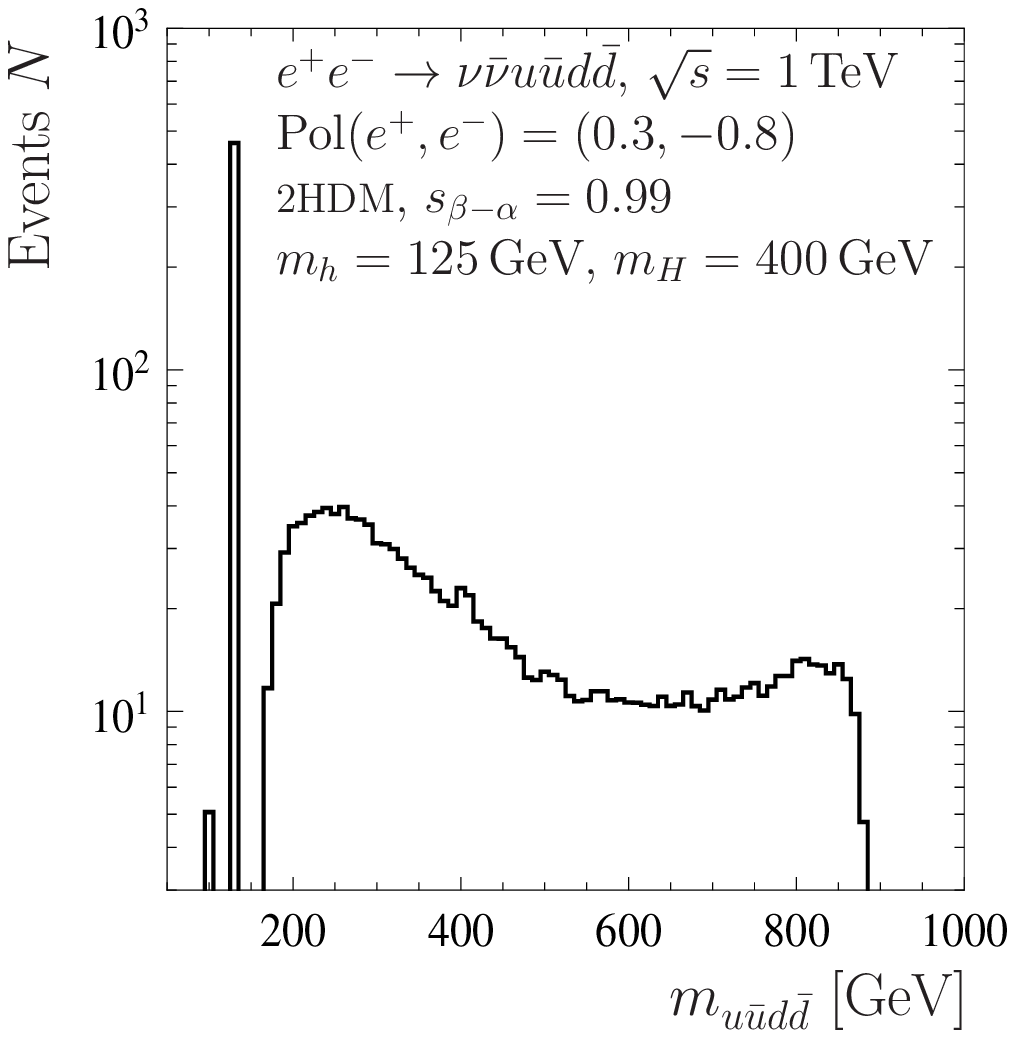} &
\includegraphics[width=0.4\textwidth]{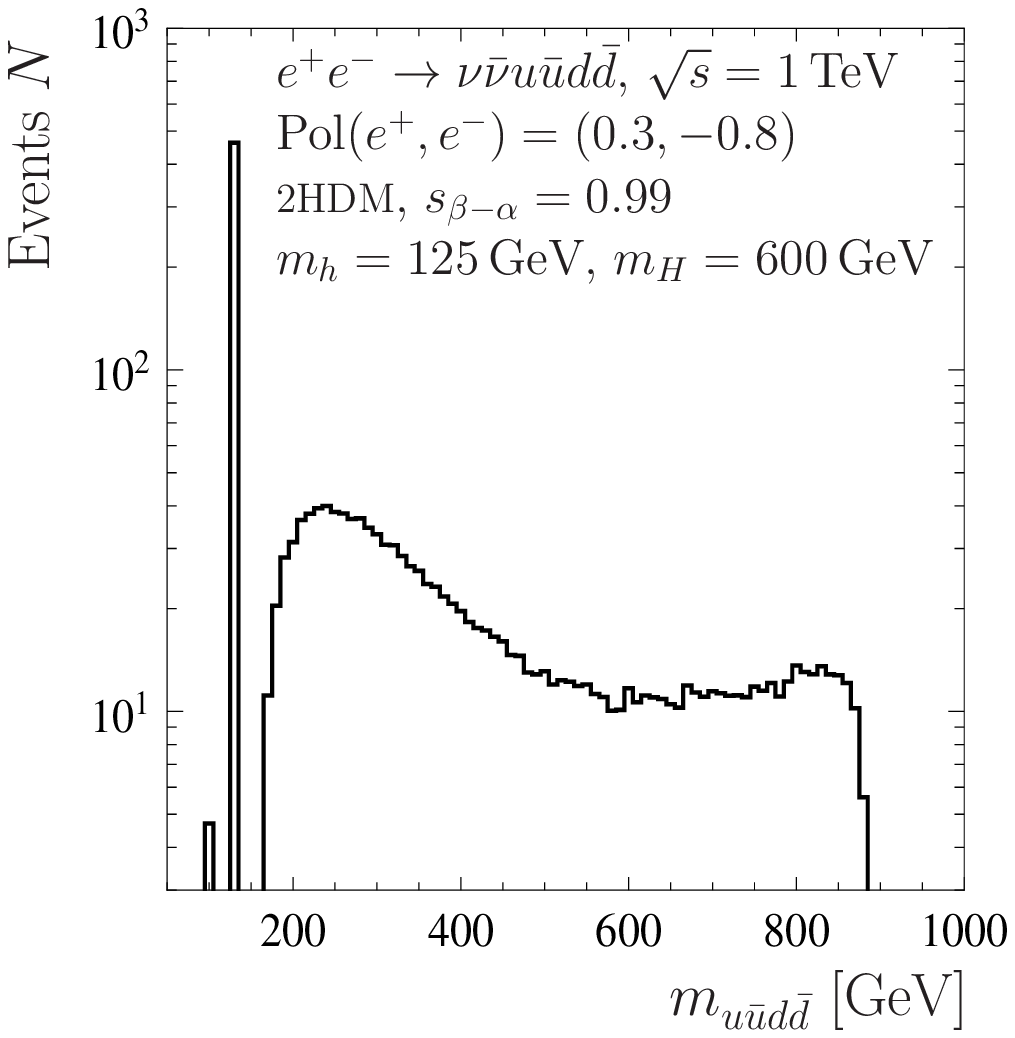} \\[-0.5cm]
 (e) & (f) \\
\end{tabular}
\end{center}
\vspace{-0.6cm}
\caption{Event rates for $e^+e^-\rightarrow \nu\bar\nu u\bar u d\bar d$ for $\sqrt{s}=1$\,TeV 
and $\lumi=500$\,fb$^{-1}$
after the cut $\ptj>75$\,GeV
as a function of the invariant mass of the $4$\,jets $m_{u\bar u d\bar d}$
in the context of a type II \thdm{} with $\tan\beta=1$ for
different values of (a,b) $s_{\beta-\alpha}:=\sin(\beta-\alpha)=0.95$;
(c,d) $s_{\beta-\alpha}=0.98$ and (e,f) $s_{\beta-\alpha}=0.99$ and the two
mass scenarios (a,c,e) $m_{H}=400$\,GeV and (b,d,f) $m_{H}=600$\,GeV.
}
\label{fig:2HDM2} 
\end{figure}

In order to discuss the prospects at the linear collider for this
scenario in a more quantitative way, we now also incorporate background
contributions into our analysis. Specifically we consider
the process $e^+e^-\rightarrow \nu\bar\nu u\bar u d\bar d$, being
a subprocess of $e^+e^-\rightarrow \nu\bar\nu +4$\,jets which includes
$\lbrace h,H\rbrace\rightarrow ZZ^{(*)}/WW^{(*)}\rightarrow u\bar u d\bar d$. 
Similarly to \sct{sec:example1}
we cut on the transverse momentum of the $4$\,jets and require it
to be larger than $75$\,GeV. The result of our study can be found in \fig{fig:2HDM2}.
The narrow peak for $m_{H}=400$\,GeV can be seen by eye in all
cases $\sin(\beta-\alpha)=\lbrace 0.95,0.98,0.99\rbrace$.
For $m_{H}=600$\,GeV again the peak broadens out. However,
in all cases a side-band analysis of the background should have good
prospects to reveal the heavy Higgs peak
and the effect of the $h-H$ interference.
The sensitivity for detecting an additional Higgs boson with suppressed
couplings to gauge bosons at a linear collider can of course be enhanced by
taking into account fermionic decays (and production modes where
the additional Higgs boson is radiated off a fermion). We leave this topic
for future work.

\section{Conclusions}
\label{sec:conclusions}

We have investigated the impact of off-shell effects for Higgs
production at a linear collider via the two production 
processes $e^+e^-\rightarrow ZH$ and $e^+e^-\rightarrow \nu\bar\nu H$
and the decay into a pair of gauge bosons, 
$H\rightarrow VV^{(*)}$ with $V\in\lbrace Z,W\rbrace$,
for different \cms{} energies and polarisations. The signal contributions
involving a \sm{}-like Higgs boson at $125$\,GeV
have been compared with background yielding the same final
state.
We have performed numerical simulations of the full 
processes $e^+e^-\rightarrow 6$\,fermions using
{\tt MadGraph5\_aMC@NLO} and we have discussed the possible impact of
initial state radiation and higher-order effects.

The fact that the mass of the observed Higgs boson of about $125$\,GeV
is far below the threshold for on-shell $W^+W^-$ and $ZZ$ production has the
consequence that the decay $H\rightarrow VV^{*}$ of an on-shell Higgs
boson suffers from a significant phase-space suppression. This implies on
the one hand that the partial width $H\rightarrow VV^{*}$, where $H$ is
on-shell, depends very sensitively on the precise numerical value of the 
Higgs-boson mass. On the other hand, contributions of an off-shell Higgs
where $VV$ are both on-shell are relatively large. This qualitative feature
is reflected in our numerical analysis. The relative importance of
contributions of an off-shell Higgs boson increases with increasing
\cms{} energy. For $\sqrt{s}>500$\,GeV those off-shell contributions 
to the total Higgs induced cross section are of
$\mathcal{O}(10\%)$. The dependence on the precise numerical value of $m_H$
is much diminished in the off-shell contributions as compared to the case of
an on-shell Higgs.

Accordingly, for the extraction of Higgs couplings to gauge bosons from 
branching ratios of $H\rightarrow VV^{*}$ a very precise measurement
of the Higgs-boson mass is needed, preferably to an accuracy of
better than $100$\,MeV, and for higher 
\cms{} energies it is important to take off-shell contributions into account.
As expected, we find that at low \cms{} energies~$\sqrt{s}$,
i.e.\ close to the production threshold, the effects of 
off-shell contributions are insignificant for the extraction of
Higgs couplings. At higher $\sqrt{s}$, however, for an accurate
determination of Higgs couplings the off-shell contributions
need to be incorporated. Those contributions can furthermore be 
utilised to extract the Higgs to gauge boson
couplings in different kinematical regimes, to check the destructive
interference with the background or to set bounds on effective operators 
and test their kinetic dependences.

A particular focus of our analysis has been on the determination of the
total width of the Higgs boson at a linear collider. We have investigated
two aspects in this context. On the one hand, we have analysed to what
extent the standard method at a linear collider, which is based on the 
$Z$ recoil method providing absolute measurements of Higgs 
branching ratios in combination with an appropriate determination of a
partial width, is affected by off-shell contributions. We have found that
at low \cms{} energies the effect of the off-shell
contributions in $H\rightarrow VV^{(*)}$ is at the sub-permil level.
At higher energies the off-shell effects can be somewhat larger and need to
be properly taken into account and/or reduced by appropriate cuts. 
In this context our analysis adds to the motivation
for performing the cross-section determination via the $Z$ recoil method
close to threshold, i.e.\ at about $\sqrt{s}=250-350$\,GeV, rather than at
higher energies where the off-shell effects become relevant.
On the other hand, we have investigated the constraints on the total width
that can be obtained from a comparison of on-shell and off-shell
contributions. At a linear collider those constraints are 
complementary to the determination of the Higgs width via the $Z$ recoil
method. However, the method based on the comparison of on-shell and off-shell
contributions has several draw-backs. Besides relying heavily on theoretical 
assumptions, this method requires very high statistics and is limited by 
the negative interference term. We therefore conclude that the standard
method at a linear collider based on the 
$Z$ recoil method is far superior for determining the Higgs
width, both because of its model-independence and the
much higher achievable precision. We have also discussed the corresponding
method at the \lhc{} and we have pointed out that the destructive
interference contribution between the Higgs-induced contributions and the
background will make it difficult to reach the sensitivity to
the SM value of the width even for high statistics.

As an example of the relevance of off-shell effects in the context of an
extended Higgs sector, we discussed the case of a 2-Higgs-Doublet model with
a \sm{}-like Higgs at $125$\,GeV and an additional heavier neutral \cp{}-even
Higgs boson with suppressed couplings to gauge bosons. We demonstrated the
importance of the interference between off-shell contributions of the light
Higgs and the on-shell contribution of the heavy Higgs. If the suppression
of the couplings of the heavy Higgs boson to gauge bosons is not too strong,
the $H\rightarrow VV^{(*)}$ channel can in this way lead to the detection of
a heavy Higgs boson at a linear collider, even beyond the kinematic limit
for producing a pair of heavy Higgs bosons, $H$ and $A$.

The analyses performed in this paper can be improved in several
respects, in particular regarding the inclusion of 
initial state radiation, beamstrahlung and higher-order effects.
Concerning the latter, we demonstrated that they can be important, in
particular in view of their dependence on the
invariant mass of the two gauge bosons.
Moreover future work might incorporate more sophisticated methods to 
identify intermediate particles or optimized cuts and statistical analyses 
to improve the sensitivity to off-shell
effects and to discriminate between the various processes.

\vspace{-1mm}
\section*{Acknowledgments}
The authors thank Nikolas Kauer for helpful comments on the manuscript.
The authors acknowledge support by
the DFG through the SFB~676 ``Particles, Strings and the Early Universe''.
This research was supported in part by the European Commission through the 
``HiggsTools'' Initial Training Network PITN-GA-2012-316704.

\end{document}